\documentclass[pdflatex,sn-mathphys]{sn-jnl}% Math and Physical Sciences Reference Style
\usepackage[numbers,sort&compress]{natbib}
\usepackage{comment}
\usepackage[font=footnotesize,labelfont=bf]{caption}
\usepackage[font=footnotesize,labelfont=bf]{subcaption}
\usepackage{booktabs} %For \toprule & \midrule
\usepackage{soul,color}
\usepackage{xcolor}
\usepackage{mhchem}%for chemical formula in bibliography
\usepackage{adjustbox}

\jyear{2021}%

\theoremstyle{thmstyleone}%
\theoremstyle{thmstyletwo}%

\theoremstyle{thmstylethree}%

\begin{document}

\title[Formation and stability of oxide perovskites]{Predicting the formation and stability of oxide perovskites by extracting underlying mechanisms using machine learning}


\author*[1]{\fnm{George Stephen} \sur{Thoppil}}\email{george.ts@iitb.ac.in}

\author[1]{\fnm{Alankar} \sur{Alankar}}\email{alankar.alankar@iitb.ac.in}
\equalcont{These authors contributed equally to this work.}


\affil*[1]{\orgdiv{Department of Mechanical Engineering}, \orgname{Indian Institute of Technology Bombay}, \orgaddress{\street{Powai}, \city{Mumbai}, \postcode{400076}, \state{Maharashtra}, \country{India}}}

%\affil[2]{\orgdiv{Department}, \orgname{Organization}, \orgaddress{\street{Street}, \city{City}, \postcode{10587}, \state{State}, \country{Country}}}

%\affil[3]{\orgdiv{Department}, \orgname{Organization}, \orgaddress{\street{Street}, \city{City}, \postcode{610101}, \state{State}, \country{Country}}}

\abstract{The optimization of properties of perovskite oxides has drawn interest on account of their diverse areas of application. In this work, the hierarchical clustering technique is used to reduce the multi--collinearity among selected features from literature that are reported to have an effect on perovskite formation and stability. Operating on the vast composition space of double oxide perovskite compositions available in literature and online repositories, in this manuscript, an attempt has been made to extract the relationship between the composition and structure to predict their formability and stability. Machine learning (ML) classifiers are trained on these datasets to predict novel stable perovskite compositions. The study uses a vast feature space to narrow down the most important factors affecting the formability and stability in perovskite compounds. It also identifies stable compositions that have band gaps suitable for photovoltaic and photocatalytic applications. The developed random forest (RF)--based models may be extended to include the implications beyond photosensitive applications by focusing on the physico--chemical mechanisms driving the phenomena behind each application.}




\maketitle
\section{\label{sec:intro}Introduction}
Amidst the ever increasing demand of energy for sustaining the affirmative industrial growth, the drive to cut fossil fuel emissions has sparked off a search for sustainable, scaleable and cost--effective solutions to generate and store energy. Among the renewable energy sources in focus, solar energy is the most plentifully available -- yet for technologies seeking to harness it, conversion efficiency and storage capacity are still areas attracting intense research and innovation. At the heart of such efforts is a class of materials known as perovskites that find use in photovoltaic cells, fuel cells, memory devices, energy--conversion catalysis, water splitting, photoelectronic devices and superconductors \cite{grinberg2013, oka2008, amgar2016, fu2019, liu2019, liu2018, nguyen2020, zhang2021, khalesi2008a, khalesi2008b}. Such varied applicability is due to exceptional physico--chemical properties such as thermal stability, redox behavior and electron mobility \cite{zhu2015}.
%%%
\par
\begin{figure}[!htbp]
    \centering
        \begin{subfigure}{0.3\textwidth}
        \centering
        \includegraphics[width=\linewidth]{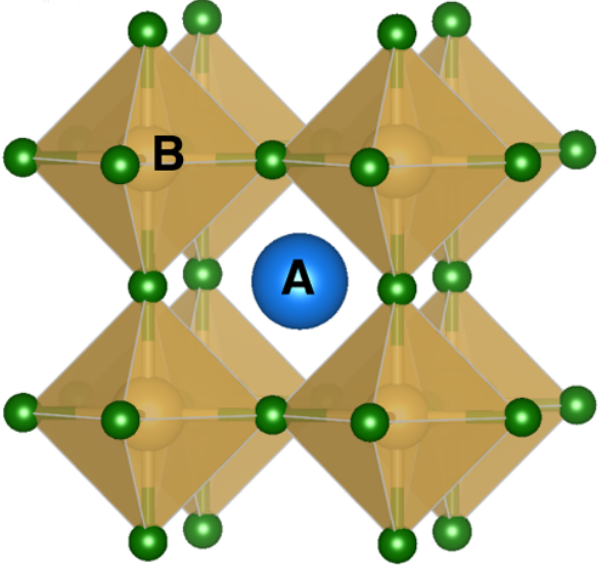}
        \vspace{-6mm}
        \caption{}\label{fig:ABX3}
        \end{subfigure}
    \hspace{5mm}
        \begin{subfigure}{0.3\textwidth}
        \centering
        \includegraphics[width=\linewidth]{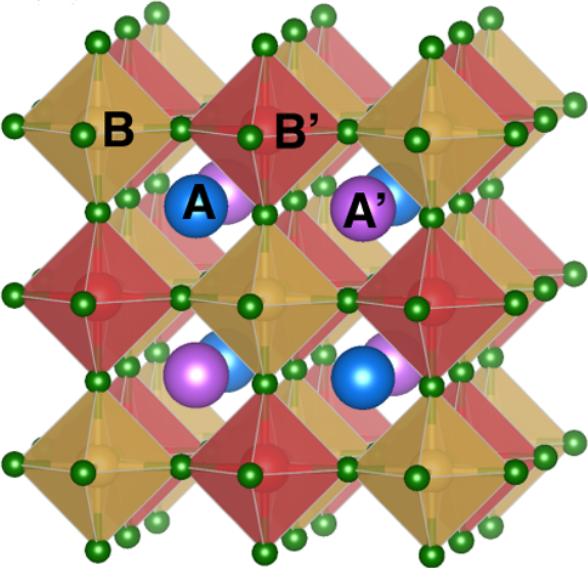}
        \vspace{-6mm}
        \caption{}\label{fig:AA'BB'X6}
        \end{subfigure}
    \caption{Crystal structure of a single \ce{ABX_3} is shown in (a) and that of a double \ce{AA$'$BB$'$X_6} perovskite having different cations in the A-- and B--sublattices \cite{talapatra2021} is shown in (b). Reprinted with permission from ACS Publications: Chemistry of Materials.}
    \label{fig:FeatureImportancesOriginalFormability}
\end{figure}

 Perovskites are, by definition, materials with a crystal structure similar to \ce{CaTiO_3}, \ce{CaSiO_3}, \ce{SrTiO_3} or \ce{BaTiO_3} \cite{nkwachukwu2021}. The perovskite mineral was identified in 1839 by German crystallographer Gustav Rose in samples from the Urals sent by Russian mineralogist Alexander K{\"a}mmerer and named in honor of Russian mineralogist Count von Perovski \cite{park2016, katz2020}. Typically, these compounds are with an $\ce{ABX_3}$ stoichiometry, where $A$ could be an alkaline, alkaline--earth or lanthanide cation, $B$ is a metal with 3d, 4d or 5d configuration and $X$ is a halogen, nitrogen or oxygen \cite{arandiyan2018}. The ideal perovskite crystal structure has a cubic 3--dimensional framework with corner--sharing $\ce{BX_6}$ octahedra \cite{talapatra2021} as shown in Fig. \ref{fig:FeatureImportancesOriginalFormability}. Differences in the cation radii can cause distortion in the idealized primitive cubic structure involving octahedral tilting of the $BX_6$ units \cite{kong2019}. Also, populating the A and B sites with different metal ions can cause perovskite oxides to behave as proton conductors, mixed ionic--electronic conductors, oxygen conductors or catalysts for oxygen reduction, oxygen evolution, hydrogen evolution, multi--functional or redox reactions \cite{zhang2021}.  Altering the composition by either replacing A or B or both, allows for the optimization of physical properties towards a wide range of applications.
%%%
%%%
The spacegroup of an ideal perovskite structure is $Pm-3m$ (221) and multiple crystallographic variants of this basic structure occur when a symmetry operation in the spacegroup is constrained. This could occur due to one of the following mechanisms \cite{sun2020}:
\begin{enumerate}
    \item Octahedral distortion: The metallic ion's electronic instability can lead to Jahn--Teller type distortion in the $BX_6$ octahedron \cite{okazaki1960}.
    \item Cation displacement within the octahedra: due to similar electron instability of the B metallic ion \cite{shirane1957}.
    \item Octahedral tilting or rotation: due to large disparity in the sizes of the A, B and X ions.
\end{enumerate}
%%%
Goldschmidt \cite{goldschmidt1926} proposed a tolerance factor $t$, to quantify the disparity in ion sizes, which is the ratio of the average distances A–X and B–X given by $t = (r_A + r_X)/\sqrt{2} (r_B + r_X)$ where $r_A$, $r_B$ and $r_X$ are the ionic radii of the A--site cation, B--site cation and X anion respectively. A perfect cubic structure is obtained when $t=1$ and distortions occur when $t<1$. Most perovskites reported in literature have $0.75<t<1.00$. For describing the perovskite structure further, Li et al. \cite{li2008} proposed the octahedral factor $\mu = r_B/r_X$.
%%%
\par Structure maps using $t$,  $\mu$ and bond--lengths have been commonly used to predict the formability of $ABX_3$ perovskites \cite{lufaso2001, li2004, zhang2007}. Filip and Giustino \cite{filip2018} were able to predict the formability of $A_2BB'X_6$--type perovskites with an accuracy of 80\% using an additional parameter $\bar{\mu}_B = \lvert r_B - r_{B'}\rvert/2r_X$. Bartel et al. \cite{bartel2019} proposed a 1--dimensional tolerance factor $\tau = \frac{r_X}{r_B} - n_A \left (n_A - \frac{r_A/r_B}{\ln{r_A/r_B}}\right)$ which was able to predict perovskite formability with 92\% accuracy using a dataset of 576 $ABX_3$ materials, $n_A$ being the oxidation state of ion $A$, $r_A$ an  d $r_B$ are the ionic radii of ions $A$ and $B$ ($r_A > r_B$).
%%%
\par The concept of materials informatics have enabled the exploration of hyper--dimensional structure maps to predict formability in perovskites \cite{tao2021a, talapatra2021, tao2021b, ihalage2021, li2021}. Morgan et al. \cite{morgan2018} performed ML based studies on the stability of perovskite oxides by training \ce{ABX3} compositions against the energy above convex hull data obtained from the Materials Project database \cite{matproj2013}. In their study the compositions within a $\pm$28 $meV/atom$ window were classified as stable. Talapatra et al. \cite{talapatra2021} used a feature set of 28 atom--specific and  geometrical properties, including the ones mentioned already. The authors trained the models to classify an exhaustive general dataset of $AA'BB'O_6$ compounds into formable and stable oxide perovskites. The energy--above--convex--hull data for the stability criterion was obtained from the Materials Project database.
%%%
\par In this work we attempt a more general approach to study the stability of perovskites based on a feature set generated from elemental properties and empirical rules proposed in literature. Our focus is on phase stability dependent on structural parameters. A novel feature analysis technique has also been deployed in order to extract hidden structure--property linkages by addressing the collinearity that may exist among the features. An effort has also been made to predict the formability of stable oxide compounds with the perovskite motif and then identify compositions that are suitable for photovoltaic and photocatalytic applications based on the ab--initio values of band--gap available in materials data repositories.
%%%
\par The outline of this paper is as follows. The dataset and the preprocessing steps to generate the feature space is described in Section \ref{sec:dataset} with the extraction of features delineated in Section \ref{sub_sec:features} and the feature importance computed using tree and permutation methods in Section \ref{sub_sec:featureimportances} onwards. The list of all features used in this work is condensed in Table \ref{tab:featurestable}. The results of the classifier model performances across different feature sets is presented in Section \ref{sec:results}. The identification of stable oxide--perovskites for some applications is shown in Section \ref{sub_sec:pv_pec} followed by the conclusions drawn from this study.
%%%
\section{{\label{sec:dataset}}Dataset and preprocessing}
\par In this work an attempt has been made to apply a novel feature analysis technique to a vast array of features to characterise the behavior of perovskites and identify the most important ones towards the prediction of formability and stability of oxide perovskites. The data used in the present work has been obtained from literature \cite{balachandran2018, vasala2015, talapatra2021} and the Materials Project database \cite{matproj2013}. 
%%%
\subsection{{\label{sub_sec:features}}Feature Extraction}
A general and `easily available' materials descriptor space was constructed considering structural, thermodynamic and elemental information. Elemental properties include intrinsic properties of elements like atomic mass, heuristic quantities such as electronegativity, atomic radius and valence electron concentration as well as physical properties such as melting temperature. The latter represent a general analogue of the bonding and electronic properties of materials.
\par Thermodynamic parameters such as enthalpy, mixing entropy and free energy were used to represent the stability among competing phases. Structural parameters such as lattice parameter, metallic radii, arrangement of the atoms in the lattice among others have also been used to provide positional information of each element in a multi--component crystalline system. These quantities are readily available on online repositories such as Materials Project \cite{matproj2013} and have been used as descriptors to improve the performance of ML models \cite{zhang2020descriptors}. From the above datasets the compositions were extracted using a material parser \cite{kononova2019} and the thermodynamic and elemental features were extracted as described in Table \ref{tab:featurestable}.
\par Features from perovskite literature that were specific to the component elements were included in the database. These perovskite--specific features are electron affinity, electronegativity, ionization energy, Zunger’s pseudopotential radius, the highest occupied molecular orbital (HOMO) energy and the lowest unoccupied molecular orbital (LUMO) energy. Taking into account the additional symmetries in double perovskites, antisymmetric and symmetric compound features are also considered, as demonstrated by Talapatra et al. \cite{talapatra2021}.
\par For the $AA'BB'O_6$ perovskite configuration, the antisymmetric compound feature  may be calculated as  $P^{B-}$= ($P_B - P_{B'}$)/2 and the symmetric compound feature as $P^{B+}$= ($P_B+P_{B'}$)/2 for the B--site, where $P_B$ and $P_{B'}$ are the elemental properties of $B$ and $B'$ for a given property $P$. 
\par Structural mapping factors introduced in Section \ref{sec:intro}, such as mismatch factors ($\overline{\mu}_A$, $\overline{\mu}_B$), octahedral factor ($\mu$) and tolerance factor ($t$) have also been included in the feature set. Additionally a larger set of generic elemental features that have been reported in material informatics literature towards the prediction of various material properties, were also included in the scope of study. This was done to distinguish any unidentified underlying mechanisms. Such generic elemental features are listed in Table \ref{tab:featurestable}.
%%%
% \subsubsection*{{\label{sub_sub_sec:elementalfeatures}}Rule--of-Mixtures (RoM) elemental properties}
\par The elemental values of the lattice constant ($pm$), metallic radius ($pm$), formation energy ($kJ/mol$), melting temperature ($K$), density ($g/cc$), electronegativity (Pauling EN) and valence electron concentration ($VEC$) of the constituent elements of the perovskites were used to calculate their respective RoM values $\bar{F}={\Sigma_{i} c_i F_i}/{\Sigma_{i} c_i}$, where $c_i$ is the molar composition and $F_i$ is the corresponding feature value. The divergence of the elemental parameters from their RoM values was also taken into consideration to account for the variance among the properties of the constituent elements. The RSSD value is computed as $\Delta F=\sqrt{\sum_{i=1}^{n} (F_i-F_m)^2}$, where $F_i$ is the elemental feature value and $F_m$ is the corresponding RoM value of that feature.
%%%
% \subsubsection*{{\label{sub_sub_sec:thermodynamicfeatures}}\hl{Thermodynamic properties}}
\par The selection of a phase during alloy formation and its subsequent stability is dictated by the energetics of formation as expressed in the quantities: enthalpy of mixing ($\Delta H_{mix}$) and entropy of mixing ($\Delta S_{mix}$). These competing thermodynamic mechanisms can be reduced to the expression for free energy $\Delta G_{mix} =  \Delta H_{mix} - T\Delta S_{mix}$, which is the determining factor for phase stability among competing phases. $\Delta H_{mix}$ was calculated using Miedema's semiemperical formulation \cite{miedema1988} from the established behaviour of the binary phase diagrams as $\Delta H_{mix} = \sum_{i,j,i\neq j} (4\Delta H_{mix}^{AB})c_ic_j$. The values for $\Delta H_{mix}^{AB}$ were obtained from the work by Takeuchi and Inoue \cite{takeuchiinoue2005}. The configurational entropy was used to approximate the entropy of mixing as $\Delta S_{mix} \simeq \Delta S_{config} = -R\sum_{i=1}^n X_i \ln X_i$.
%%%
% \subsubsection*{{\label{sub_sub_sec:elementalfeatures}}Symmetry and spacing in the elemental lattices}
\par As the properties of a material are determined by its structure, parameters like angular and radial distribution functions and coordination number can be used to represent different structures. Due to lack of simplicity, however, they have not been used in ML based methods to predict properties \cite{zhang2020descriptors}. The symmetry notation and the lattice parameter were combined to form a categorical tuple (e.g: \textit{Cu\_tuple\_('m--3m', 361.0)}). The tuple information was included in the features as the symmetry and spacing indicators of the component elements. Empirical rules formulated to predict phase formation \cite{singh2014, yang2012} such as a geometrical parameter $\lambda$, and a solid solution formation parameter $\Omega$ -- as described in Table \ref{tab:featurestable}, were also included.
%%%
\begin{table}[!htbp]
\scriptsize
\centering
\caption{Features used in the present work.}
\begin{adjustbox}{width=0.99\textwidth}
\begin{tabular}{p{0.45\linewidth}p{0.35\linewidth}p{0.15\linewidth}} %{llr}
\toprule
%\toprule
\multicolumn{1}{c}{Feature} & \multicolumn{1}{c}{Description} & \multicolumn{1}{c}{Descriptor in model}\\
\midrule
% \midrule
\multicolumn{3}{l}{\textbf{Perovskite--specific features}}\\
\midrule
$t = (r_A + r_X)/{\sqrt{2} (r_B + r_X)}$ & Tolerance factor & $t$\\
$\mu = r_B / r_X$ & Octahedral factor & $\mu$\\
$\overline{\mu}_A = \lvert r_A - r_{A'}\rvert/2r_X$ & Mismatch factor & $\overline{\mu}_A$\\
$\overline{\mu}_B = \lvert r_B - r_{B'}\rvert/2r_X$ & Mismatch factor & $\overline{\mu}_B$\\
\midrule
$E_{ea}$ & Electron affinity & $E_{ea}^{A+}, E_{ea}^{A-}, E_{ea}^{B+}, E_{ea}^{B-}$\\
$\chi$ & Electronegativity & $\chi^{A+}, \chi^{A-}, \chi^{B+}, \chi^{B-}$\\
$I_1$ & $1^{st}$ Ionization energy & $I_1^{A+}, I_1^{A-}, I_1^{B+}, I_1^{B-}$\\
$r_P$ & Pseudopotential radius & $Z_{rad}^{A+}, Z_{rad}^{A-}, Z_{rad}^{B+}, Z_{rad}^{B-}$\\
Lowest unoccupied molecular orbital & LUMO energy & $LUMO^{A+...A-...B+...B-}$\\
Highest occupied molecular orbital & HOMO energy & $HOMO^{A+...A-...B+...B-}$\\
\midrule
% \midrule
\multicolumn{3}{l}{\textbf{Generic features}}\\
\midrule
$r_m=\sum_{i=1}^{n}c_ir_i$ & RoM metallic radius & r\_RoM\\
$a_m=\sum_{i=1}^{n}c_ia_i$ & RoM lattice constant & a\_RoM\\
$T_m=\sum_{i=1}^{n}c_iT_i$ & RoM melting temperature & Tm (K)\_RoM\\
$\rho_m=\sum_{i=1}^{n}c_i\rho_i$ & RoM density & $\rho$\_RoM\\
$H^f_m=\sum_{i=1}^{n}c_iH^f_i$ & RoM formation energy & Hf\_RoM\\
$\chi_m=\sum_{i=1}^{n}c_i\chi_i$ & RoM electronegativity & $\chi$\_RoM\\
$VEC_m=\sum_{i=1}^{n}c_iVEC_i$ & RoM VEC & VEC\_RoM\\
\midrule
$\delta=\sqrt{\sum_{i=1}^{n} (1-\frac{r_i}{r_m})^2}$ & RSSD of metallic radii & $\delta$\\
$\Delta a=\sqrt{\sum_{i=1}^{n} (a_i-a_m)^2}$ & RSSD of lattice constants & $\Delta a$\\
$\Delta T=\sqrt{\sum_{i=1}^{n} (T_i-T_m)^2}$ & RSSD of melting temperatures & $\Delta Tm$\\
$\Delta \rho=\sqrt{\sum_{i=1}^{n} (\rho_i-\rho_m)^2}$ & RSSD of densities & $\Delta \rho$\\
$\Delta H^f=\sqrt{\sum_{i=1}^{n} (H^f_i-H^f_m)^2}$ & RSSD of formation energies & $\Delta H^f$\\
$\Delta \chi=\sqrt{\sum_{i=1}^{n} (\chi_i-\chi_m)^2}$ & RSSD of electronegativities & $\Delta \chi$\\
$\Delta VEC=\sqrt{\sum_{i=1}^{n} (VEC_i-VEC_m)^2}$ & RSSD of VECs & $\Delta VEC$\\
\midrule
$\Delta H_{mix}=\sum_{i=1, i\neq j}^{n} (4\Delta H_{mix}^{AB})c_ic_j$ & Mixing enthalpy by Miedema's rule \cite{miedema1988, takeuchiinoue2005} & $\Delta H_{mix}$\\
$\Delta S_{mix}=-R\sum_{i=1}^{n} (c_i\ln{c_i})$ & Mixing entropy approximated to configurational entropy & $\Delta S_{mix}$\\
\midrule
Elemental symmetry \& lattice parameter tuples & e.g.: Cu\_tuple\_ (`m--3m', 361.0) & \\
\midrule
$\mathit{\lambda} = \frac{\Delta S_{mix}}{\delta^2}$ & Geometrical parameter for phase formation \cite{singh2014} & $\mathit{\lambda}$\\
$\mathit{\Omega} = \frac{T_m\Delta S_{mix}}{\lvert \Delta H_{mix}\rvert}$ & Parameter for predicting solid--solution formation \cite{yang2012} & $\mathit{\Omega}$\\
\bottomrule
\end{tabular}
\end{adjustbox}
\label{tab:featurestable}
\end{table}
%%%
%\vspace{1cm}
\subsection{{\label{sub_sec:featureimportances}}Feature Importance (FI)}
The feature importance towards the prediction of each target variable was computed in order to confirm established theoretical linkages and possibly unearth novel Process-Structure-Property (PSP) relations, in addition to helping eliminate redundant ones.
\par To extract the importance of each feature used in the database, a model examination technique know as Permutation Feature Importance (PFI) was used in addition to the default Tree Feature Importance (TFI) of the Random Forest (RF) classifier model. The PFI method quantifies the reduction in a model's accuracy score when the values of a single feature are shuffled randomly, breaking the relation between the target variable and the features. This drop in a model's score indicates the model's dependence on that particular feature. This method is model agnostic and is permutated multiple times to adequately evaluate the significance of a feature. 
%%%
\par However, the model could suffer if there are correlated features in the dataset as PFI has access to the correlated features even as the feature in question is being shuffled. To overcome this shortcoming, hierarchical clustering is performed on the features’ Spearman rank--order correlations. Spearman's correlation is defined as the Pearson correlation coefficient between the ranked feature columns. The Pearson's correlation coefficient \cite{pearson1895} given by $r=\Sigma (x_i - \bar{x}) (y_i - \bar{y})/\sqrt{\Sigma (x_i - \bar{x})^2 (y_i - \bar{y}^2)}$ provides a measure of linear correlation between two sets of data $x$ and $y$ with means $\bar{x}$ and $\bar{y}$ respectively.
\par The features are then clustered hierarchically using Ward's minimum variance method which finds the pair of clusters leading to a minimum increase in variance within the cluster after merging. The increase in variance is quantified as the Euclidean distance between cluster centers. Once all the features have been clustered, the condensed distance matrix can be plotted as a dendrogram \cite{chehreghani2020} with Euclidean distance versus the features. In a dendogram, features get clustered as the distance value increases. While iterating over the cluster distance, a threshold is picked which optimises the accuracy of predictions towards a particular property and in the process filters out the most correlated features affecting the accuracy, from the dataset. These methods have been implemented in this work using open source SciPy tools \cite{scipy}.
\par The RF models have also been tuned for the optimum model hyperparameters and the partial dependencies of the features plotted similar to an earlier work \cite{revi2021} that explored the use of ML models to determine elasticity constants of multi--component alloys.
%%%
\subsection{FI towards formability}
%%%
\subsubsection{Using perovskite--specific features}
%%%
\begin{figure}[ht]
%%%%%RIGHT MINIPAGE
    \begin{minipage}[!b]{0.57\linewidth}
    \centering
        \begin{subfigure}{\textwidth}
        \centering
        \includegraphics[width=\linewidth]{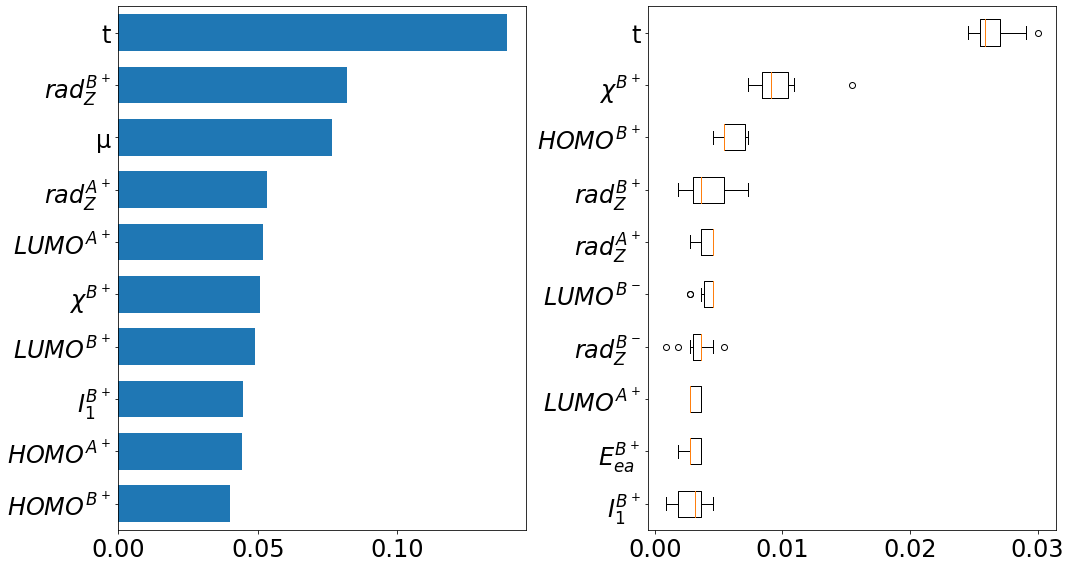}
        \vspace{-5mm}
        \caption{}\label{fig:FormOriginalFI_pre}
        \end{subfigure}
    \centering
        \begin{subfigure}{\textwidth}
        \centering
        \includegraphics[width=\linewidth]{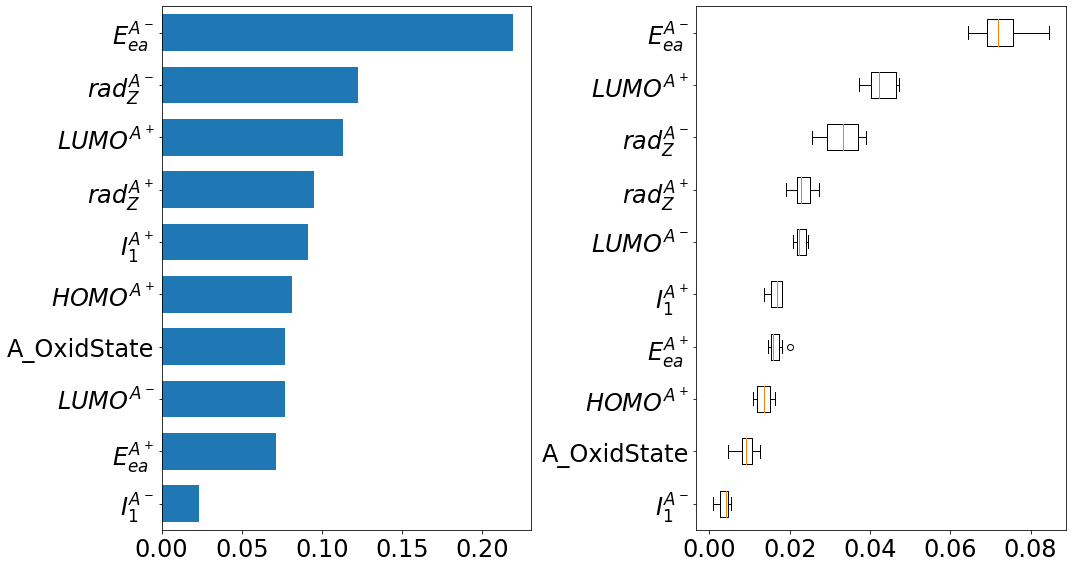}
        \vspace{-5mm}
        \caption{}\label{fig:FormOriginalFI_post}
        \end{subfigure}
    \end{minipage}
% \hspace{0.5cm}
%%%%%RIGHT MINIPAGE
    \begin{minipage}[!h]{0.41\linewidth}
        \begin{subfigure}{0.45\textwidth}
        \centering
        \includegraphics[width=\linewidth]{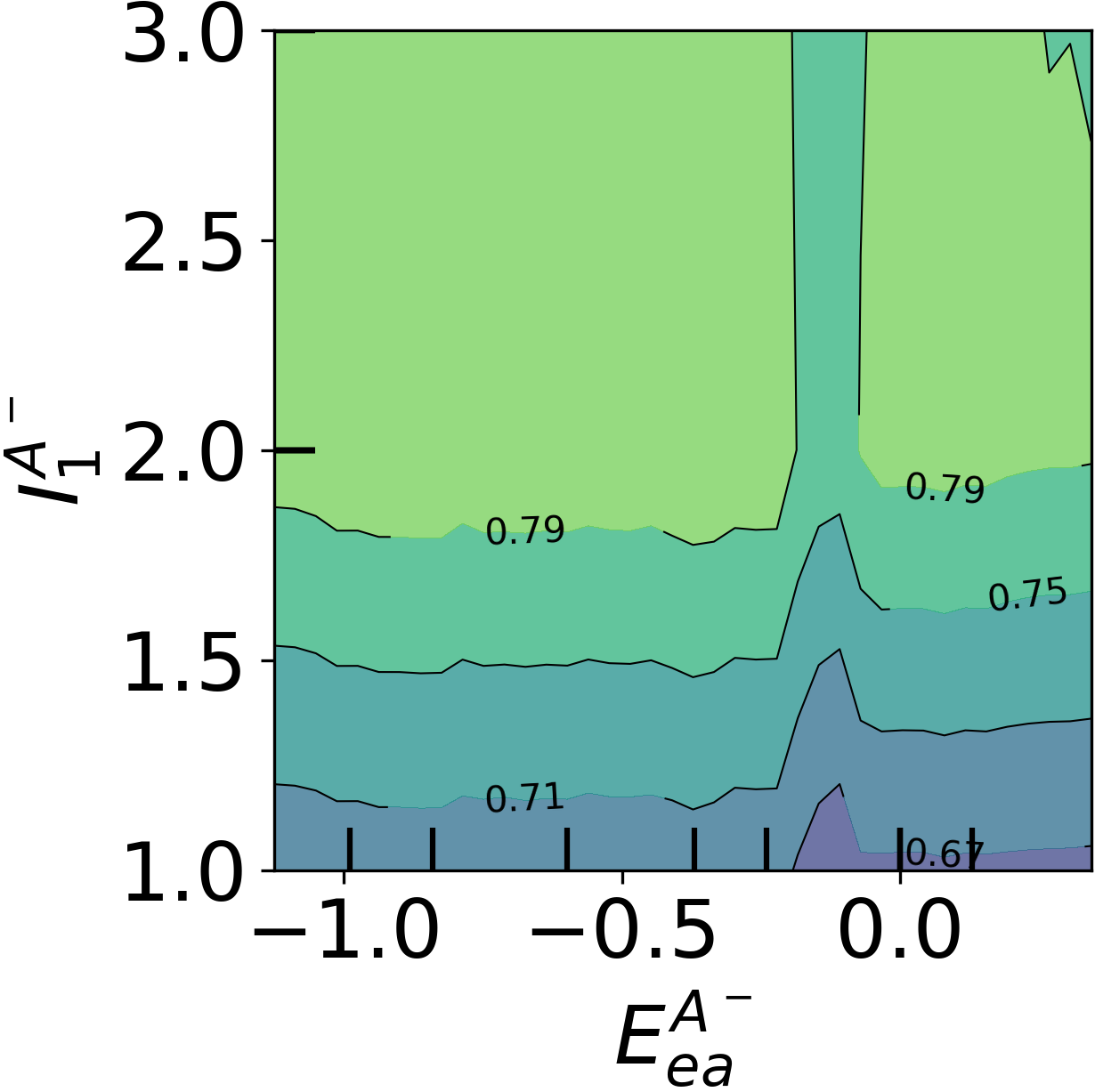}
        \vspace{-5mm}
        \caption{}\label{fig:PartDep_OriginalForm6}
        \end{subfigure}
    \hspace{2mm}
        \begin{subfigure}{0.44\textwidth}
        \centering
        \includegraphics[width=\linewidth]{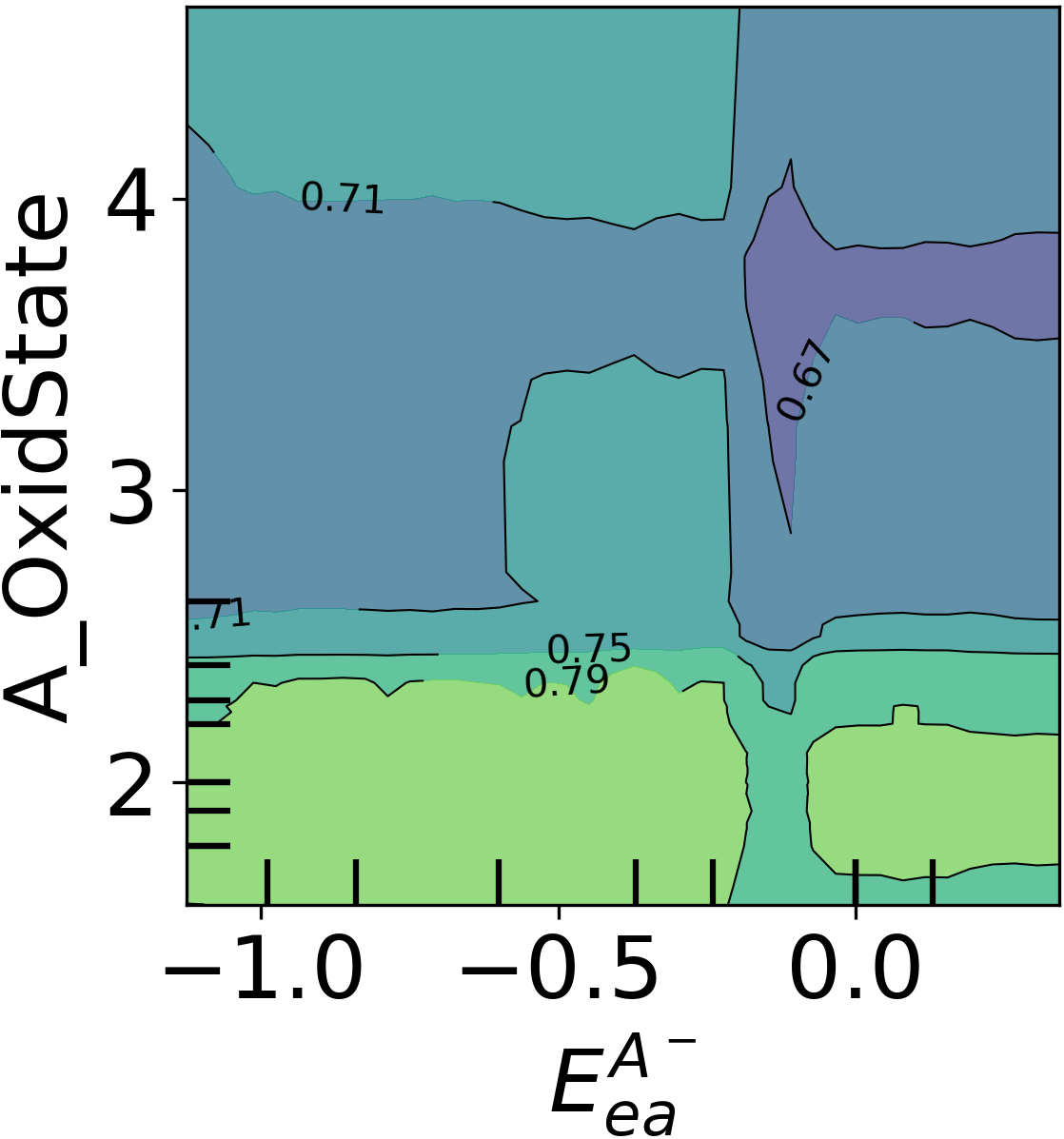}
        \vspace{-5mm}
        \caption{}\label{fig:PartDep_OriginalForm5}
        \end{subfigure}
    \hspace{2mm}
        \begin{subfigure}{0.44\textwidth}
        \centering
        \includegraphics[width=\linewidth]{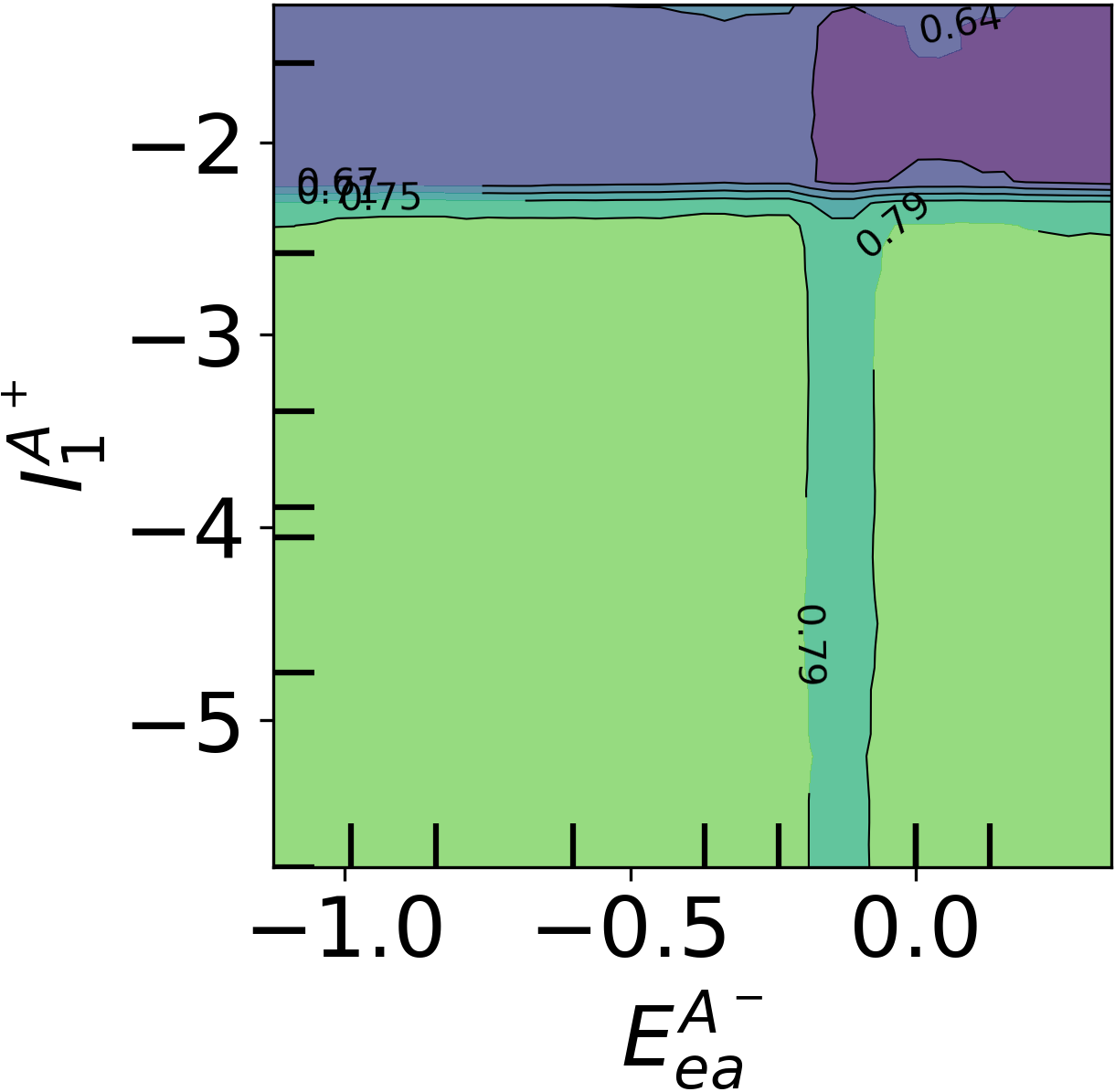}
        \vspace{-5mm}
        \caption{}\label{fig:PartDep_OriginalForm4}
        \end{subfigure}
    \hspace{2mm}
        \begin{subfigure}{0.45\textwidth}
        \centering
        \includegraphics[width=\linewidth]{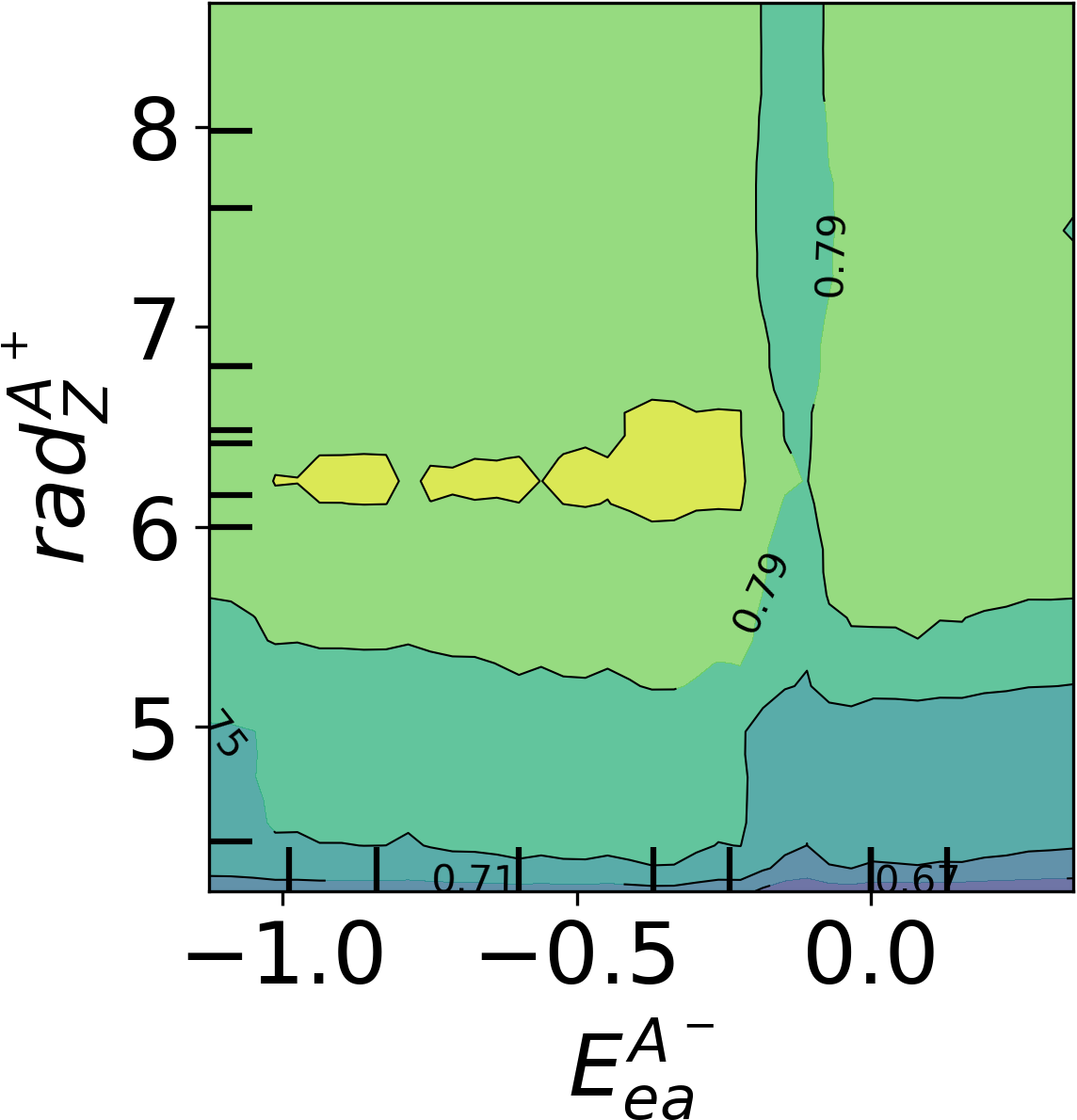}
        \vspace{-5mm}
        \caption{}\label{fig:PartDep_OriginalForm3}
        \end{subfigure}
    \hspace{2mm}
        \begin{subfigure}{0.43\textwidth}
        \centering
        \includegraphics[width=\linewidth]{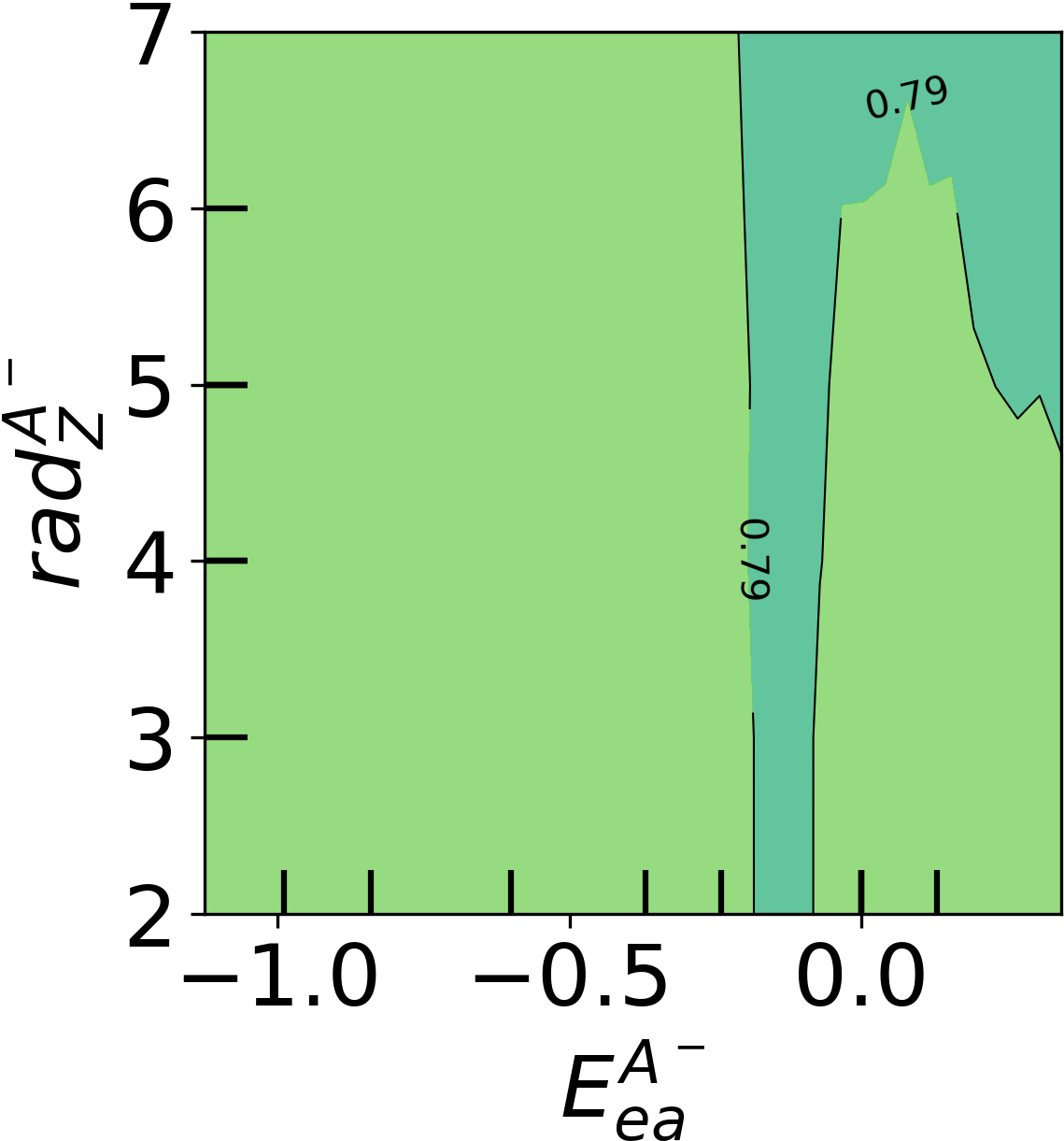}
        \vspace{-5mm}
        \caption{}\label{fig:PartDep_OriginalForm2}
        \end{subfigure}
    \hspace{2mm}
        \begin{subfigure}{0.46\textwidth}
        \centering
        \includegraphics[width=\linewidth]{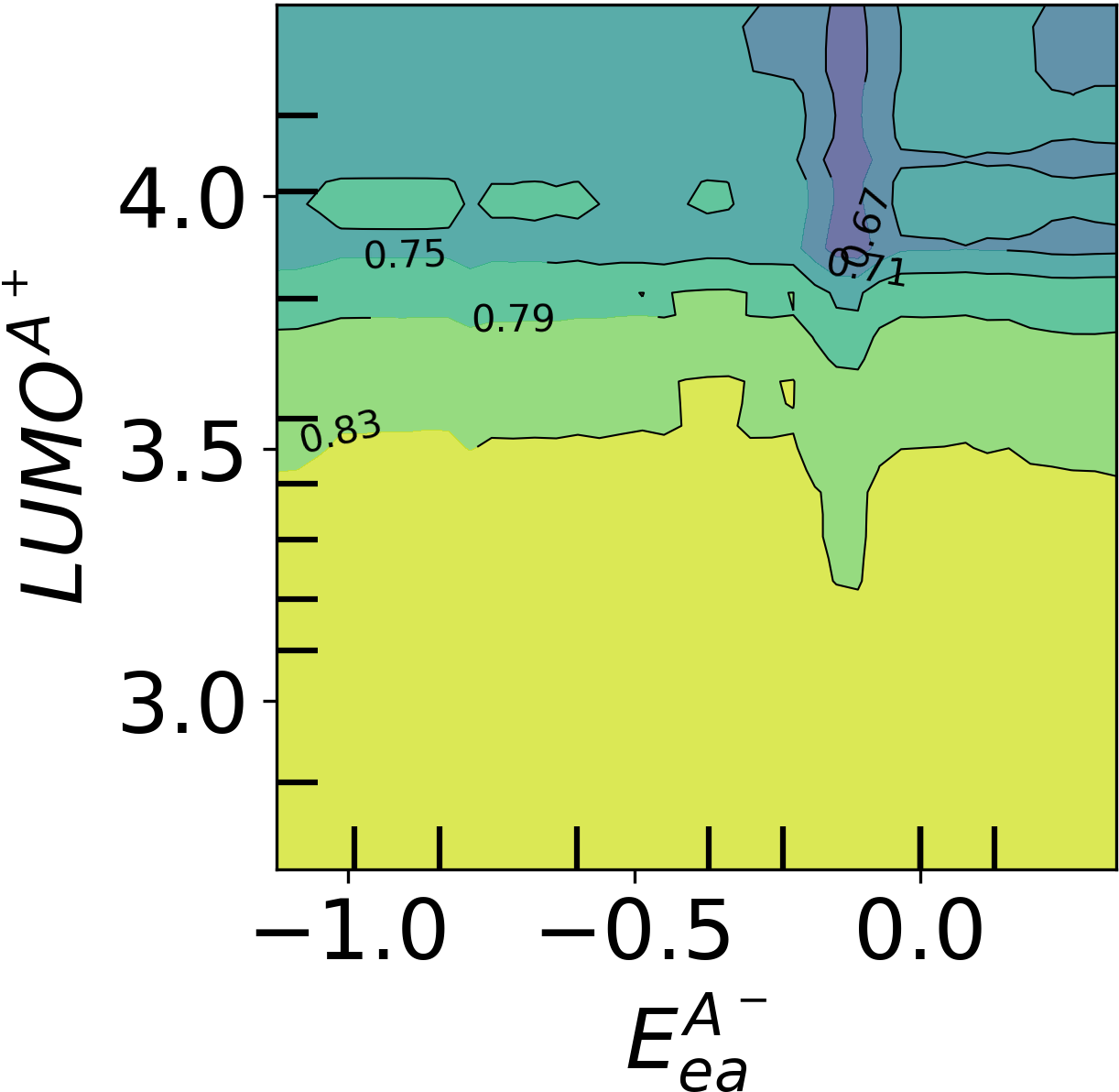}
        \vspace{-5mm}
        \caption{}\label{fig:PartDep_OriginalForm1}
        \end{subfigure}
    \end{minipage}
\centering
\vspace{-3mm}
\caption{TFI (left) and PFI (right) (a) before  hierarchical clustering and (b) after hierarchical clustering of perovskite--specific features towards prediction of formability. The hierarchical clustering dendrogram for Spearman rank--order correlations is shown in \ref{sub_sec:clusteringFormability}. Only the features that survive this hierarchical clustering are included in the feature importance graphic in (b). Partial dependency plots of formability on perovskite features (c) \textit{LUMO}$^{A+}$ (d) \textit{rad}$_{Z}^{A-}$ (e) \textit{rad}$_{Z}^{A+}$ (f) \textit{I}$_{1}^{A+}$ (g) \textit{A\_OxidState} (h) \textit{I}$_{1}^{A1}$ plotted against the highest importance feature $E_{ea}^{A-}$.}
\label{fig:FIandPartDep_OriginalForm}
\end{figure}
%%%
The features from the perovskite database were first analysed to study their importance towards prediction of the formability of perovskite structures. In addition to the numerical value features used in the original work, some of the categorical variables also were included such as the elements occupying the A, A$'$, B and B$'$ sites of the oxide perovskite and whether the perovskites were the single-- or double--oxide types. 
\par The TFI and PFI shown in Fig. \ref{fig:FormOriginalFI_pre} indicate that Goldschmidt's tolerance factor $t$ is the most important contributing factor to the formability of perovskite oxides followed by the electronegativity, pseudopotential radius and ionisation energy of the B--site atom.
%%%
\begin{figure}[!h]%tbp]
    \centering
        \begin{subfigure}{0.4\textwidth}
        \centering
        \includegraphics[width=\linewidth]{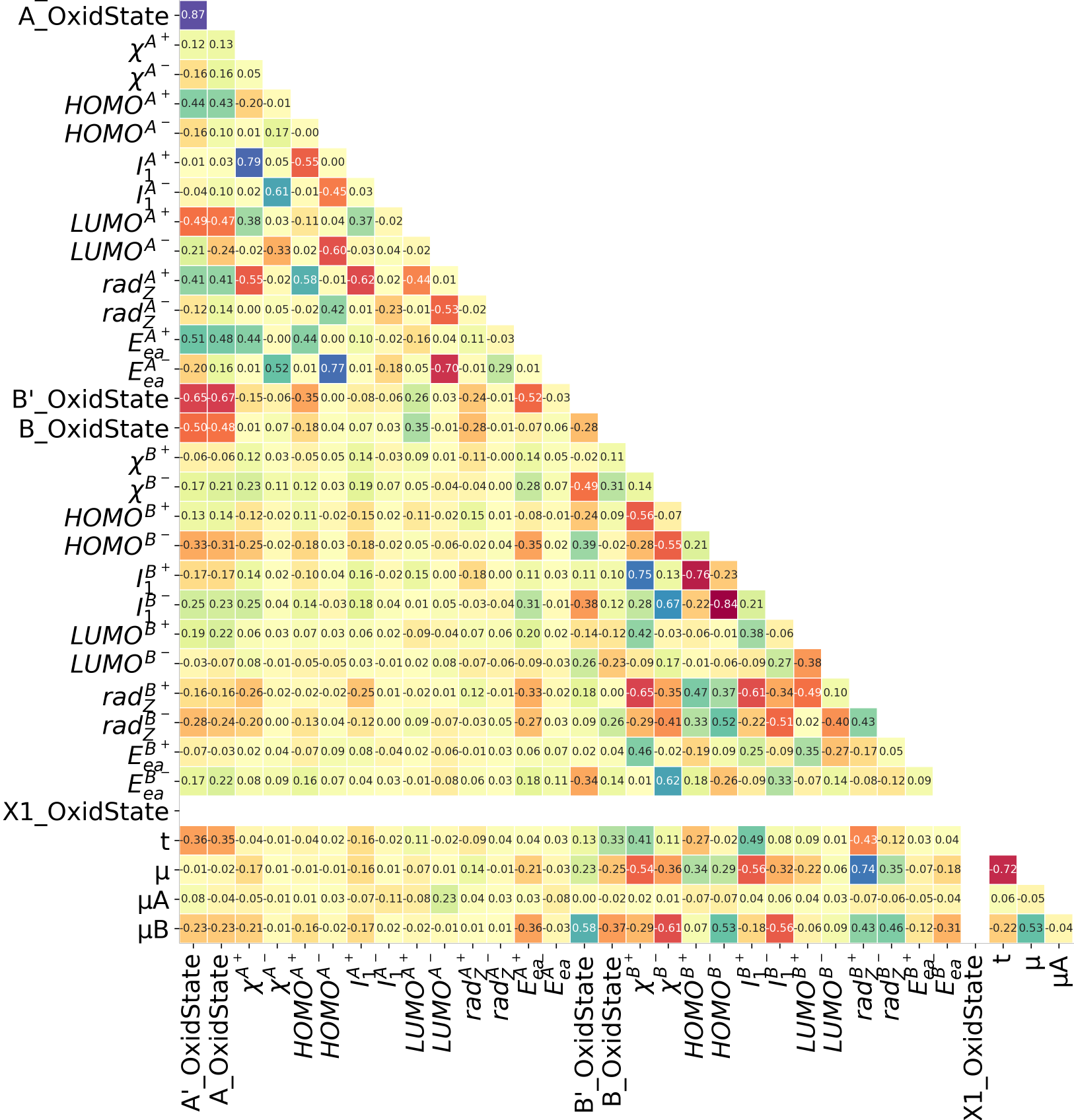}
        \vspace{-5mm}
        \caption{}\label{fig:OriginalForm_PearsonsFull1}
        \end{subfigure}
    \hspace{5mm}
        \begin{subfigure}{0.4\textwidth}
        \centering
        \includegraphics[width=\linewidth]{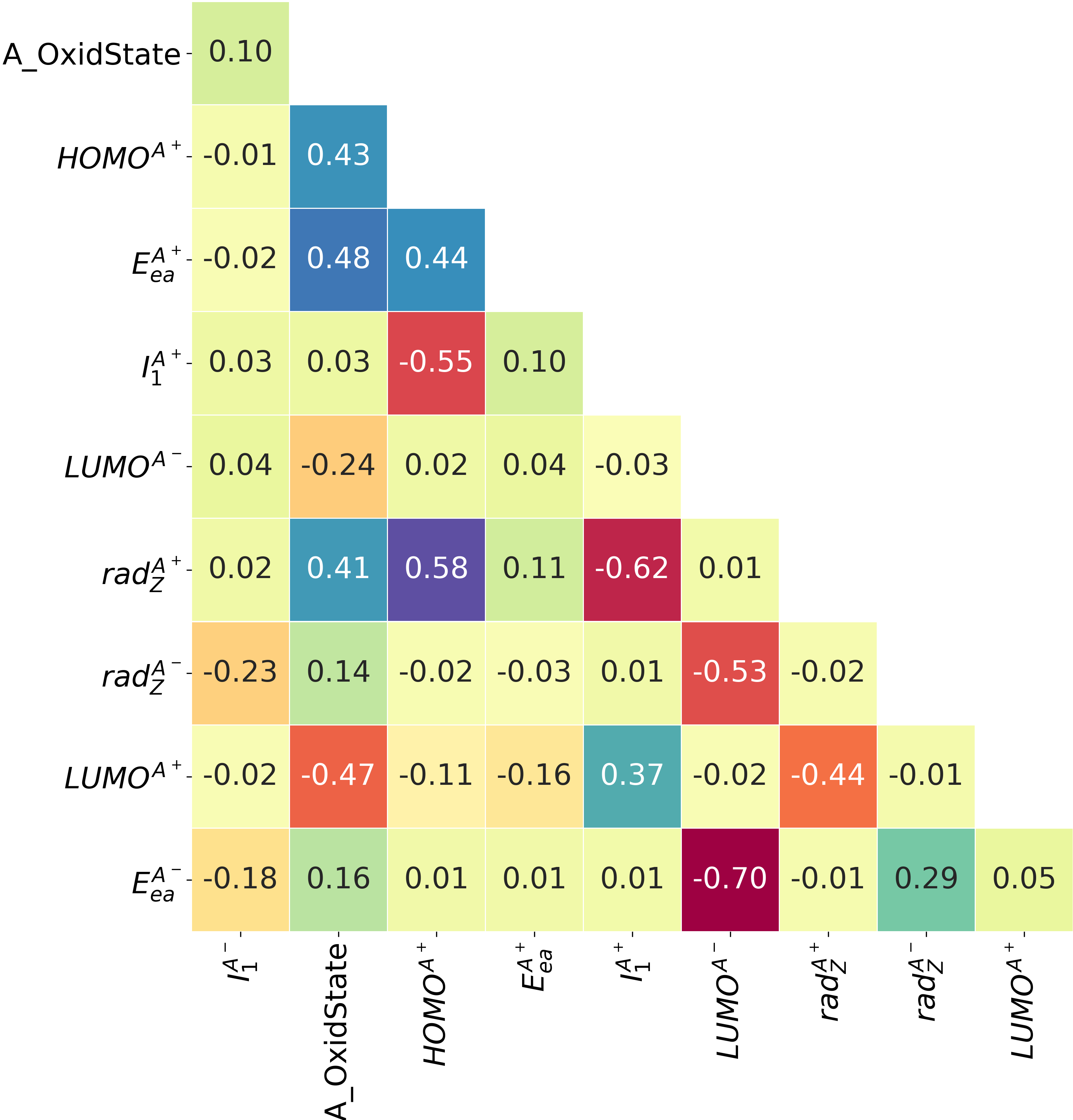}
        \vspace{-5mm}
        \caption{}\label{fig:OriginalForm_PearsonsReduced1}
        \end{subfigure}
    \vspace{-3mm}
    \caption{Pearson's correlation heatmaps for perovskite specific features (a) before and (b) after hierarchical clustering towards prediction of formability.}
    \label{fig:}
\end{figure}
%%%
\par The Pearson's correlation matrices, before and after hierarchical clustering are depicted as heatmaps in Figs. \ref{fig:OriginalForm_PearsonsFull1} and \ref{fig:OriginalForm_PearsonsReduced1} respectively. The Pearson's correlation coefficients are indicative of the degree of collinearity among the features spanning the dataset. The heatmaps depict a reduction in the correlation scores between features after hierachical clustering, as described in Section \ref{sub_sec:featureimportances}.
After eliminating the multicollinear features based on the optimum cutoff shown in Fig. \ref{fig:FormOriginalFI_dendro} in Appendix \ref{sub_sec:clusteringFormability}, the electron affinity, Zunger's pseudopotential radius and LUMO energy of the A--site atom are seen to have the highest feature importance towards predicting formability as shown by the TFI and PFI in Fig. \ref{fig:FormOriginalFI_post}. This indicates that while the tolerance factor $t$ was found to be highly correlated to B--site atom features in the dataset, the readiness of the A--site atom to provide an electron is also an important underlying factor towards the formability of perovskites \cite{Tao2021}. 
\par The partial dependencies of the important features after hierarchical clustering shown in Fig. {\ref{fig:FIandPartDep_OriginalForm}} also enable some interesting observations. When the \textit{LUMO}$^{A+}$ is less than 3.5 eV (Fig. {\ref{fig:PartDep_OriginalForm6}}) the probability of forming a perovskite is greater than 0.8 irrespective of the $E_{ea}^{A-}$ value. A similar probability inference may also be made on the basis of \textit{rad}$A_Z^-$ and \textit{rad}$A_Z^+$ values greater than 5 (Figs. {\ref{fig:PartDep_OriginalForm5}} and {\ref{fig:PartDep_OriginalForm4}}) except around $E_{ea}^{A-}$ values close to 0.0 eV. 6 out of the 10 highest--importance--features were chosen to plot partial dependencies of each versus the most important feature. The features that displayed the most `interesting' variations were chosen -- without any implications of change in the order of importance among the top 10 features for the RF classifiers.
%%%
\begin{figure}[!h]
%%%%%LEFT MINIPAGE
    \begin{minipage}[!b]{0.57\linewidth}
    \centering
        \begin{subfigure}{\textwidth}
        \centering
        \includegraphics[width=\linewidth]{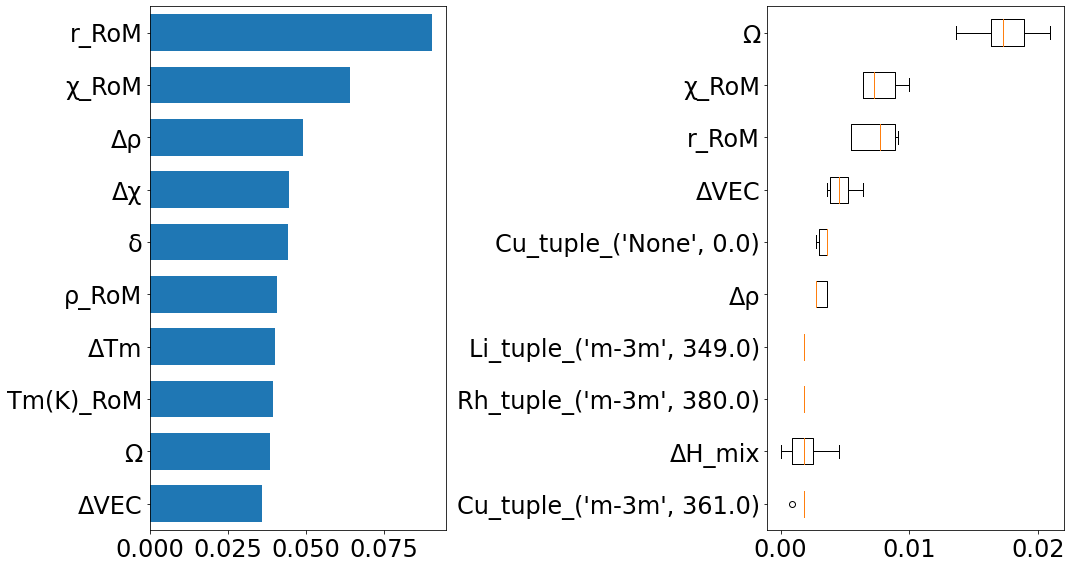}
        %\vspace{-8mm}
        \caption{}\label{fig:FormNewFI_pre}
        \end{subfigure}
    \centering
        \begin{subfigure}{\textwidth}
        \centering
        \includegraphics[width=\linewidth]{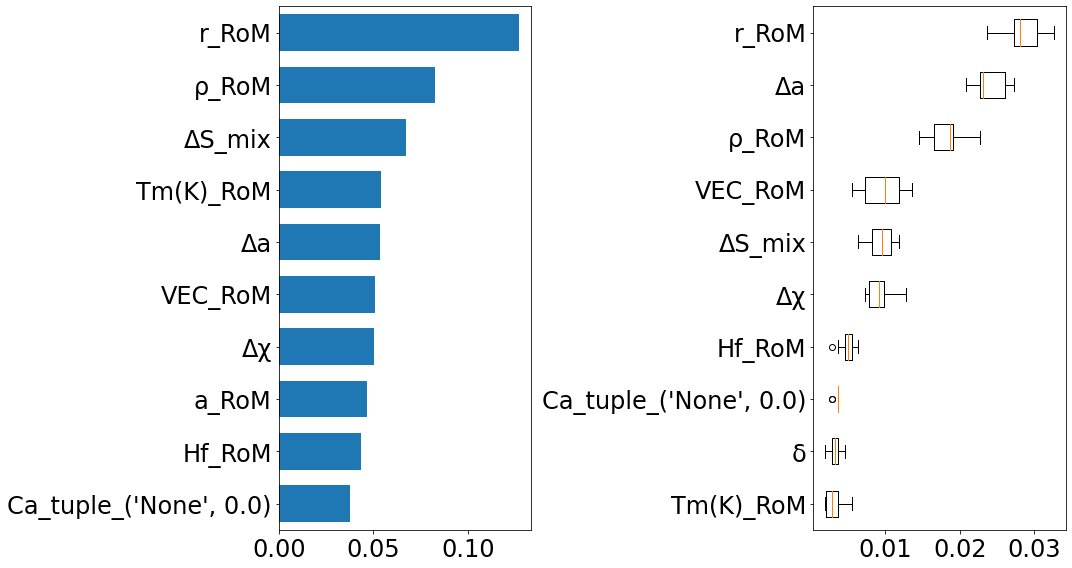}
        %\vspace{-8mm}
        \caption{}\label{fig:FormNewFI_post}
        \end{subfigure}
    \end{minipage}
\hspace{2mm}
%%%%%RIGHT MINIPAGE
    \begin{minipage}[!b]{0.41\linewidth}
    \centering
        \begin{subfigure}{0.45\textwidth}
        \centering
        \includegraphics[width=\linewidth]{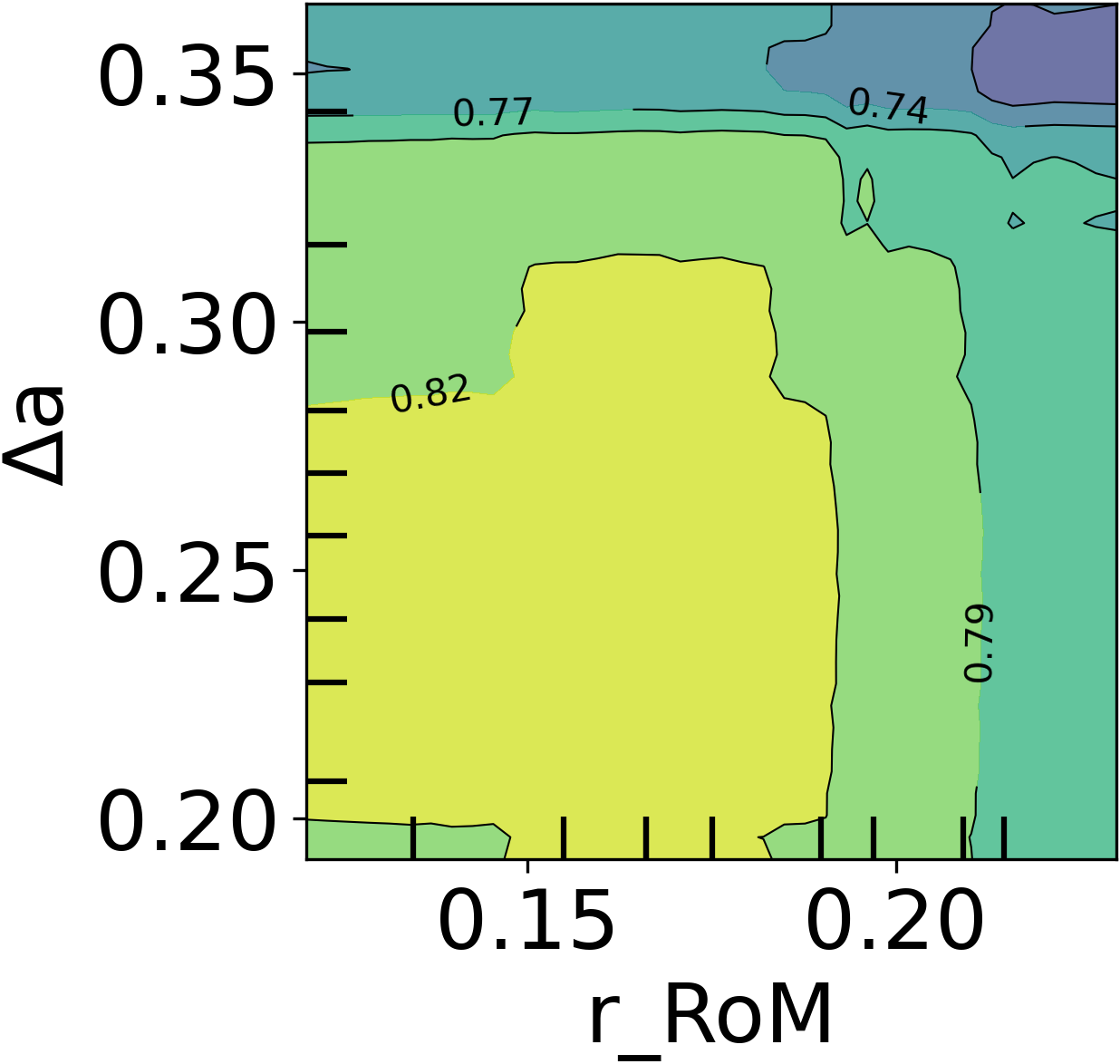}
        %\vspace{-5mm}
        \caption{}\label{fig:PartDep_GenericFormability1}
        \end{subfigure}
    % \hfill
        \begin{subfigure}{0.46\textwidth}
        \centering
        \includegraphics[width=\linewidth]{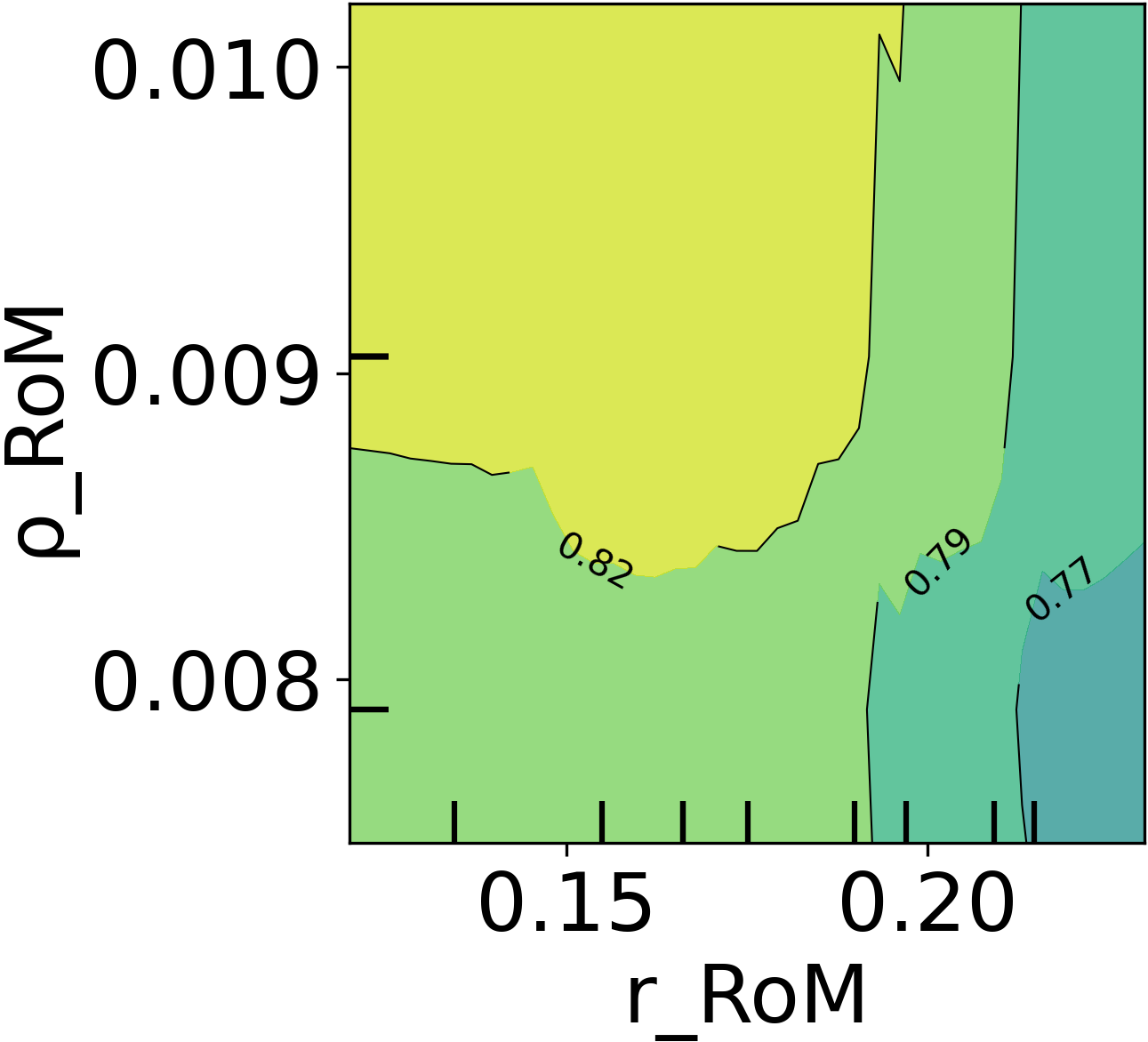}
        %\vspace{-5mm}
        \caption{}\label{fig:PartDep_GenericFormability2}
        \end{subfigure}
    % \hfill
        \begin{subfigure}{0.45\textwidth}
        \centering
        \includegraphics[width=\linewidth]{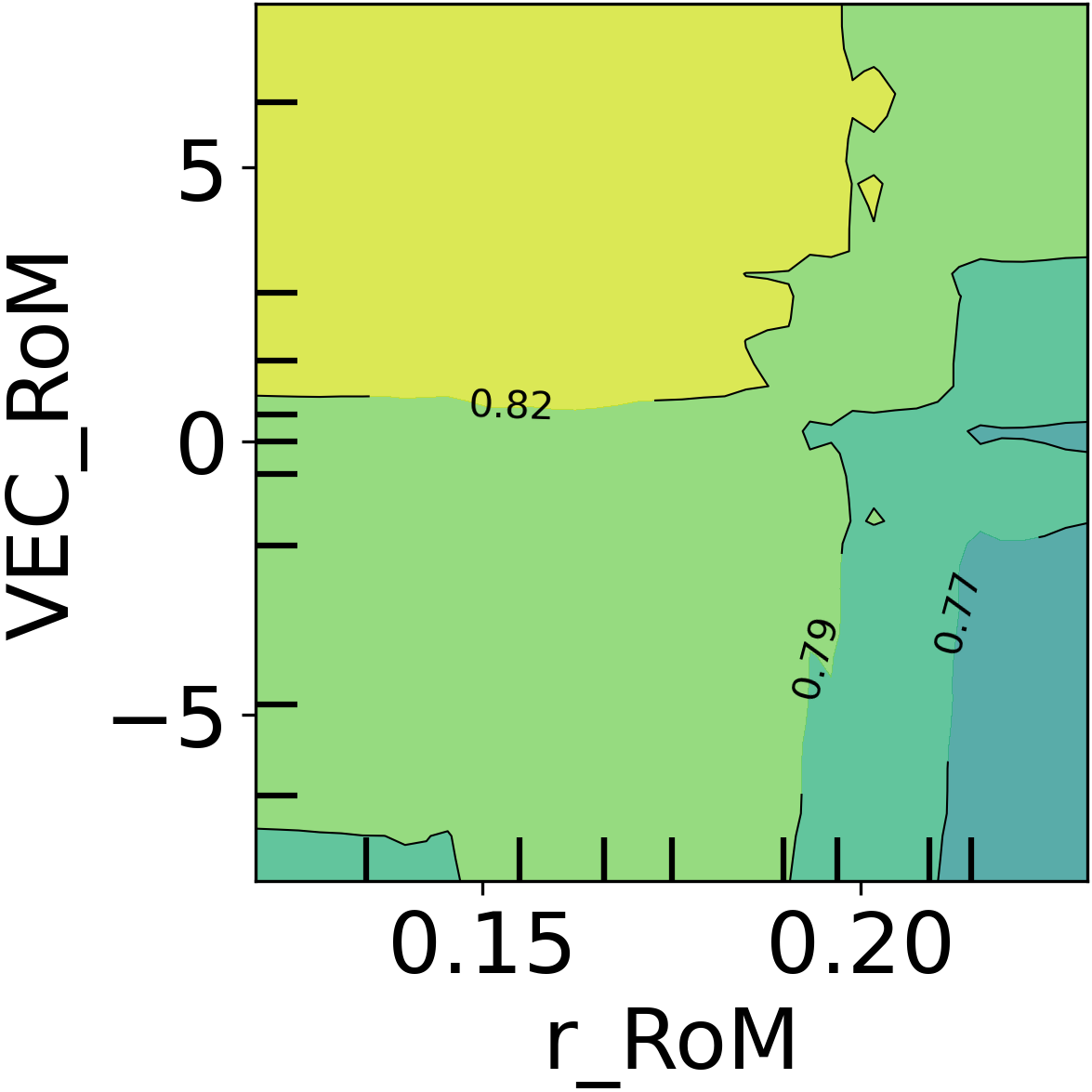}
        %\vspace{-5mm}
        \caption{}\label{fig:PartDep_GenericFormability3}
        \end{subfigure}
    % \hfill
        \begin{subfigure}{0.46\textwidth}
        \centering
        \includegraphics[width=\linewidth]{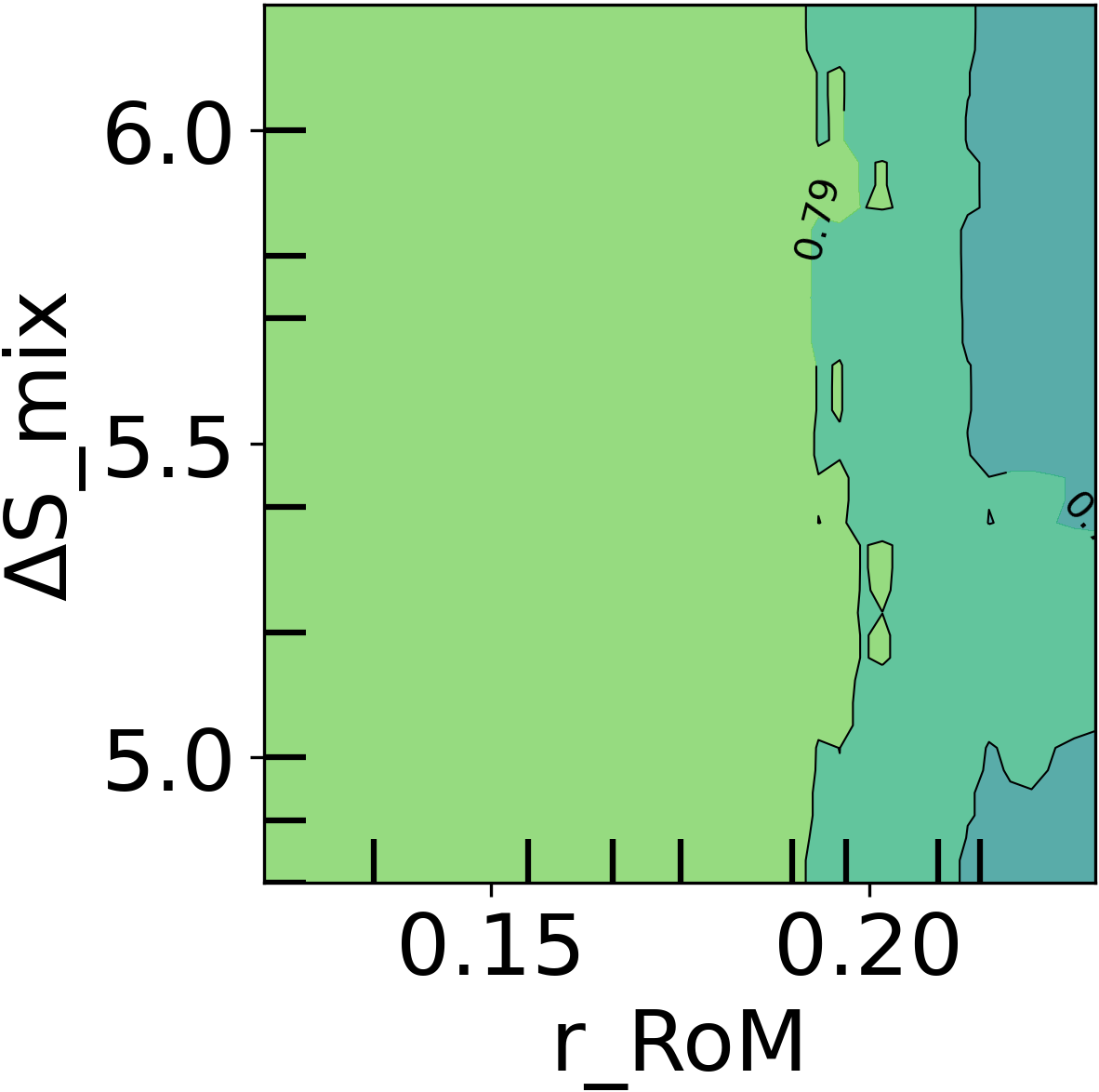}
        %\vspace{-5mm}
        \caption{}\label{fig:PartDep_GenericFormability4}
        \end{subfigure}
    % \hfill
        \begin{subfigure}{0.46\textwidth}
        \centering
        \includegraphics[width=\linewidth]{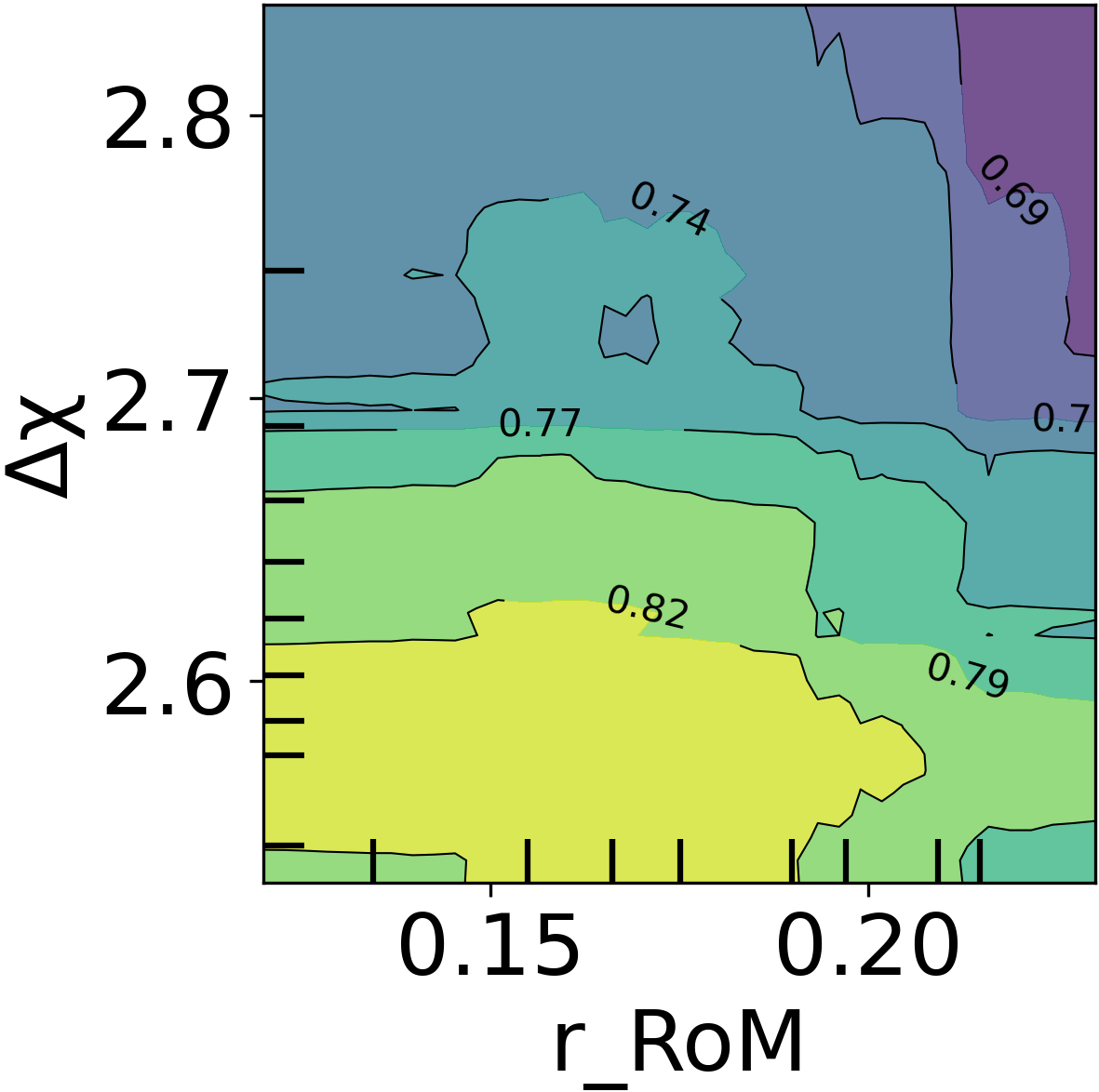}
        %\vspace{-5mm}
        \caption{}\label{fig:PartDep_GenericFormability5}
        \end{subfigure}
    % \hfill
        \begin{subfigure}{0.44\textwidth}
        \centering
        \includegraphics[width=\linewidth]{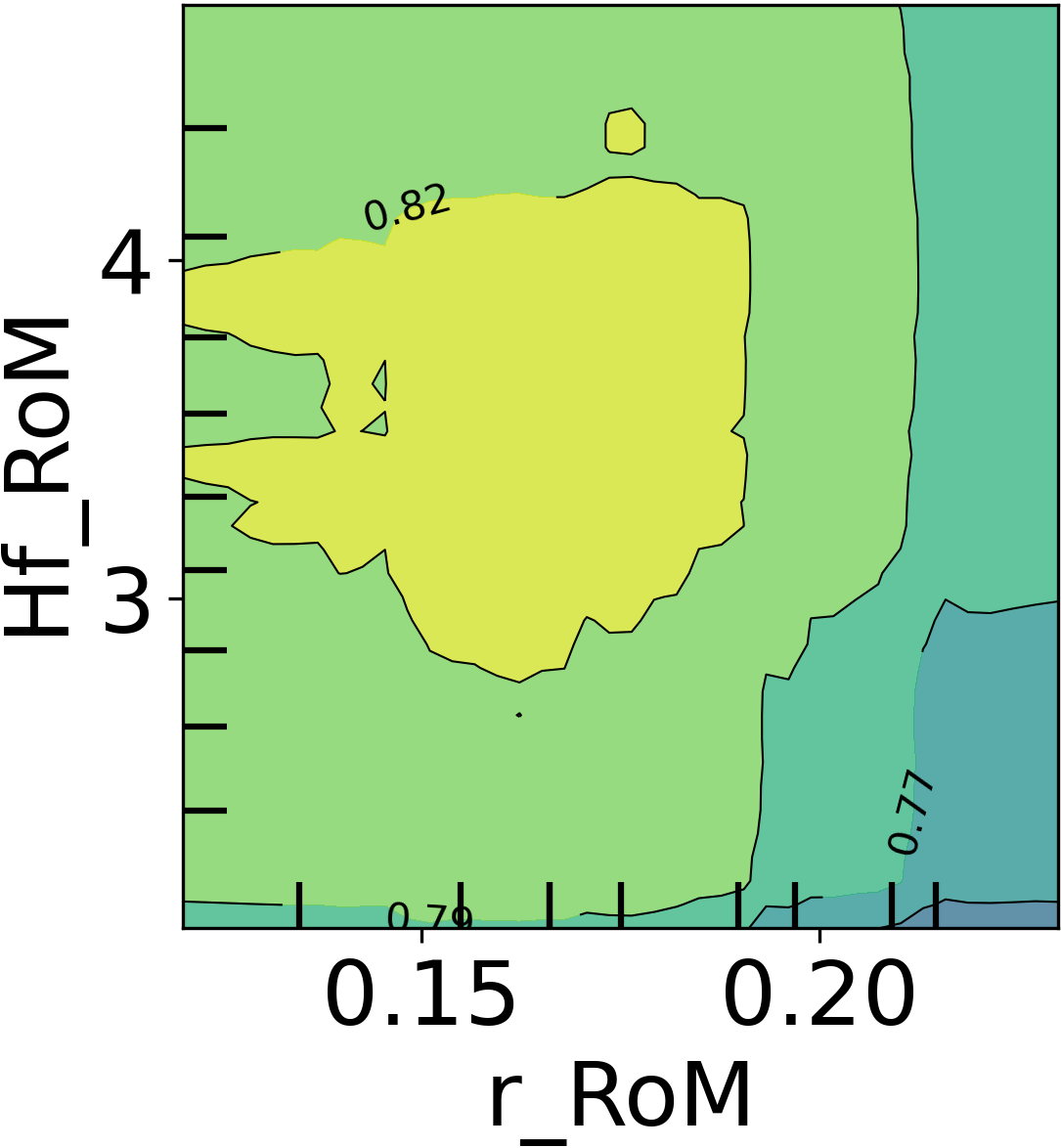}
        %\vspace{-5mm}
        \caption{}\label{fig:PartDep_GenericFormability6}
        \end{subfigure}
    \end{minipage}
    \vspace{-3mm}
    \caption{TFI (left) and PFI (right) (a) before hierarchical clustering and (b) after hierarchical clustering of novel generic features towards prediction of formability. The hierarchical clustering dendrogram for Spearman rank--order correlations is shown in Appendix \ref{sub_sec:clusteringFormability}. Partial dependency plots of formability on novel generic features (c) $\delta$ (d) $\rho$\textit{\_RoM} (e) \textit{VEC\_RoM} (f) $\Delta$\textit{S\_mix} (g) $\Delta\chi$ (h) \textit{Hf\_RoM} plotted against the highest importance feature \textit{r\_RoM}.}
    \label{fig:FIandPartDep_GenericForm}
\end{figure}
%%%
\subsubsection{Using novel generic features}
\begin{figure}[!h]%tbp]
    \centering
        \begin{subfigure}{0.4\textwidth}
        \centering
        \includegraphics[width=\linewidth]{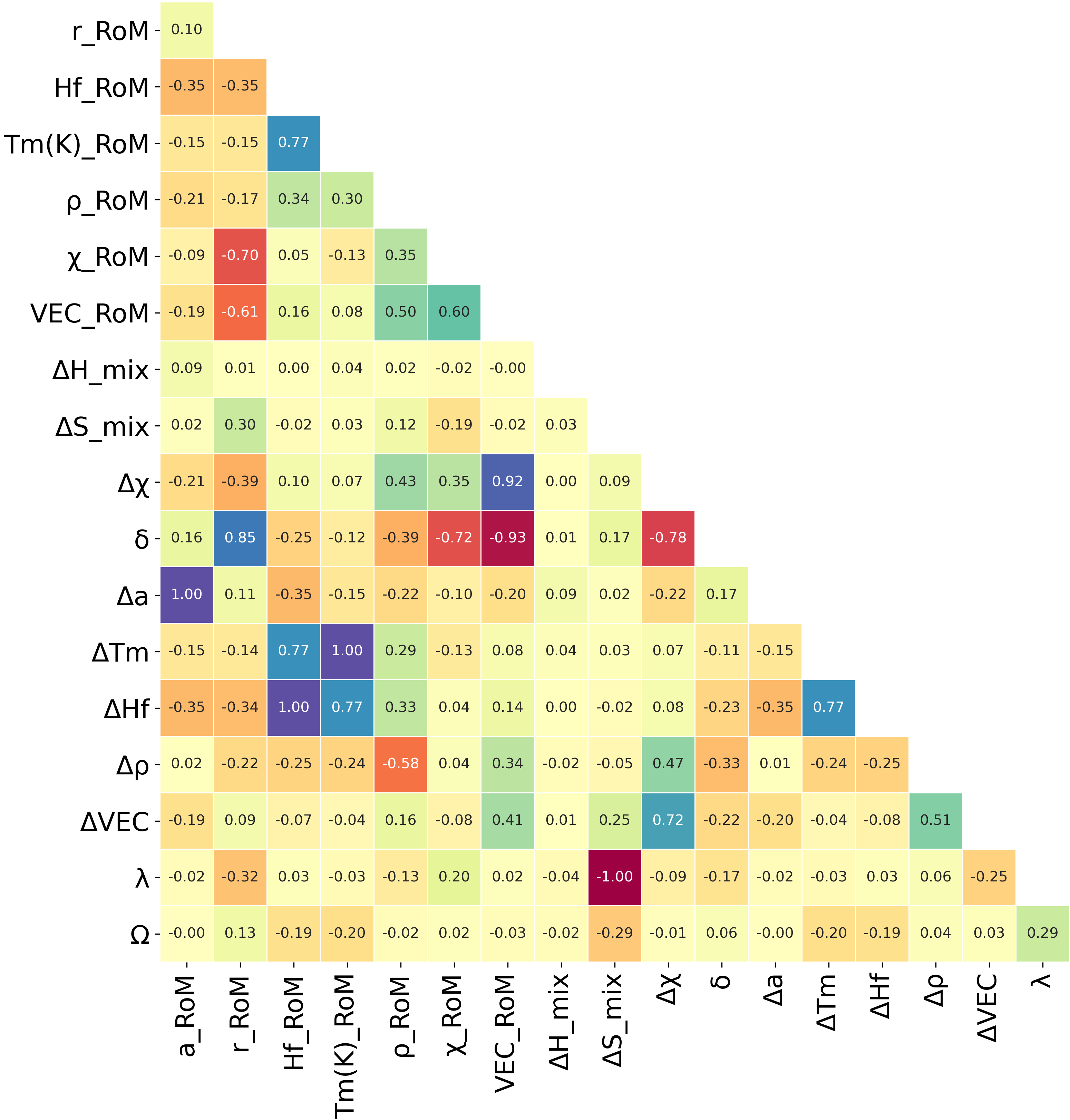}
        \vspace{-5mm}
        \caption{}\label{fig:GenericForm_PearsonsFull1}
        \end{subfigure}
    \hspace{5mm}
        \begin{subfigure}{0.4\textwidth}
        \centering
        \includegraphics[width=\linewidth]{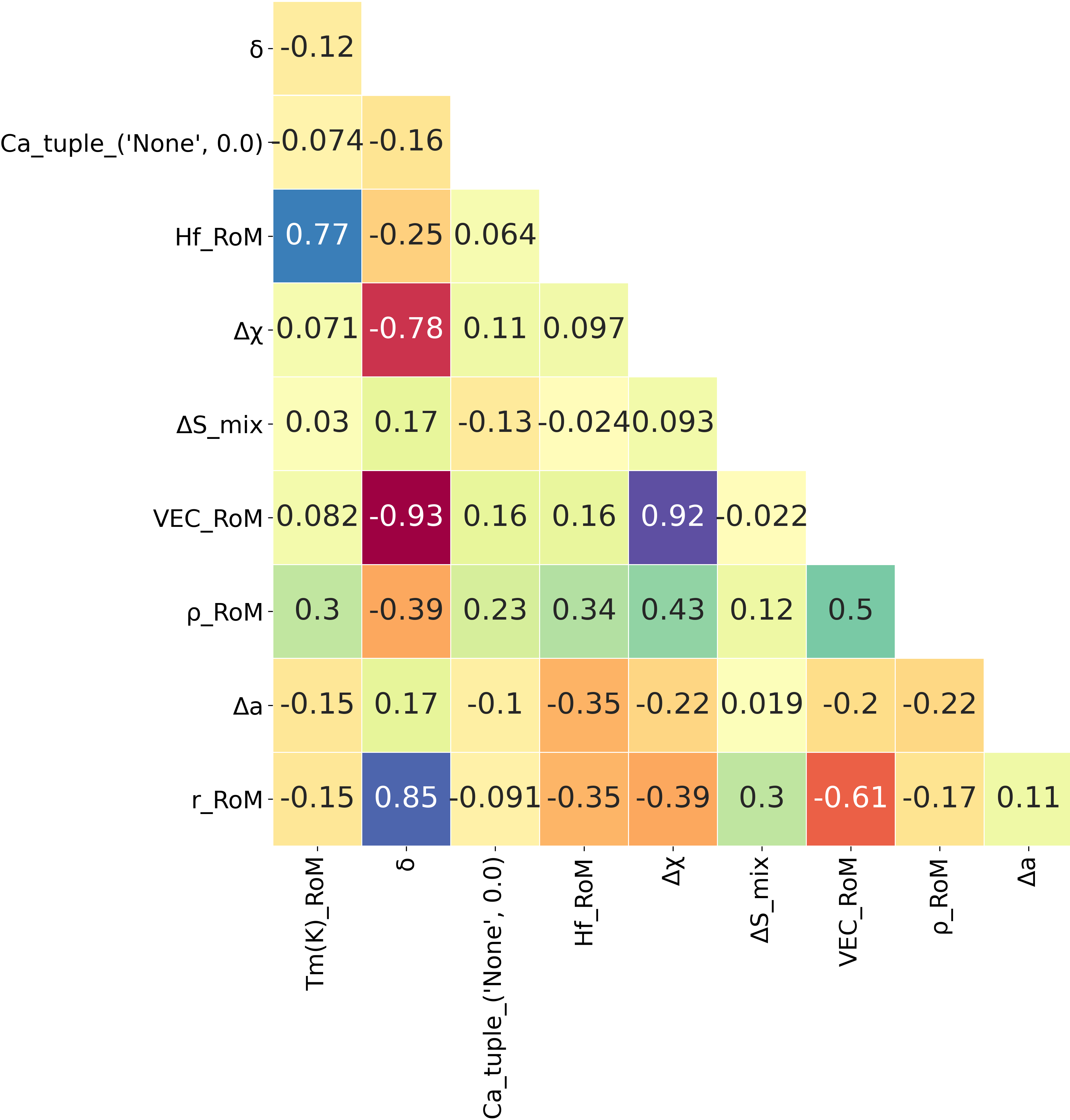}
        \vspace{-5mm}
        \caption{}\label{fig:GenericForm_PearsonsReduced1}
        \end{subfigure}
    \vspace{-3mm}
    \caption{Pearson's correlation heatmaps for generic features (a) before and (b) after hierarchical clustering towards prediction of formability.}
    \label{fig:}
\end{figure}

%%%
Novel elemental property features described in Table \ref{tab:featurestable} were analysed to study their importance towards prediction of the formability of perovskite structures. The TFI and PFI prior to hierarchical clustering indicate the importance of the $\mathit{\Omega}$ parameter for solid solution formation, electronegativity $\chi$ and radius $r$, in deciding the formability of perovskite oxides as shown in Fig. \ref{fig:FormNewFI_pre}. The Pearson's correlation heatmaps, before and after hierarchical clustering in Figs. {\ref{fig:GenericForm_PearsonsFull1}} and {\ref{fig:GenericForm_PearsonsReduced1}} respectively, depict the reduction in correlation among features.
%%%
\par The hierarchical clustering of features are optimized based on cut off distance towards the highest accuracy as shown in Fig. \ref{fig:FormNewFI_dendro}. The density $\rho$ and the RSSD value of lattice parameter $a$ are also found to have significant effect on the formability of perovskite oxide compounds as shown in Fig. \ref{fig:FormNewFI_post}.
%%%
% \clearpage
%%%
\par In the partial dependence of formability of the important novel generic features after hierarchical clustering  in Fig. {\ref{fig:FIandPartDep_GenericForm}}, it is seen that the probability of forming a perovskite is more than 0.82 when: $\Delta a$ is less than 0.28 (Fig. {\ref{fig:PartDep_GenericFormability1}}), $\rho$\textit{\_RoM} greater than 0.0088 kg m$^{-3}$ (Fig. {\ref{fig:PartDep_GenericFormability2}}) and \textit{VEC\_RoM} more than 1 (Fig. {\ref{fig:PartDep_GenericFormability3}}) except for values of \textit{r\_RoM} greater than 0.18 $\AA$.
%%%
\subsection{FI towards stability}
%%%
\subsubsection{Using perovskite--specific features}
%%%
\par
\begin{figure}[!h]
    \begin{minipage}[!b]{0.57\linewidth}
    \centering
        \begin{subfigure}{\textwidth}
        \centering
        \includegraphics[width=\linewidth]{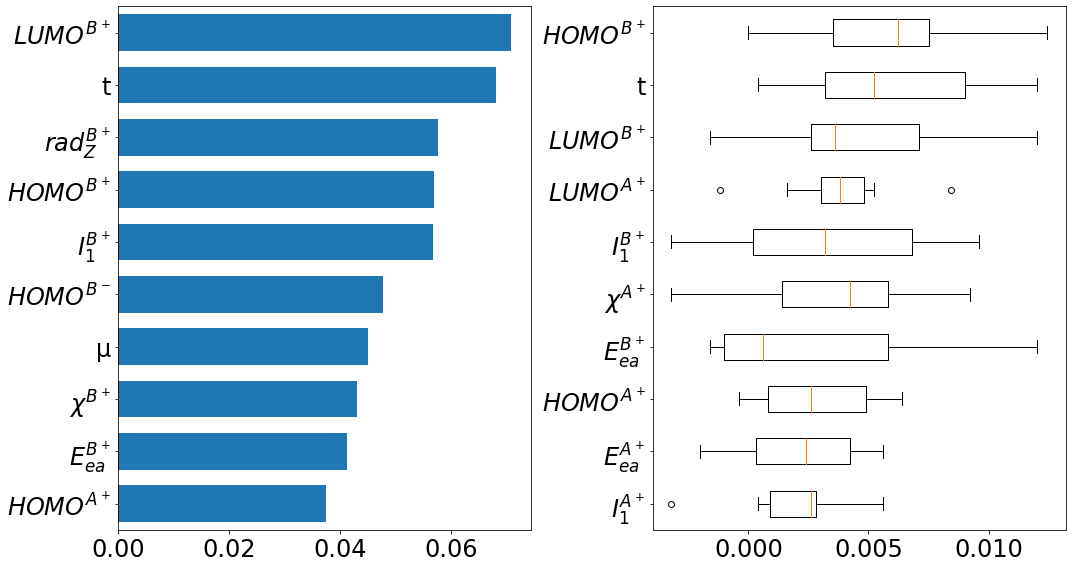}
        %\vspace{-8mm}
        \caption{}\label{fig:StabOriginalFI_pre}
        \end{subfigure}
    \centering
        \begin{subfigure}{\textwidth}
        \centering
        \includegraphics[width=\linewidth]{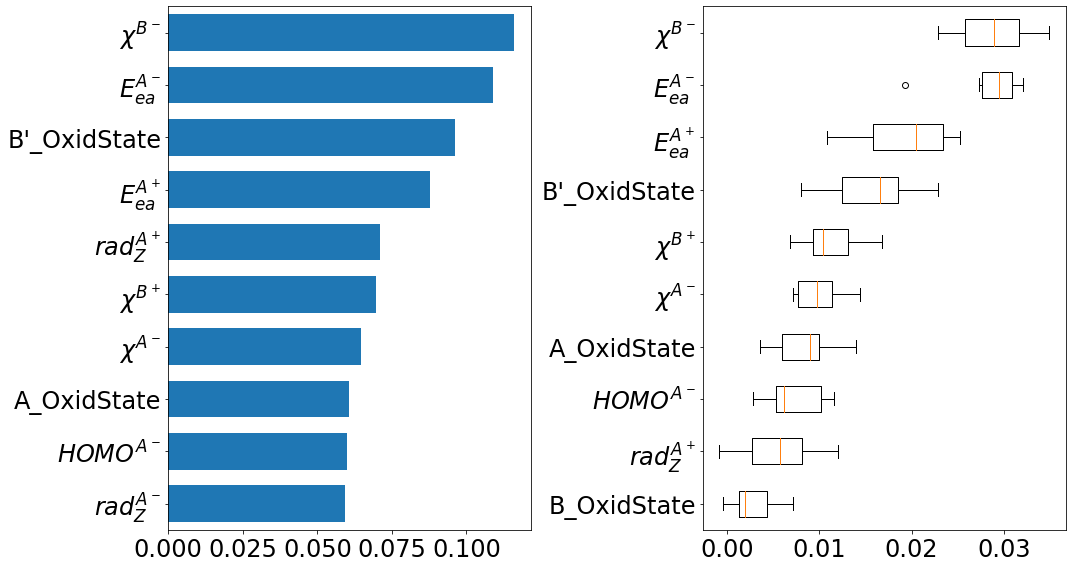}
        %\vspace{-8mm}
        \caption{}\label{fig:StabOriginalFI_post}
        \end{subfigure}
    \end{minipage}
% \hspace{0.5cm}
    \begin{minipage}[!b]{0.41\linewidth}
    \centering
        \begin{subfigure}{0.44\textwidth}
        \centering
        \includegraphics[width=\linewidth]{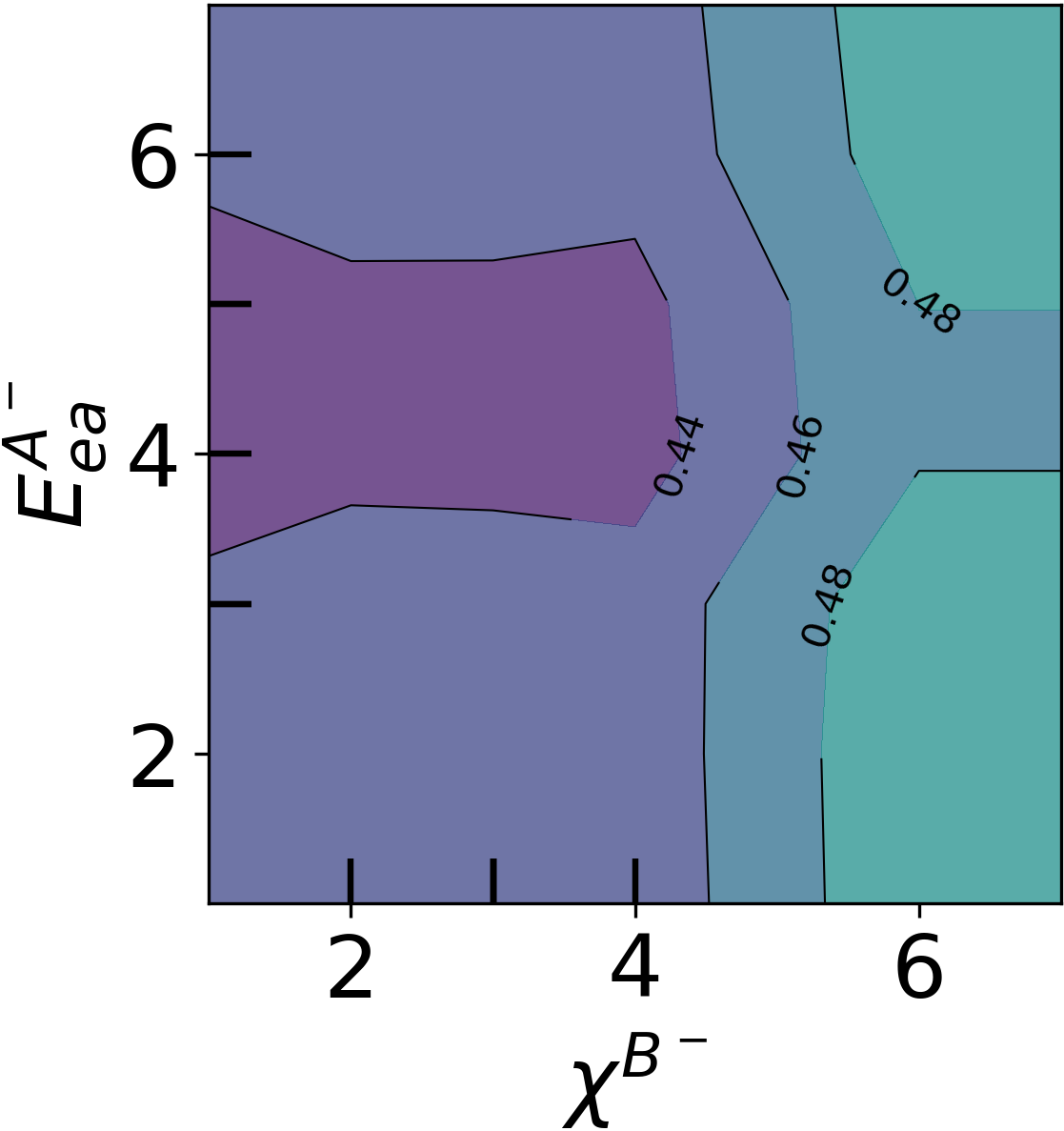}
        \vspace{-5mm}
        \caption{}\label{fig:PartDep_OriginalStability1}
        \end{subfigure}
    % \hfill
        \begin{subfigure}{0.46\textwidth}
        \centering
        \includegraphics[width=\linewidth]{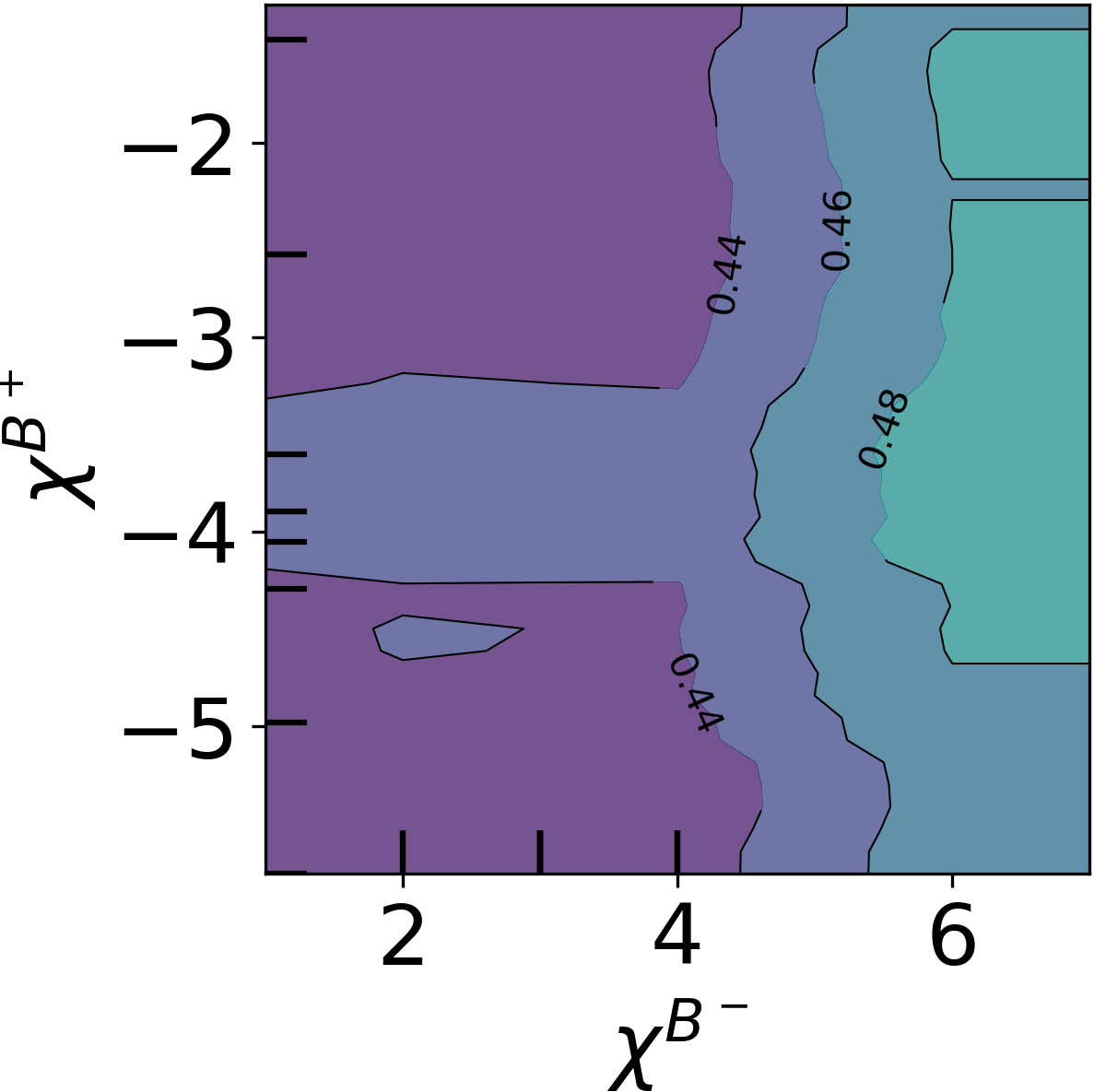}
        \vspace{-5mm}
        \caption{}\label{fig:PartDep_OriginalStability2}
        \end{subfigure}
    % \hfill
        \begin{subfigure}{0.47\textwidth}
        \centering
        \includegraphics[width=\linewidth]{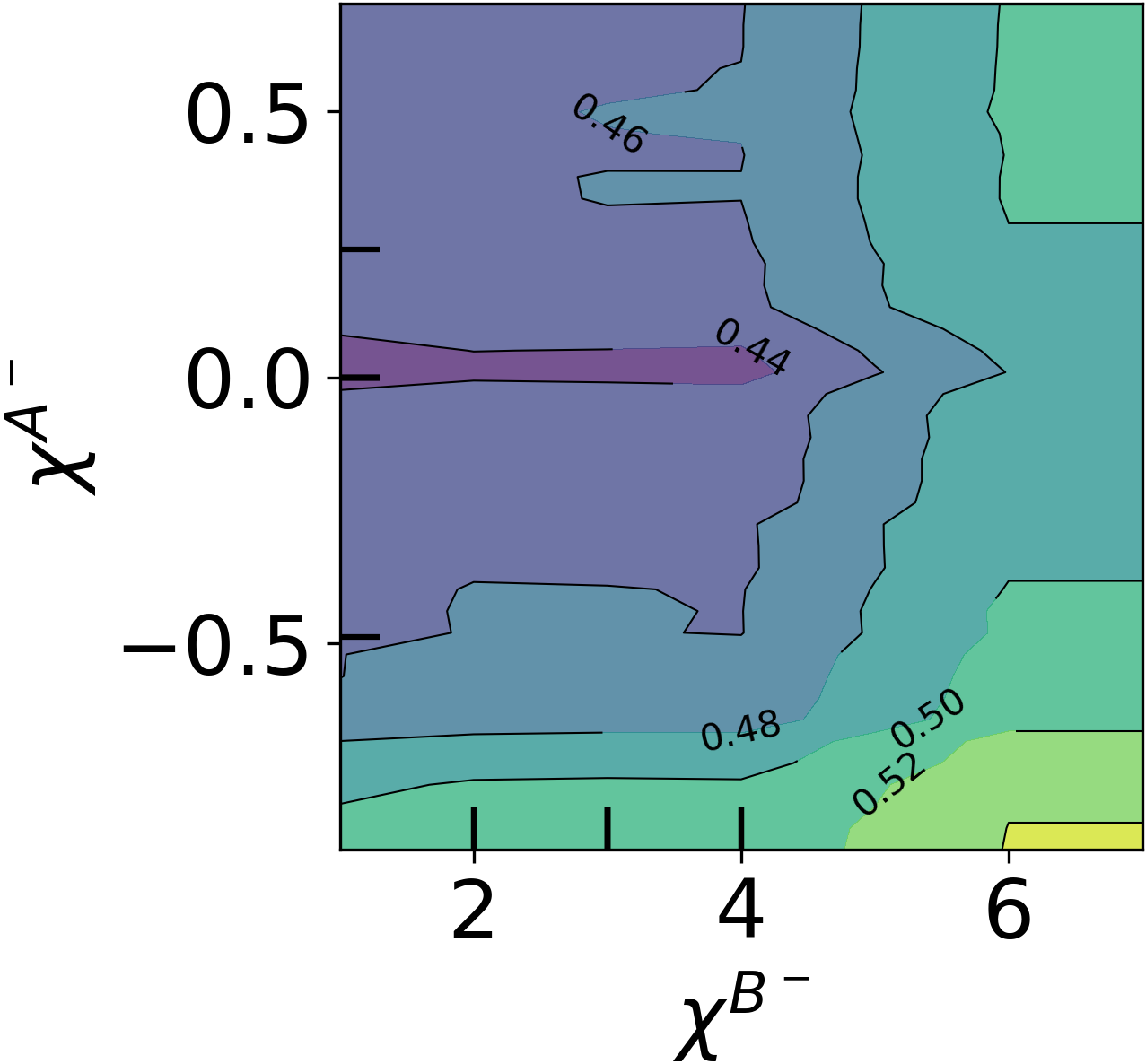}
        \vspace{-5mm}
        \caption{}\label{fig:PartDep_OriginalStability3}
        \end{subfigure}
    % \hfill
        \begin{subfigure}{0.47\textwidth}
        \centering
        \includegraphics[width=\linewidth]{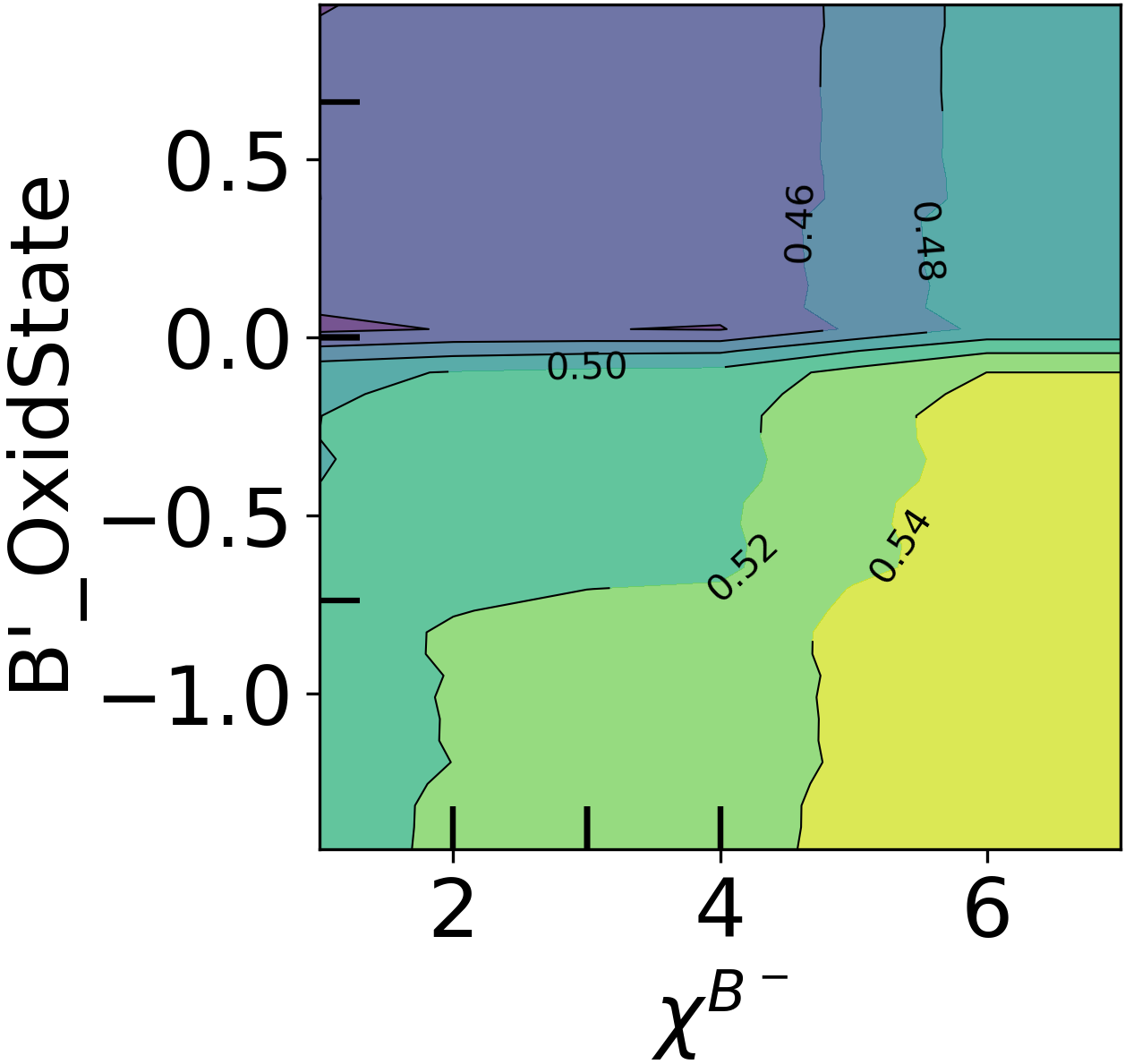}
        \vspace{-5mm}
        \caption{}\label{fig:PartDep_OriginalStability4}
        \end{subfigure}
    % \hfill
        \begin{subfigure}{0.45\textwidth}
        \centering
        \includegraphics[width=\linewidth]{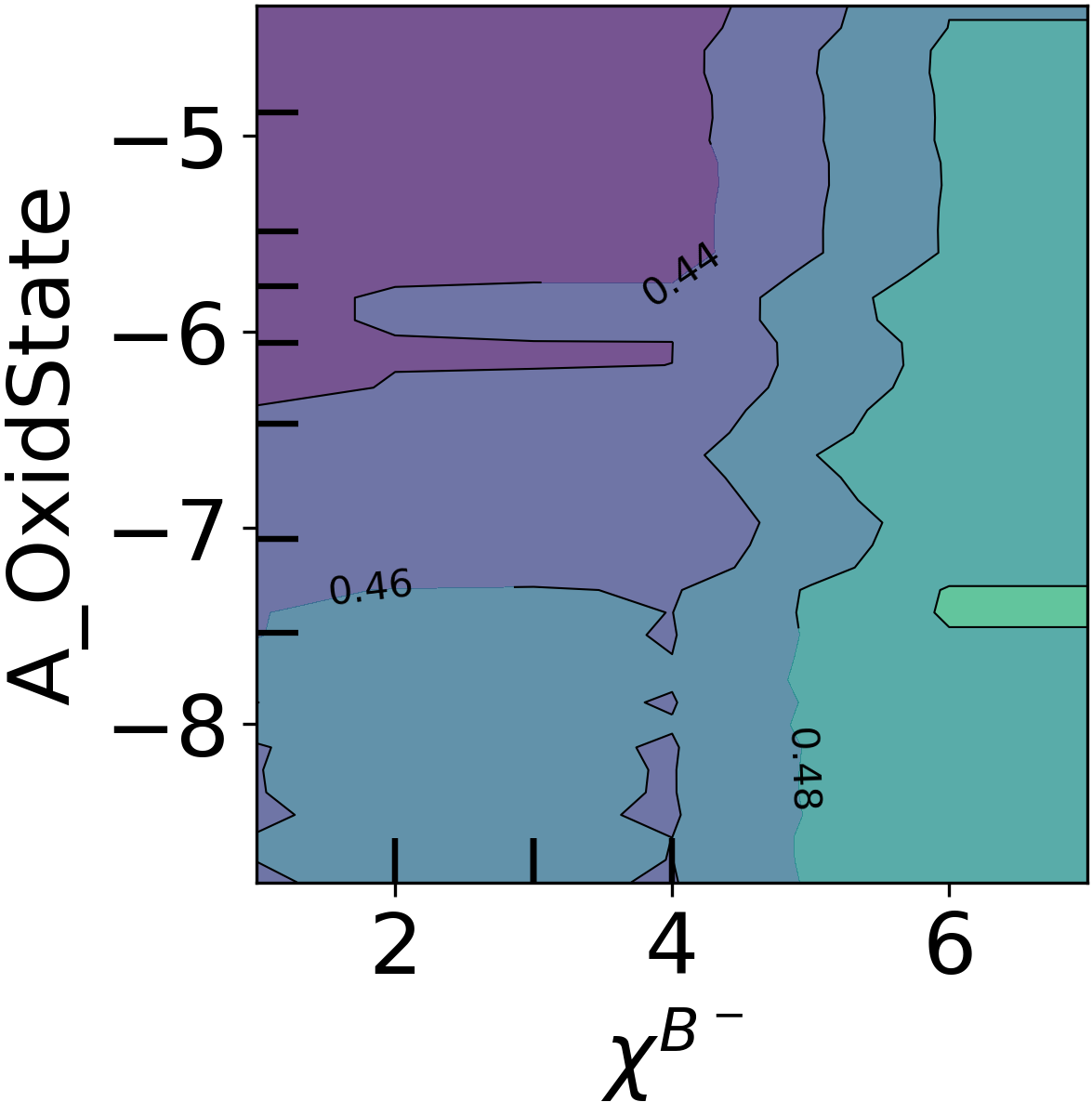}
        \vspace{-5mm}
        \caption{}\label{fig:PartDep_OriginalStability5}
        \end{subfigure}
    % \hfill
        \begin{subfigure}{0.47\textwidth}
        \centering
        \includegraphics[width=\linewidth]{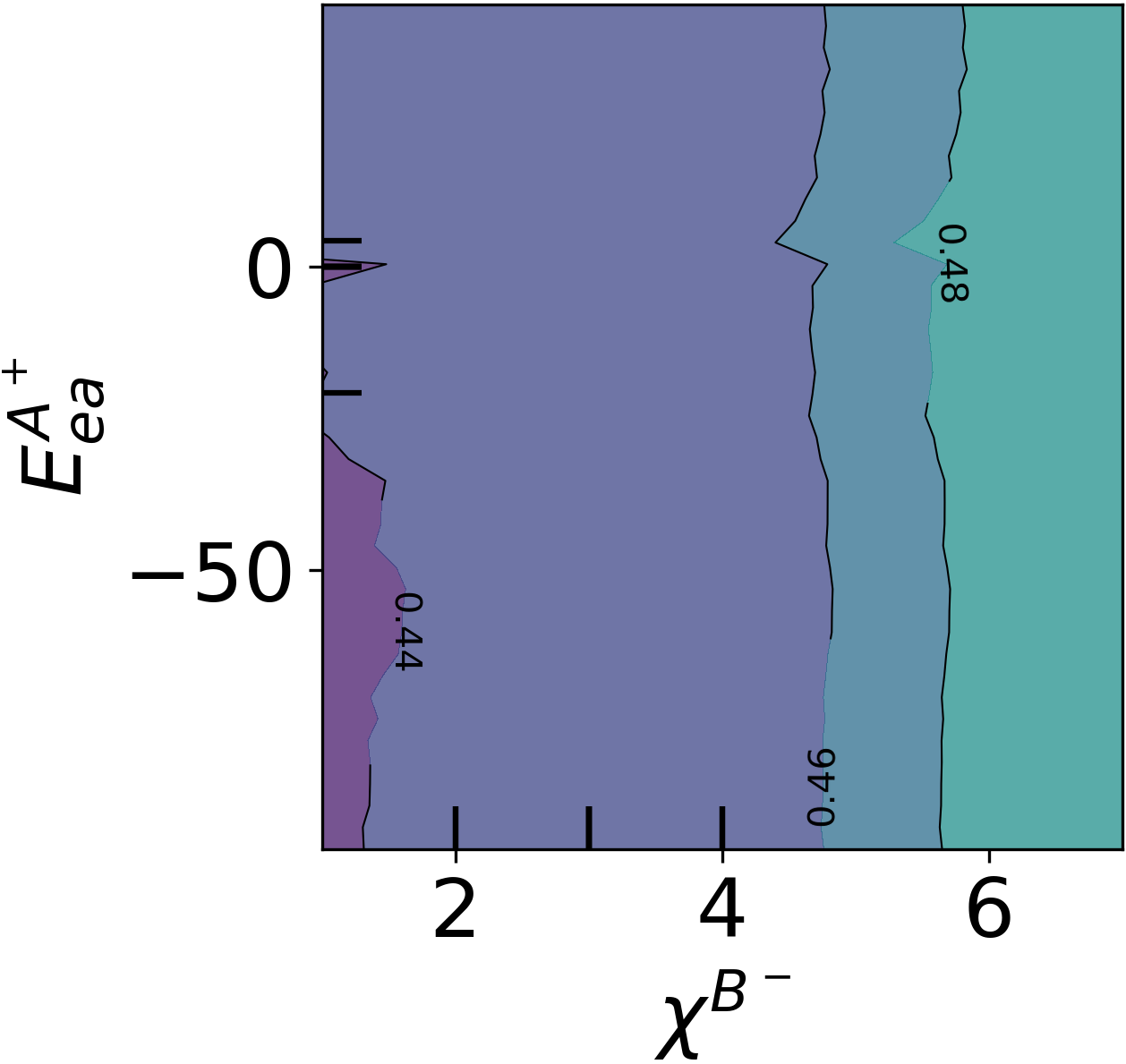}
        \vspace{-5mm}
        \caption{}\label{fig:PartDep_OriginalStability6}
        \end{subfigure}
    \end{minipage}
    \caption{TFI (left) and PFI (right). (a) Before hierarchical clustering and (b) after hierarchical clustering of perovskite--specific features towards prediction of stability. The hierarchical clustering dendrogram for Spearman rank--order correlations is shown in Appendix \ref{sub_sec:clusteringStability}. Partial dependency plots of formability on perovskite--specific features (c) $E_{ea}^{A-}$ (d) $\chi^{B+}$ (e) $\chi^{A-}$ (f) \textit{B'\_OxidState} (g) \textit{A\_OxidState} (h) $E_{ea}^{A-}$ plotted against the highest importance feature $\chi^{B-}$.}
    %\vspace{-5mm}
    \label{fig:FIandPartDep_OriginalStability}
\end{figure}

\begin{figure}[!h]%tbp]
    \centering
        \begin{subfigure}{0.4\textwidth}
        \centering
        \includegraphics[width=\linewidth]{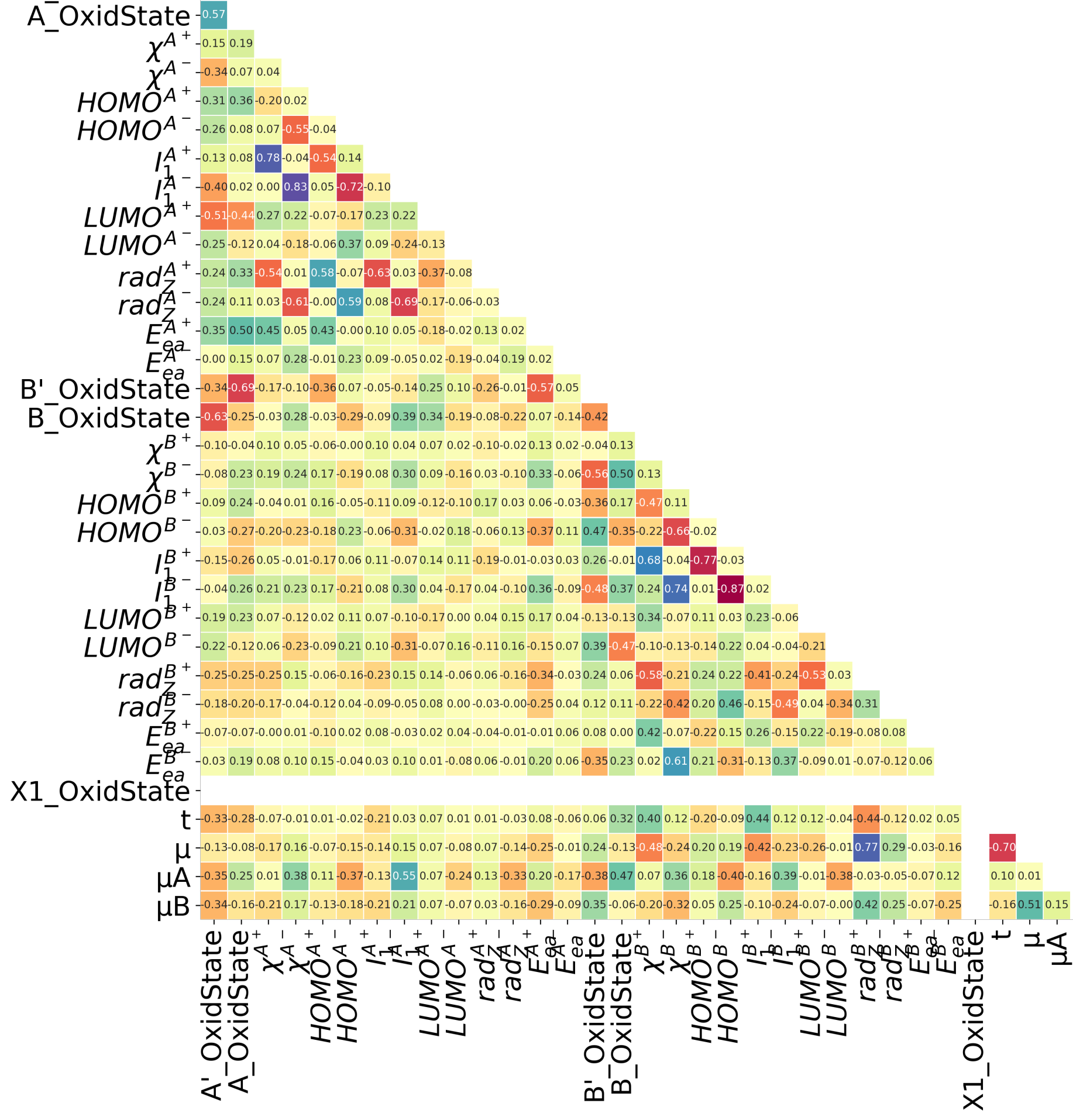}
        \vspace{-5mm}
        \caption{}\label{fig:OriginalStab_PearsonsFull1}
        \end{subfigure}
    \hspace{5mm}
        \begin{subfigure}{0.4\textwidth}
        \centering
        \includegraphics[width=\linewidth]{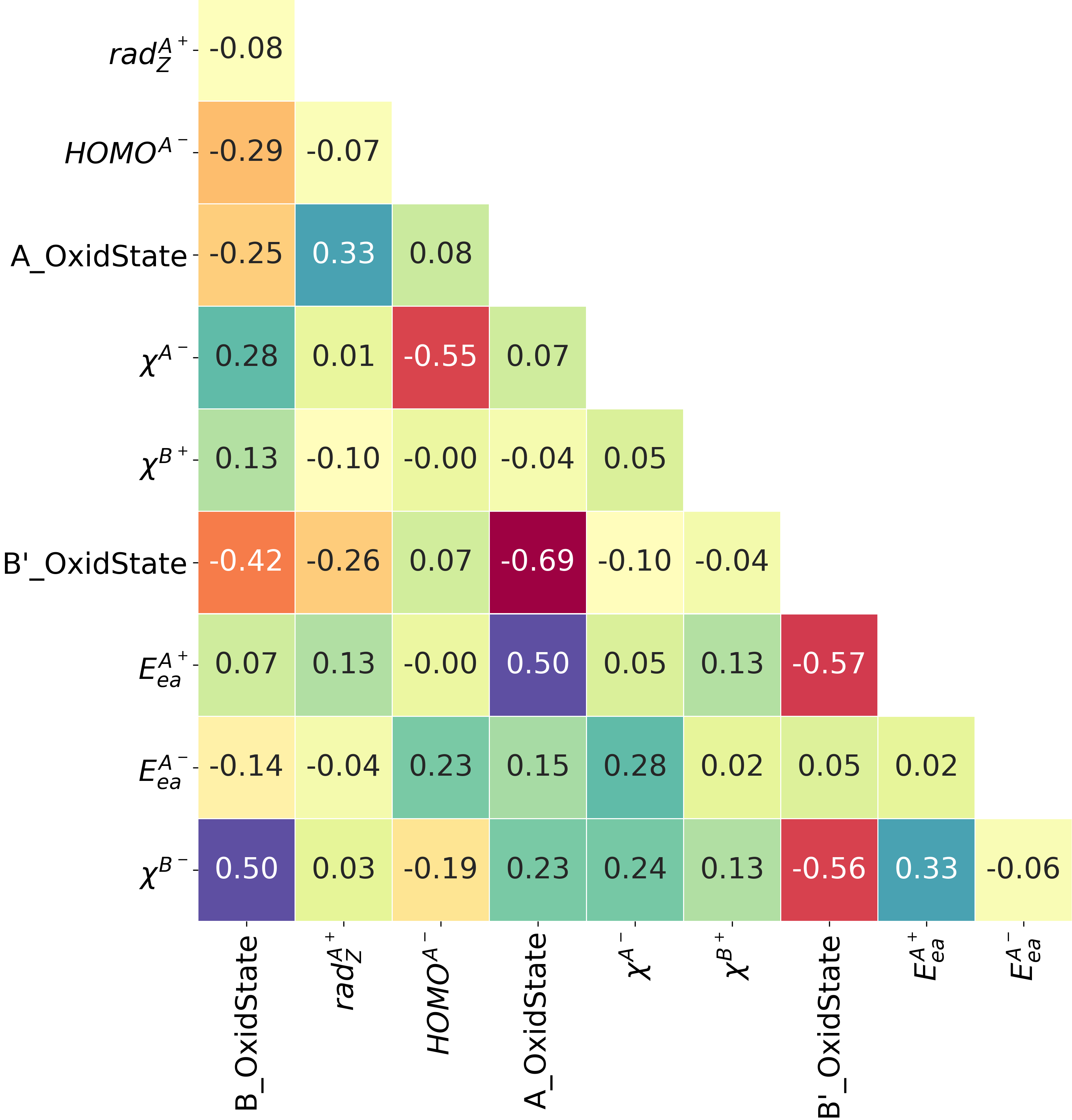}
        \vspace{-5mm}
        \caption{}\label{fig:OriginalStab_PearsonsReduced1}
        \end{subfigure}
    \vspace{-3mm}
    \caption{Pearson's correlation heatmaps for perovskite--specific features (a) before and (b) after hierarchical clustering towards prediction of stability.}
    \label{fig:}
\end{figure}

 The features of the original dataset were similarly analysed with regard to stability. The TFI and PFI initially indicated that the outer orbital energies for the B--site atom and the tolerance factor $t$ (Fig. \ref{fig:StabOriginalFI_pre}) were important in deciding the stability of perovskites. A significant lack of distinctness towards predicting the stability is also observed as evident in Fig. \ref{fig:StabOriginalFI_pre}. The multi--collinearity among the features are shown as Pearson's correlation heatmaps, before and after hierarchical clustering in Figs. \ref{fig:OriginalStab_PearsonsFull1} and \ref{fig:OriginalStab_PearsonsReduced1} respectively.
%%%
%%%
% \clearpage
\par After hierarchical clustering the features were pruned based on the optimal cutoff for accuracy of stability prediction, as shown in Fig. \ref{fig:StabOriginalFI_dendro} in Appendix \ref{sub_sec:clusteringStability}. The TFI and PFI shown in Fig. \ref{fig:StabOriginalFI_post} indicate with greater clarity that the electronegativity $\chi$ of the B--site atom and the electron affinity of the A--site atom could also be important underlying factors for stability.
%%%
\par The partial dependencies of stability on the important perovskite--specific features after hierarchical clustering are shown in Fig. {\ref{fig:FIandPartDep_OriginalStability}}. The probability of a stable composition is greater than 0.5 when the $\chi^{A-}$ is less than -0.5 (Fig. {\ref{fig:PartDep_OriginalStability3}}) and \textit{B'\_OxidState} is less than 0.0 (Fig. {\ref{fig:PartDep_OriginalStability4}}) for all $\chi^{B-}$ greater than 5.
%%%
\vspace{\topsep}
% \pagebreak
\subsubsection{Using novel generic features}
\begin{figure}[!h]
    \begin{minipage}[!b]{0.57\linewidth}
    \centering
        \begin{subfigure}{\textwidth}
        \centering
        \includegraphics[width=\linewidth]{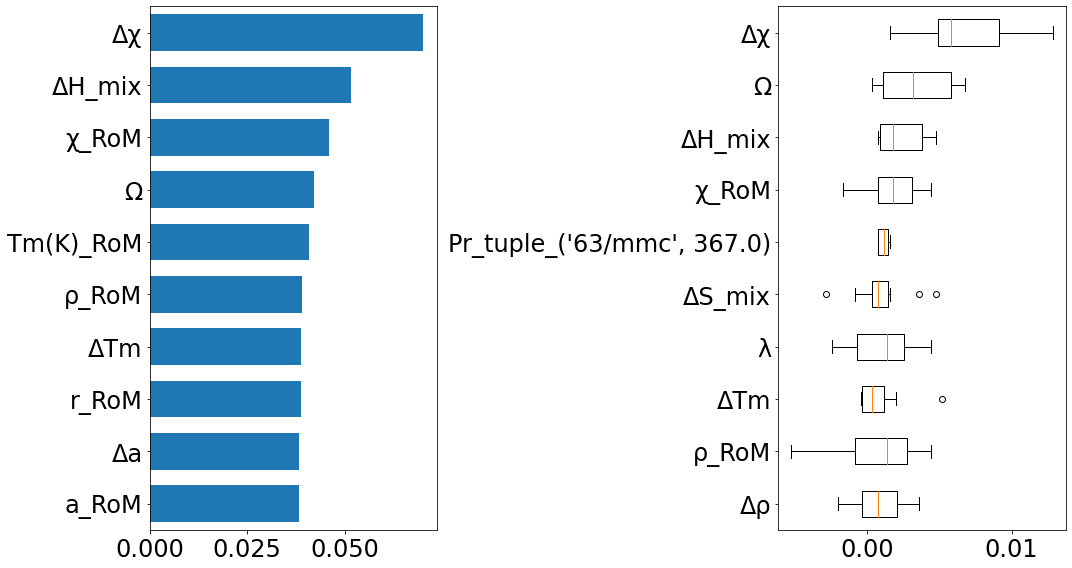}
        %\vspace{-8mm}
        \caption{}\label{fig:StabNewFI_pre}
        \end{subfigure}
    \centering
        \begin{subfigure}{\textwidth}
        \centering
        \includegraphics[width=\linewidth]{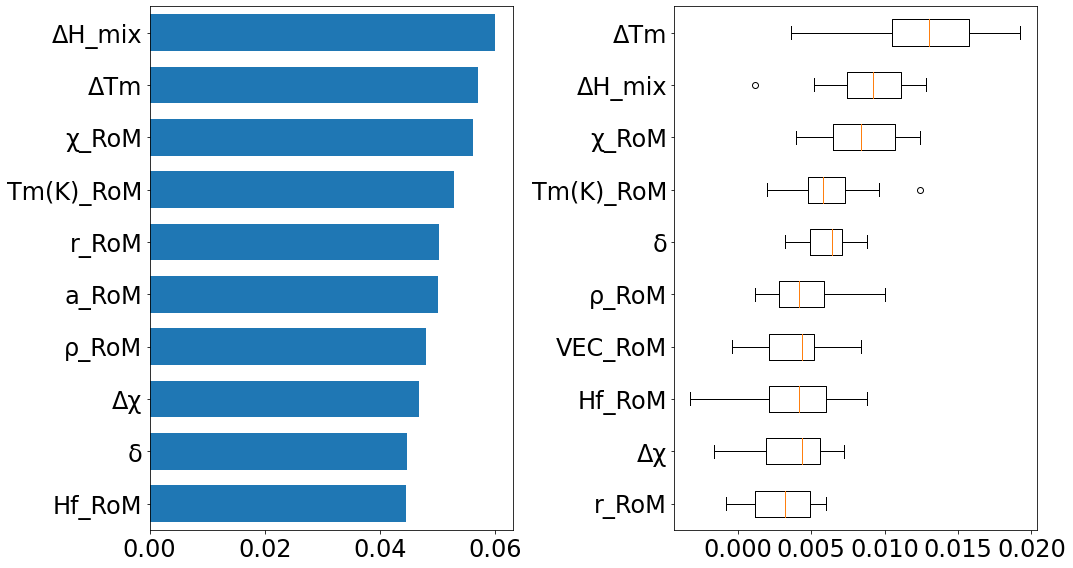}
        %\vspace{-8mm}
        \caption{}\label{fig:StabNewFI_post}
        \end{subfigure}
    \end{minipage}
% \hspace{0.5cm}
    \begin{minipage}[!b]{0.41\linewidth}
    \centering
        \begin{subfigure}{0.48\textwidth}
        \centering
        \includegraphics[width=\linewidth]{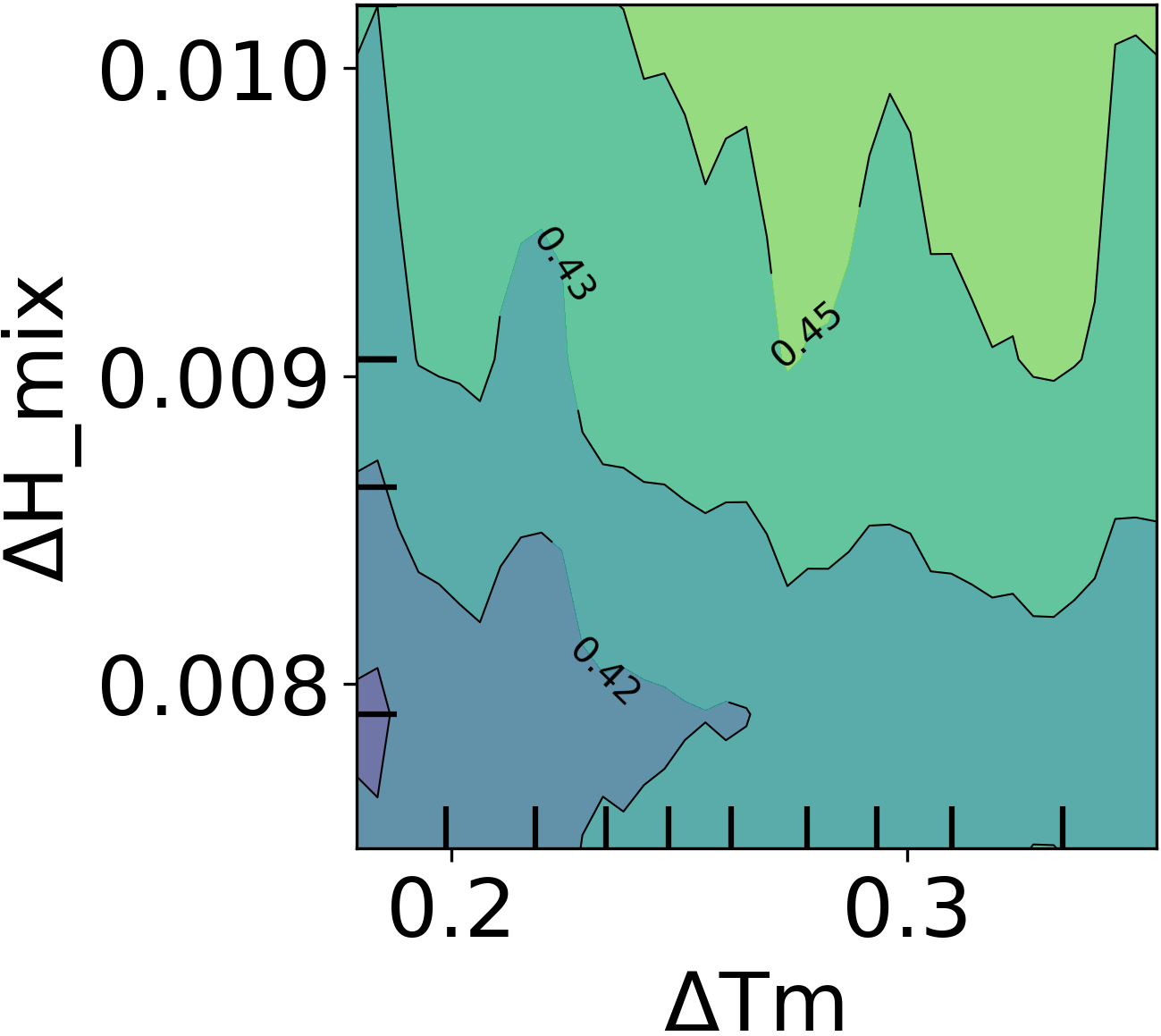}
        \vspace{-5mm}
        \caption{}\label{fig:PartDep_GenericStability1}
        \end{subfigure}
    % \hfill
        \begin{subfigure}{0.44\textwidth}
        \centering
        \includegraphics[width=\linewidth]{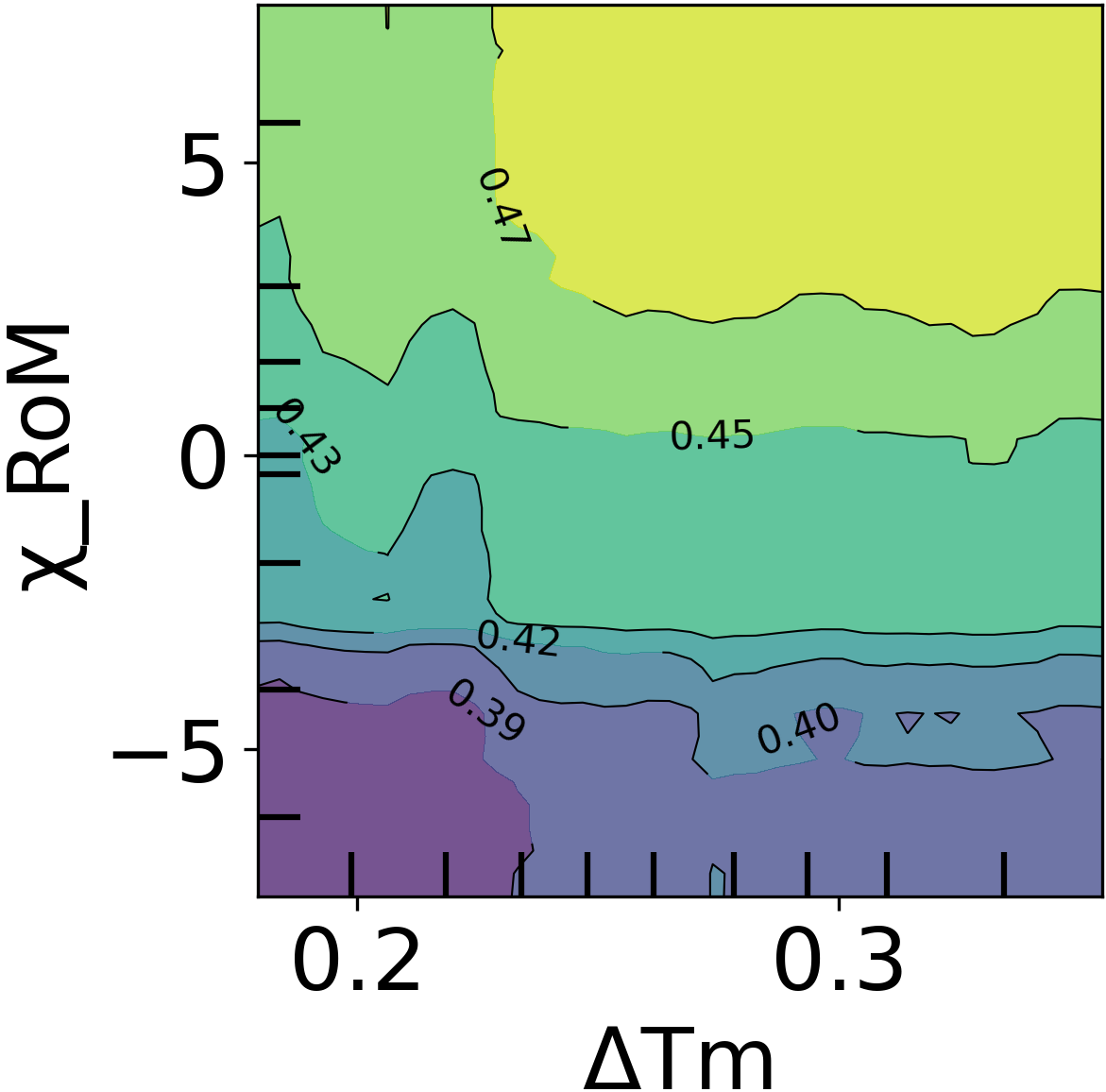}
        %\vspace{-5mm}
        \caption{}\label{fig:PartDep_GenericStability2}
        \end{subfigure}
    % \hfill
        \begin{subfigure}{0.45\textwidth}
        \centering
        \includegraphics[width=\linewidth]{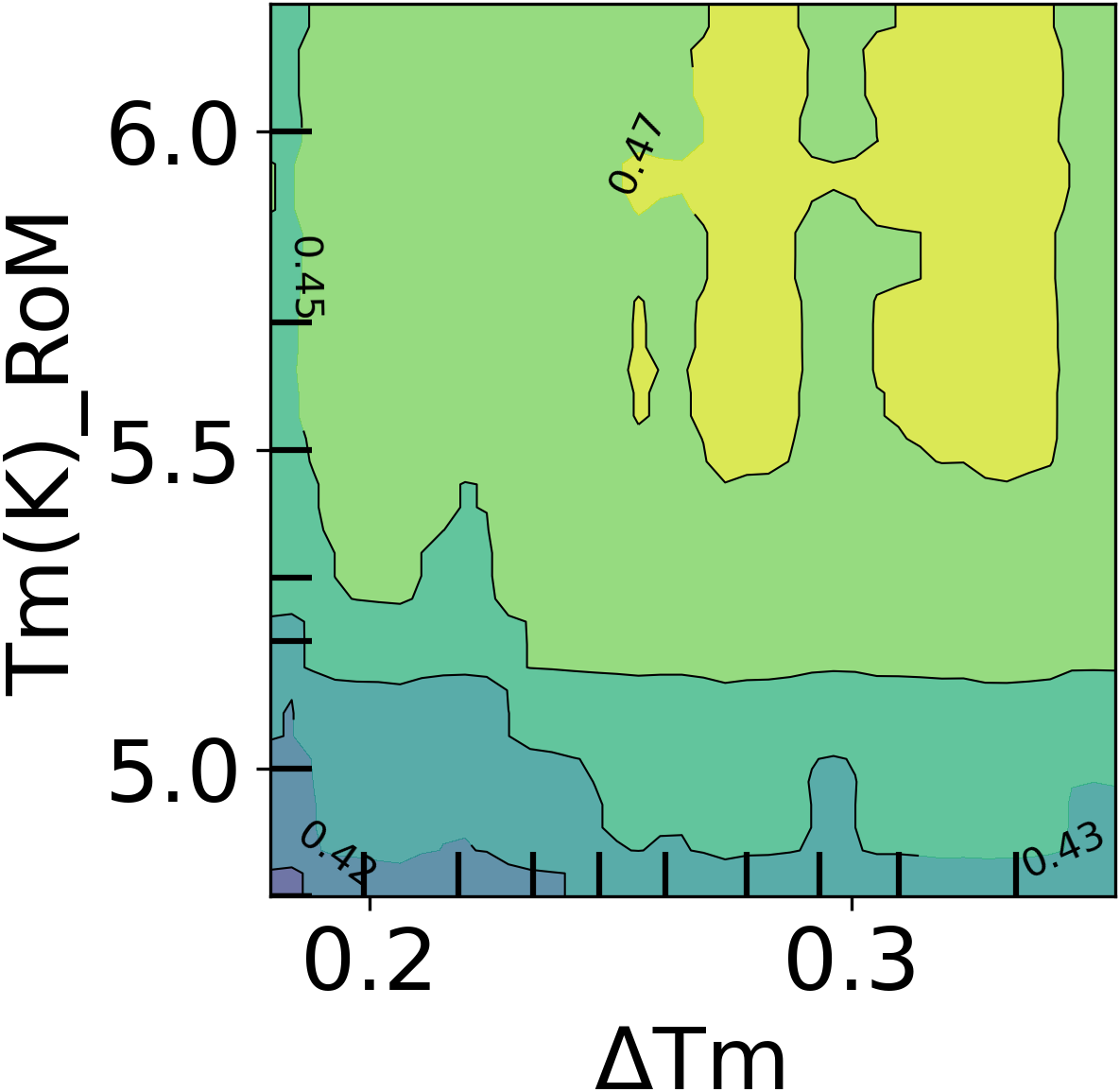}
        %\vspace{-5mm}
        \caption{}\label{fig:PartDep_GenericStability3}
        \end{subfigure}
    % \hfill
        \begin{subfigure}{0.45\textwidth}
        \centering
        \includegraphics[width=\linewidth]{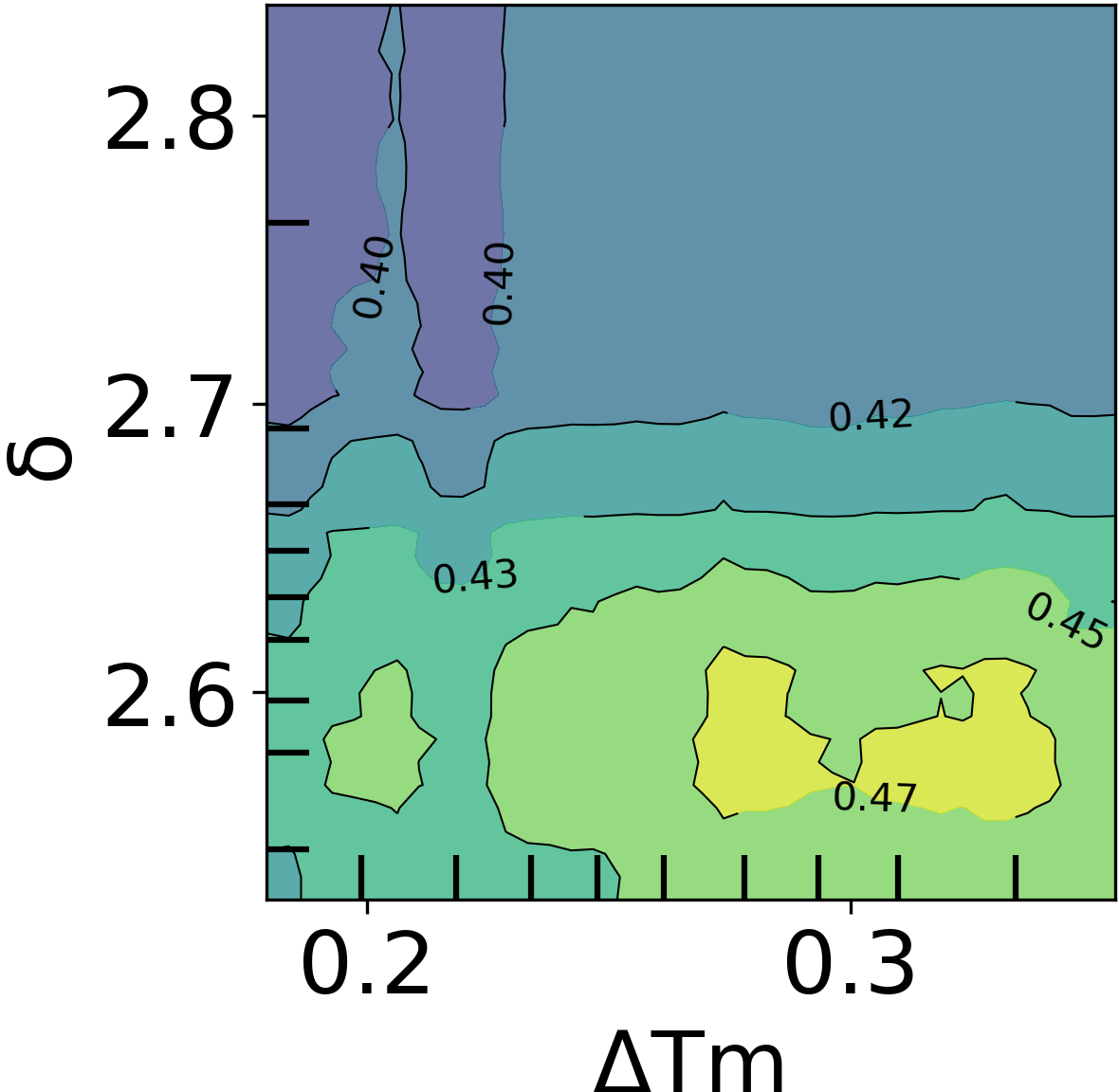}
        %\vspace{-5mm}
        \caption{}\label{fig:PartDep_GenericStability4}
        \end{subfigure}
    % \hfill
        \begin{subfigure}{0.43\textwidth}
        \centering
        \includegraphics[width=\linewidth]{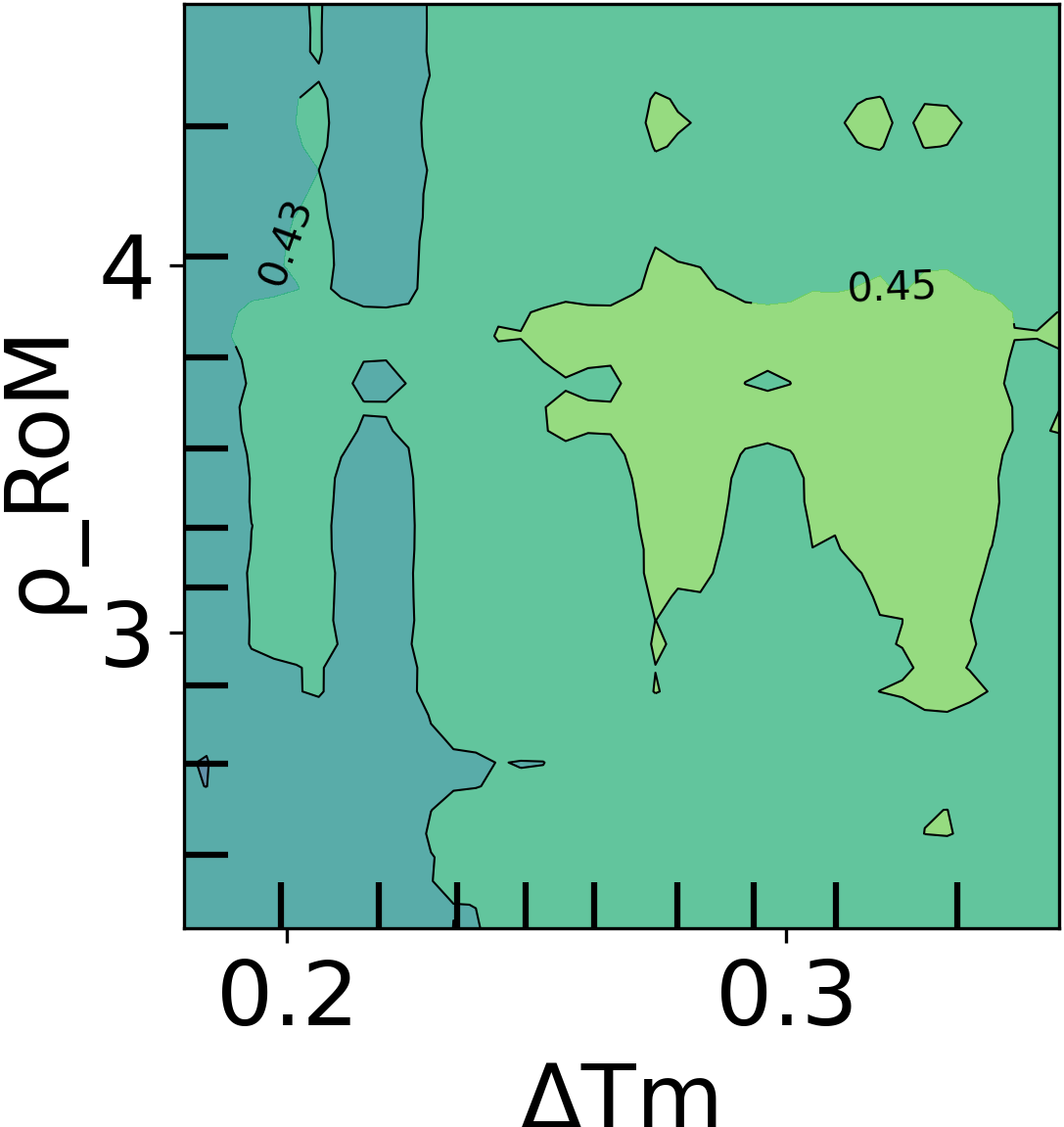}
        %\vspace{-5mm}
        \caption{}\label{fig:PartDep_GenericStability5}
        \end{subfigure}
    % \hfill
        \begin{subfigure}{0.47\textwidth}
        \centering
        \includegraphics[width=\linewidth]{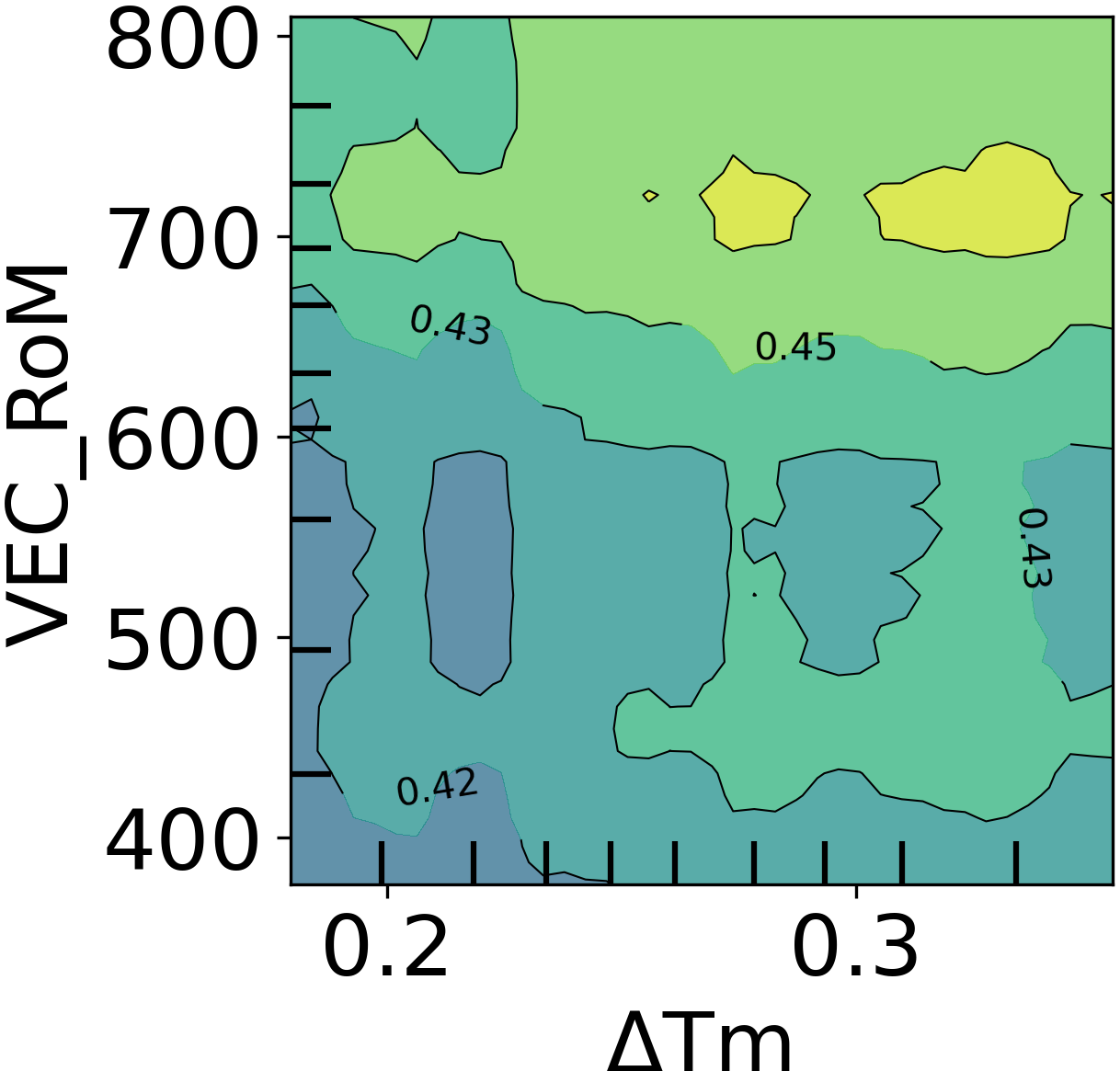}
        %\vspace{-5mm}
        \caption{}\label{fig:PartDep_GenericStability6}
        \end{subfigure}
    \end{minipage}
    \vspace{-3mm}
    \caption{TFI (left) and PFI (right) (a) before hierarchical clustering and (b) after hierarchical clustering of novel generic features towards prediction of stability. The hierarchical clustering dendrogram for Spearman rank--order correlations is shown in Appendix \ref{sub_sec:clusteringStability}. Partial dependency plots of formability on perovskite--specific features (c) $\Delta H$\textit{\_mix} (d) $\chi$\textit{\_RoM} (e) \textit{Tm (K)\_RoM} (f) $\delta$ (g) $\rho$\textit{\_RoM} (h) \textit{VEC\_RoM} plotted against the highest importance feature $\Delta$\textit{Tm}.}
    \label{fig:PartDep_GenericStability}
\end{figure}

\begin{figure}[!hhh]%tbp]
    \centering
        \begin{subfigure}{0.4\textwidth}
        \centering
        \includegraphics[width=\linewidth]{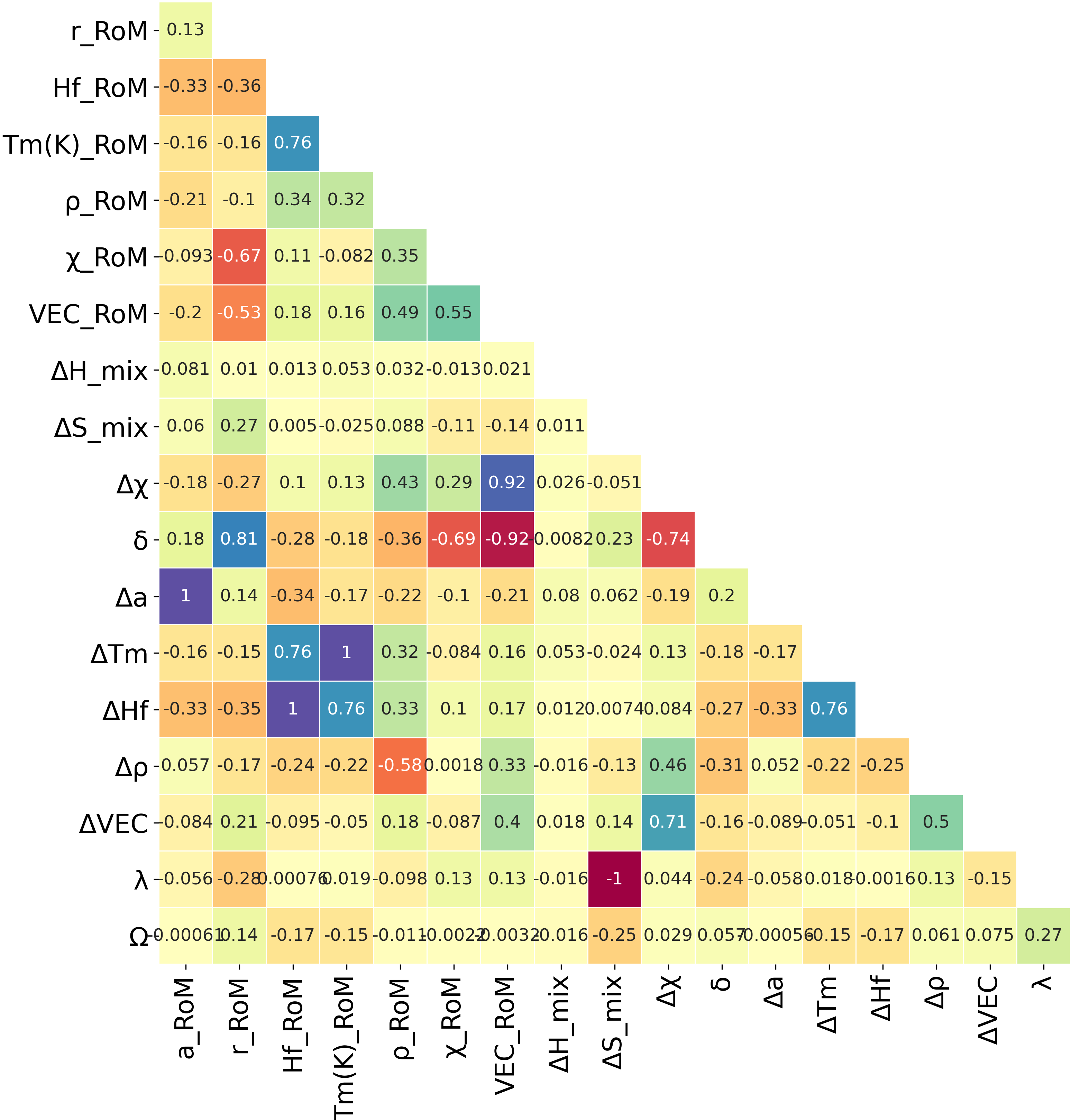}
        \vspace{-5mm}
        \caption{}\label{fig:GenericStab_PearsonsFull1}
        \end{subfigure}
    \hspace{5mm}
        \begin{subfigure}{0.4\textwidth}
        \centering
        \includegraphics[width=\linewidth]{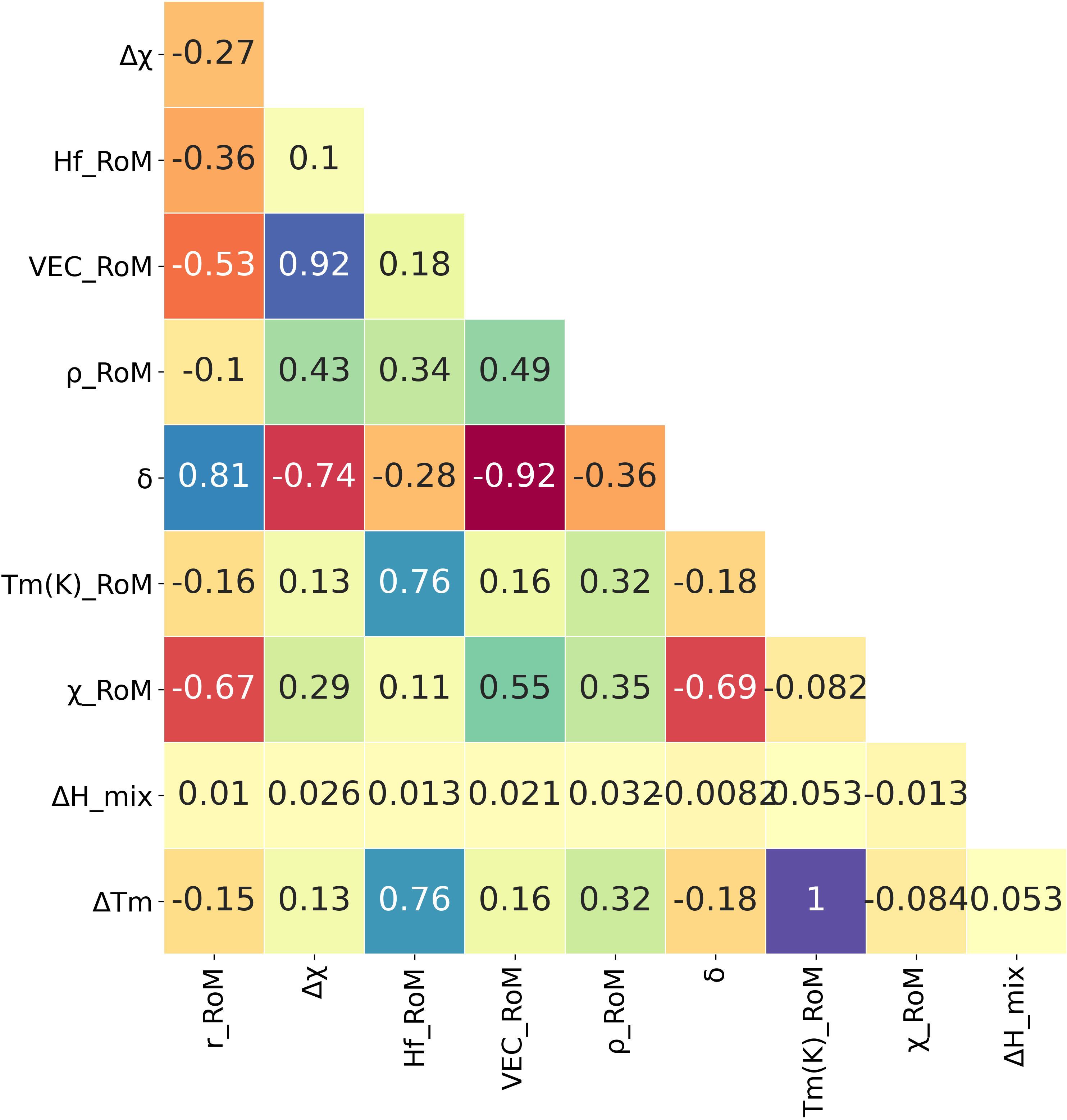}
        \vspace{-5mm}
        \caption{}\label{fig:GenericStab_PearsonsReduced1}
        \end{subfigure}
    \vspace{-3mm}
    \caption{Pearson's correlation heatmaps for generic features (a) before and (b) after hierarchical clustering towards prediction of stability.}
    \label{fig:}
\end{figure}

%%%
The novel generic features were similarly analysed with regard to stability.
The TFI and PFI (Fig. \ref{fig:StabNewFI_pre}) initially indicated a dependence on the electronegativity $\chi$, $\Omega$ parameter and mixing enthalpy $\Delta H_{mix}$. The hierarchical clustering of multi--collinear features shown in Fig. \ref{fig:StabNewFI_dendro} in Appendix \ref{sub_sec:clusteringStability} reinforces these inferences towards the same features shown by the TFI and PFI in Fig. \ref{fig:StabNewFI_post}. The Pearson's correlation matrices, before and after hierarchical clustering are depicted as heatmaps in Figs. {\ref{fig:GenericStab_PearsonsFull1}} and {\ref{fig:GenericStab_PearsonsReduced1}} respectively.
%%%
%%%
%%%
\par The partial dependencies of stability on the important generic features after hierarchical clustering is shown in Fig. {\ref{fig:PartDep_GenericStability}}. The probability of stable compositions occurring is greater than 0.47 when $\chi$\textit{\_RoM} is greater than 3 (Fig. {\ref{fig:PartDep_GenericStability2}}) and $\delta$ is around 2.6 $\AA$ for values of $\Delta$\textit{Tm} around 0.3 (Fig. {\ref{fig:PartDep_GenericStability4}}).
%%%
% \pagebreak
% \newpage
% \clearpage
%%%
\subsection{FI towards formability and stability simultaneously}
A significant overlap of features affecting the formation and stability of perovskites was observed, using both generic and perovskite--specific features. A multi--output classifier model was used to study the feature importances towards the prediction of both formability and stability simultaneously.
%%%
\subsubsection{Using perovskite--specific features}
\vspace{-\parsep}
%%%
\begin{figure}[!b]%tbp]
\centering
    \begin{subfigure}{0.45\textwidth}
        \centering
        \includegraphics[width=\linewidth]{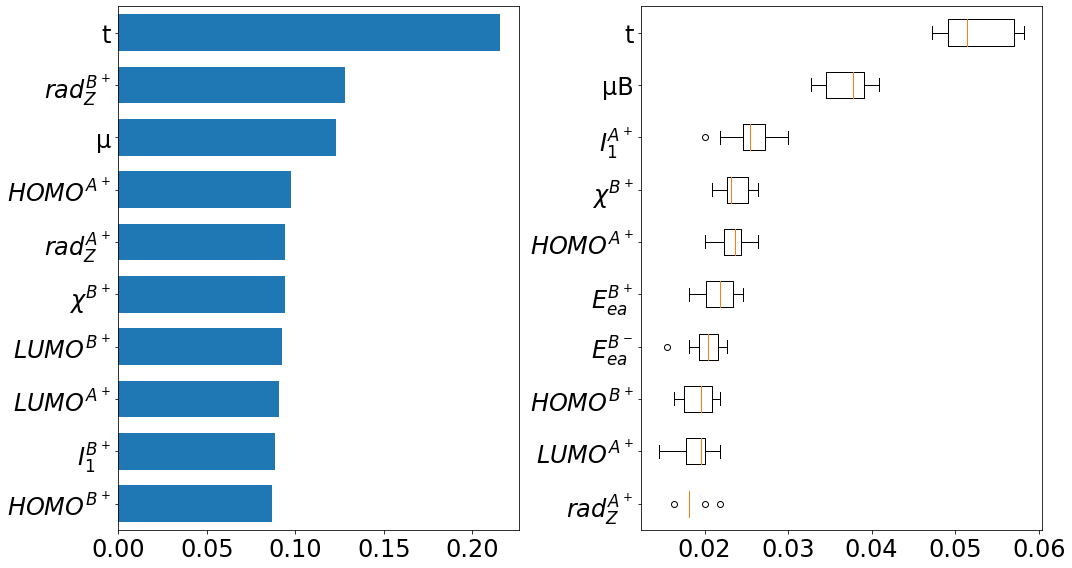}
        %\vspace{-8mm}
        \caption{}\label{fig:FormStabOriginalFI_pre}
    \end{subfigure}
    \centering
    \begin{subfigure}{0.45\textwidth}
        \centering
        \includegraphics[width=\linewidth]{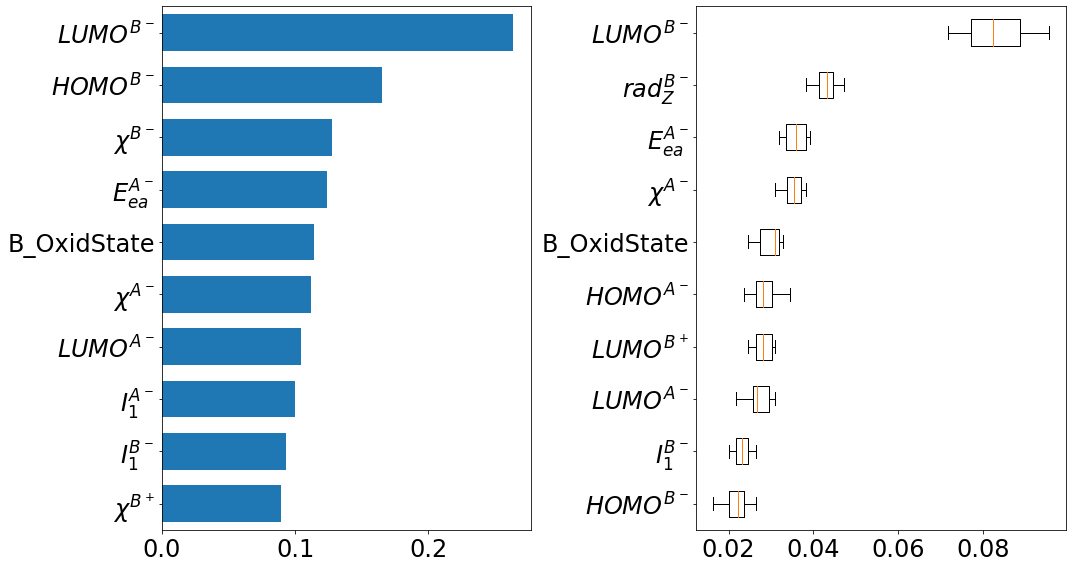}
        %\vspace{-8mm}
        \caption{}\label{fig:FormStabOriginalFI_post}
    \end{subfigure}
\caption{TFI (left) and PFI (right) (a) before hierarchical clustering and (b) after hierarchical clustering of perovskite--specific features towards prediction of formability and stability simultaneously. The hierarchical clustering dendrogram for Spearman rank--order correlations is shown in Appendix \ref{sub_sec:clusteringFormStab}.}
%\vspace{-5mm}
\label{fig:FormStabOriginalFI}
\end{figure}
%%%
\begin{figure}[!h]%tbp]
\centering
    \begin{subfigure}{0.4\textwidth}
        \centering
        \includegraphics[width=\linewidth]{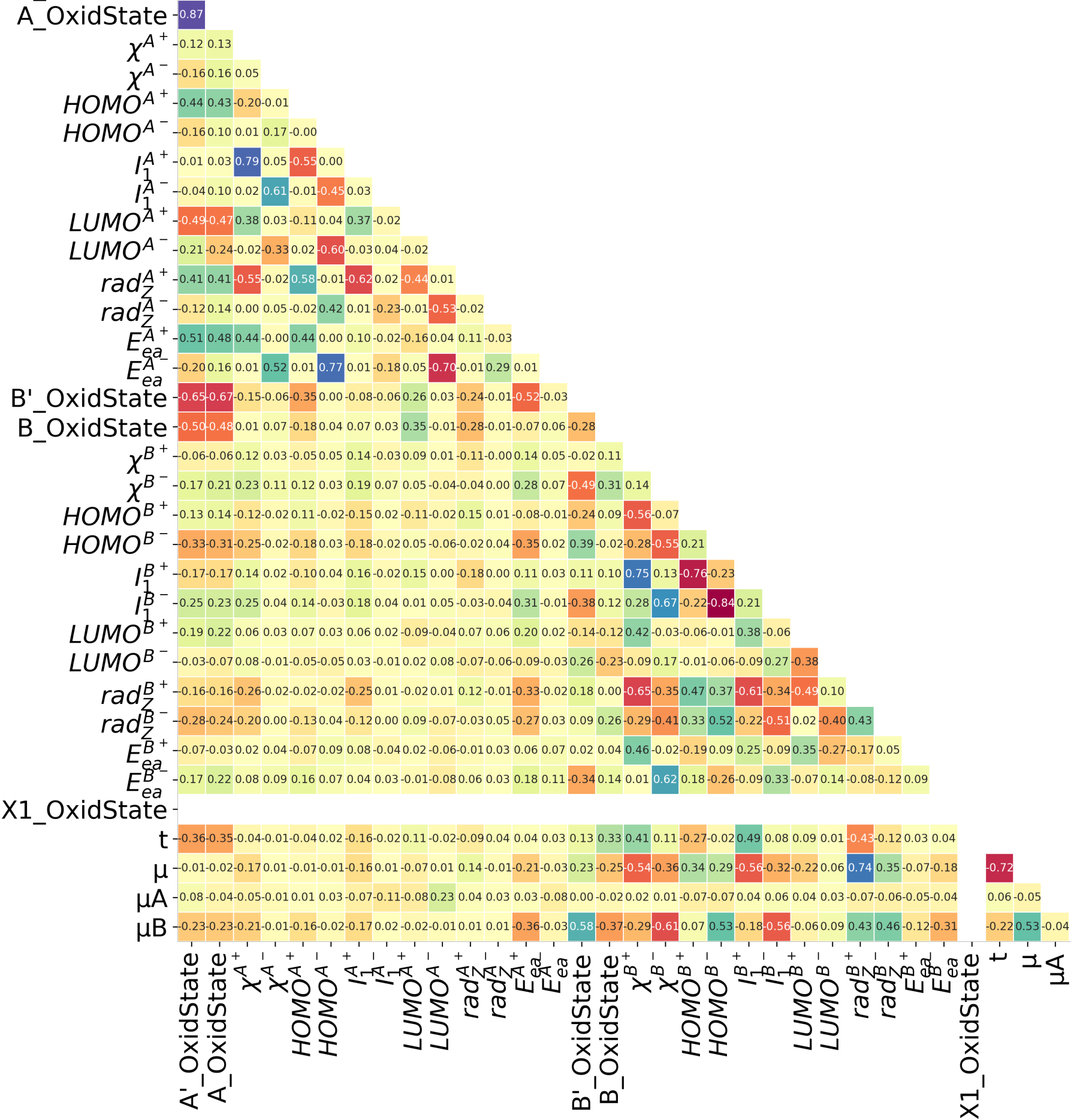}
        \vspace{-5mm}
        \caption{}\label{fig:OriginalFormStab_PearsonsFull1}
    \end{subfigure}
\hspace{5mm}
    \begin{subfigure}{0.4\textwidth}
        \centering
        \includegraphics[width=\linewidth]{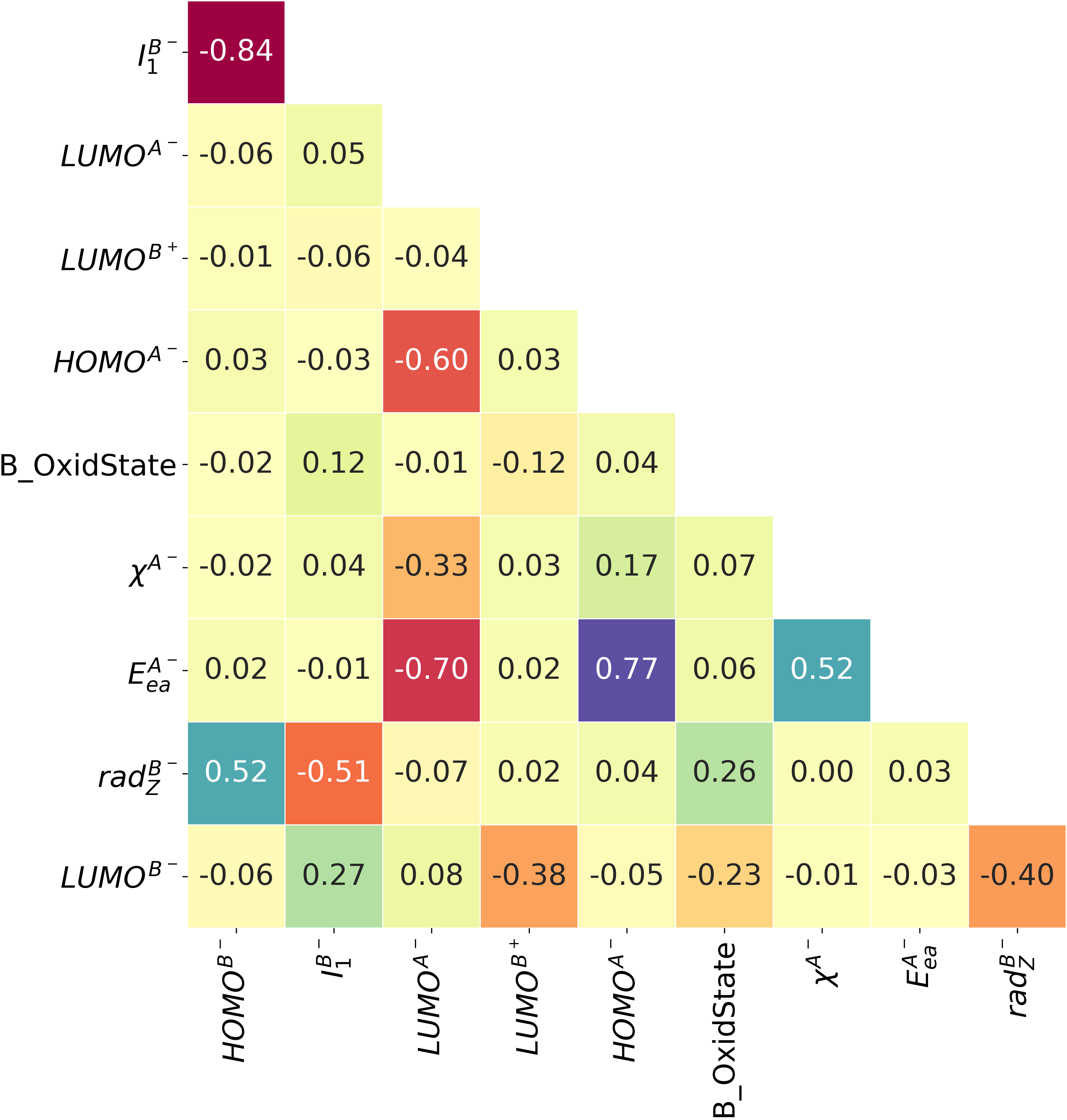}
        \vspace{-5mm}
        \caption{}\label{fig:OriginalFormStab_PearsonsReduced1}
    \end{subfigure}
\vspace{-3mm}
\caption{Pearson's correlation heatmaps for perovskite specific features (a) before and (b) after hierarchical clustering towards prediction of both formability and stability.}
\label{fig:}
\end{figure}
%%%
\vspace{-\parsep}
%%%
\par Among the perovskite--specific features of the database, the initial TFI and PFI indicated that the tolerance factor $t$ and the octahedral factors $\mu$ and $\mu_B$ have an influence on the prediction of formability and stability together, as shown in Fig. \ref{fig:FormStabOriginalFI_pre}.
%%%
\par The reduction in correlation among features are depicted as Pearson's correlation heatmaps, before and after hierarchical clustering in Figs. {\ref{fig:OriginalFormStab_PearsonsFull1}} and {\ref{fig:OriginalFormStab_PearsonsReduced1}} respectively. Following the  hierarchical clustering of multi--collinear features, as shown in Fig. \ref{fig:FormStabOriginalFI_dendro} in Appendix \ref{sub_sec:clusteringFormStab}, the lowest unoccupied molecular orbital energy for the B--site atom $LUMO^{B-}$ is seen to have a strong underlying effect in addition to the other significant parameters as observed in Fig. \ref{fig:FormStabOriginalFI_post}.
%%%
%\pagebreak
%\clearpage
%%%
\subsubsection{Using novel generic features}
\vspace{-\parsep}
\vspace{-\parsep}
\vspace{-\topsep}
%%%
\begin{figure}[!h]%htbp]
\centering
    \begin{subfigure}{0.45\textwidth}
        \centering
        \includegraphics[width=\linewidth]{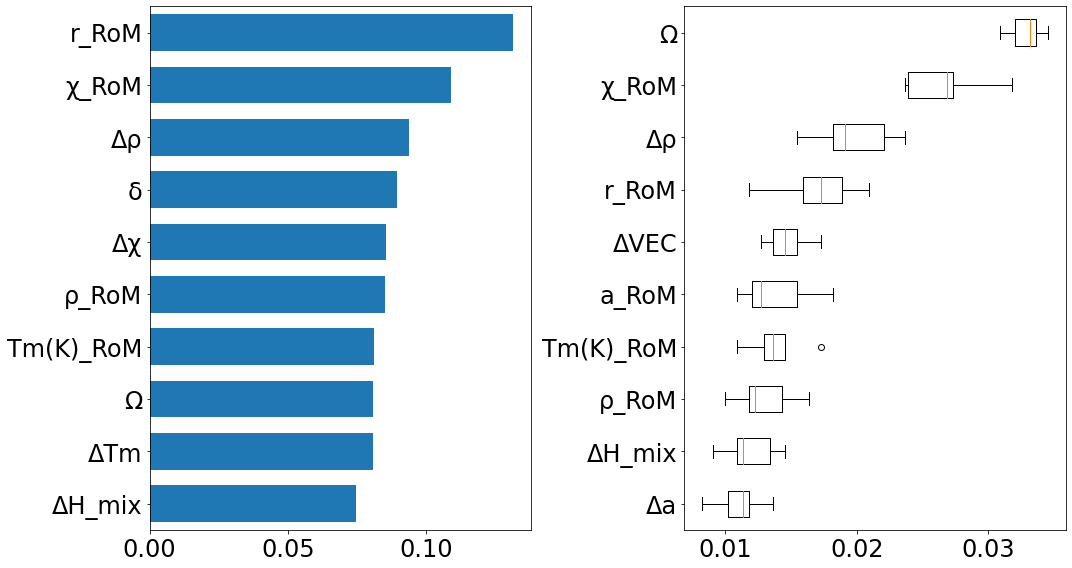}
        %\vspace{-8mm}
        \caption{}\label{fig:FormStabNewFI_pre}
    \end{subfigure}
    \centering
    \begin{subfigure}{0.45\textwidth}
        \centering
        \includegraphics[width=\linewidth]{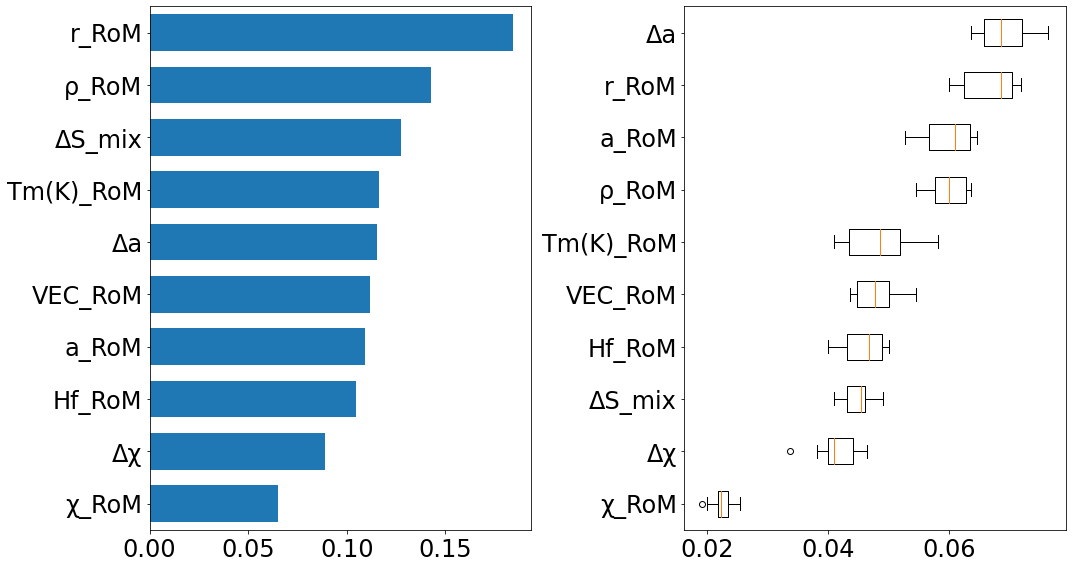}
        %\vspace{-8mm}
        \caption{}\label{fig:FormStabNewFI_post}
    \end{subfigure}
    \caption{TFI (left) and PFI (right) (a) before hierarchical clustering and (b) after hierarchical clustering of novel generic features towards prediction of formability and stability simultaneously. The hierarchical clustering dendrogram for Spearman rank--order correlations is shown in Appendix \ref{sub_sec:clusteringFormStab}.}
    %\vspace{-5mm}
    \label{fig:FormStabNewFI}
\end{figure}
%%%
\vspace{-\parsep}
\vspace{-\topsep}
%%%
\begin{figure}[!h]%tbp]
\centering
    \begin{subfigure}{0.4\textwidth}
        \centering
        \includegraphics[width=\linewidth]{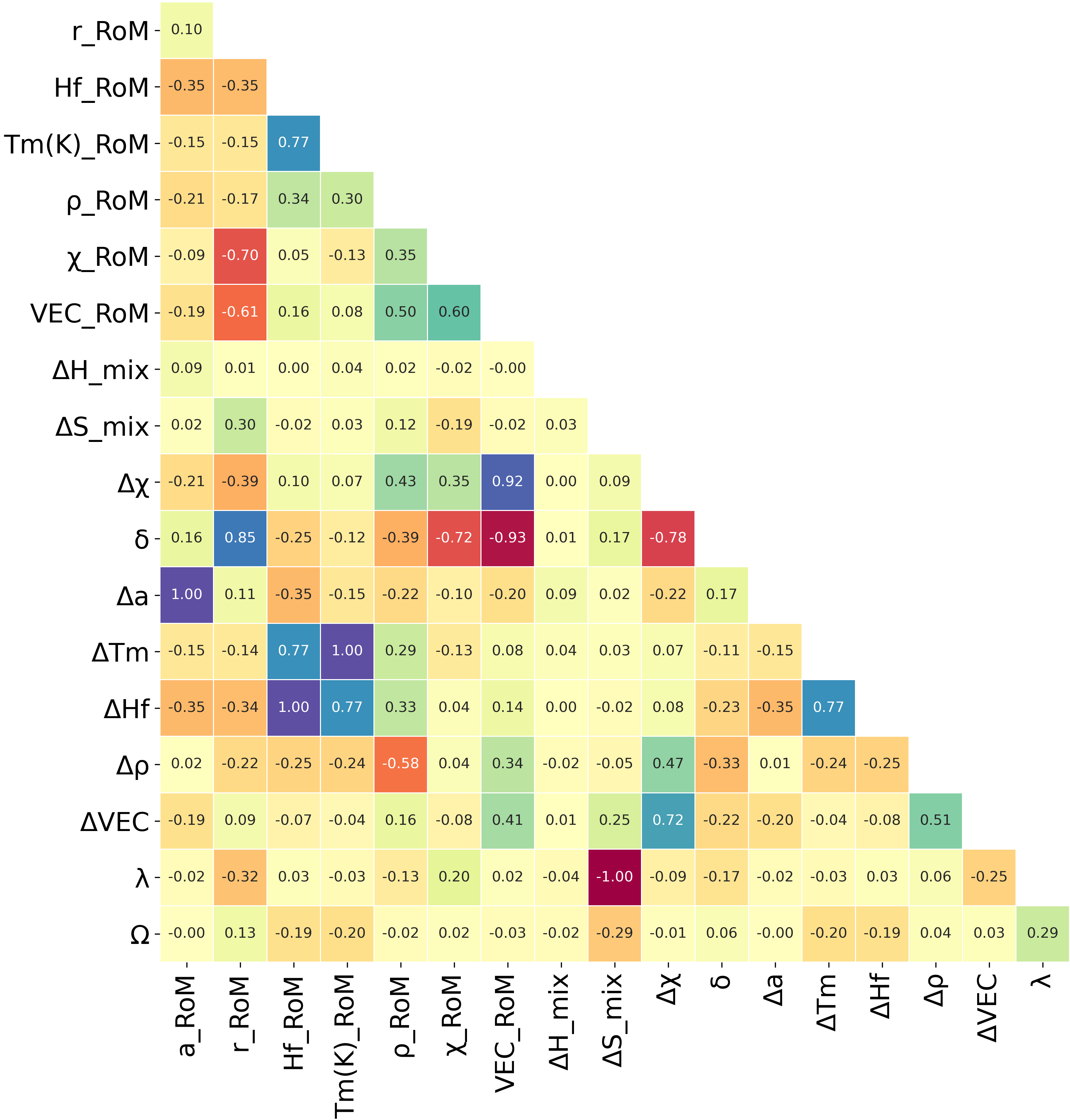}
        \vspace{-5mm}
        \caption{}\label{fig:GenericFormStab_PearsonsFull1}
    \end{subfigure}
\hspace{5mm}
    \begin{subfigure}{0.4\textwidth}
        \centering
        \includegraphics[width=\linewidth]{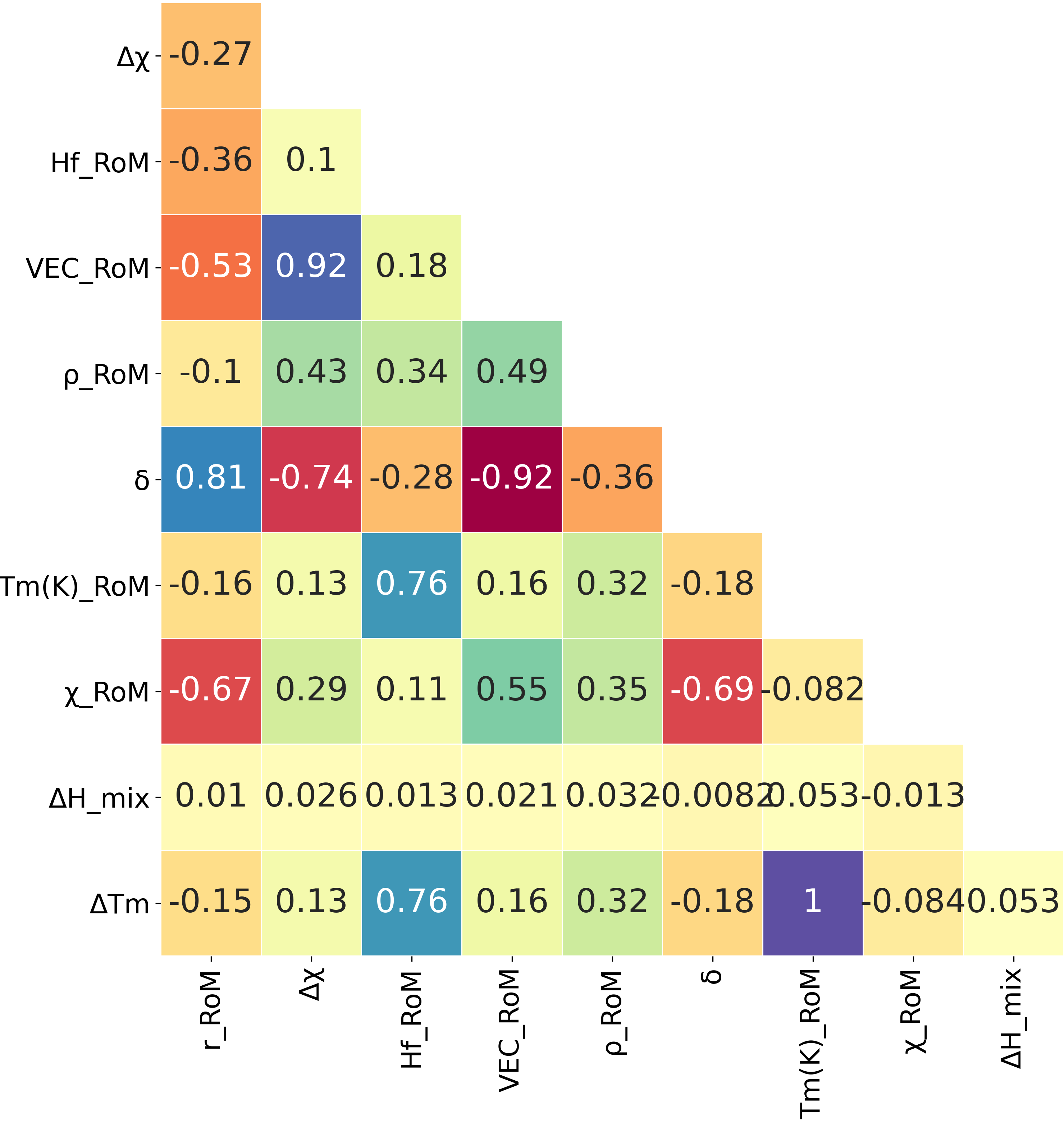}
        \vspace{-5mm}
        \caption{}\label{fig:GenericFormStab_PearsonsReduced1}
    \end{subfigure}
\vspace{-3mm}
\caption{Pearson's correlation heatmaps for generic features (a) before and (b) after hierarchical clustering towards prediction of both formability and stability.}
\label{fig:}
\end{figure}
%%%
\vspace{-\parsep}
\vspace{-\topsep}
\par The $\mathit{\Omega}$ parameter and electronegativity $\chi$ initially turned up as the important features alongwith the radius $r$ and density $\rho$ as seen in Fig. \ref{fig:FormStabNewFI_pre}. After hierarchical clustering (Fig. \ref{fig:FormStabNewFI_dendro} in Appendix \ref{sub_sec:clusteringFormStab}), the lattice parameter $a$, entropy of mixing $S_{mix}$ and melting temperature $T_m$ are also seen to exert an influence on the simultaneous prediction of formability and stability.
%%%
Pearson's correlation matrices, before and after hierarchical clustering are depicted as heatmaps in Figs. {\ref{fig:GenericFormStab_PearsonsFull1}} and {\ref{fig:GenericFormStab_PearsonsReduced1}} respectively.
%%%
%%%
% \clearpage
% \pagebreak
% \newpage
\section{\label{sec:results}Results}
%%%
\subsection{Multi--ouput classifier for formability and stability}
A multi--output RF classifier was used to predict the formability and stability simultaneously for a given perovskite oxide composition. Both the perovksite--specific and generic feature sets were used to determine the accuracies of prediction, separately.
%%%
\subsection{Using perovskite--specific features}
%%%
%\vspace{-2mm}
\begin{figure}[!h]%tbp]
    \centering
    \begin{subfigure}{0.24\textwidth}
        \centering
        \includegraphics[width=\linewidth]{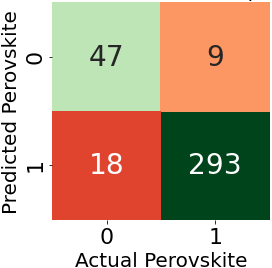}
        \vspace{-6mm}
        \caption{}\label{fig:FormStabOriginalFI_FormCMTest}
    \end{subfigure}
    \hfill
    \begin{subfigure}{0.24\textwidth}
        \centering
        \includegraphics[width=\linewidth]{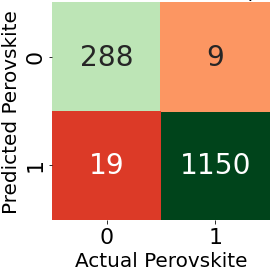}
        \vspace{-6mm}
        \caption{}\label{fig:FormStabOriginalFI_FormCMOverall}
    \end{subfigure}
    \hfill
    \begin{subfigure}{0.24\textwidth}
        \centering
        \includegraphics[width=\linewidth]{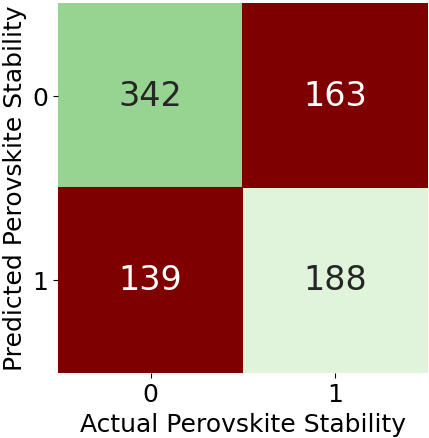}
        \vspace{-6mm}
        \caption{}\label{fig:FormStabOriginalFI_StabCMTest}
    \end{subfigure}
    \hfill
    \begin{subfigure}{0.24\textwidth}
        \centering
        \includegraphics[width=\linewidth]{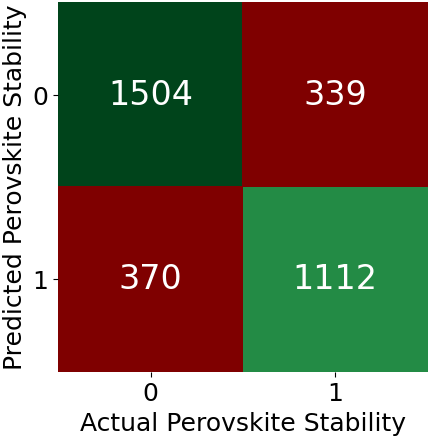}
        \vspace{-6mm}
        \caption{}\label{fig:FormStabOriginalFI_StabCMOverall}
    \end{subfigure}
    \caption{Heatmap of the confusion matrix for (a) test predictions and (b) overall predictions of formability and (c) test predictions and (d) overall predictions of stability using perovskite--specific features. The diagonal entries (green) are the correctly classified categories and the off--diagonal entries (red) are the wrongly classified ones.}
    \label{fig:}
\end{figure}
\begin{figure}[!h]%tbp]
\centering
    \begin{subfigure}{0.24\textwidth}
        \centering
        \includegraphics[width=\linewidth]{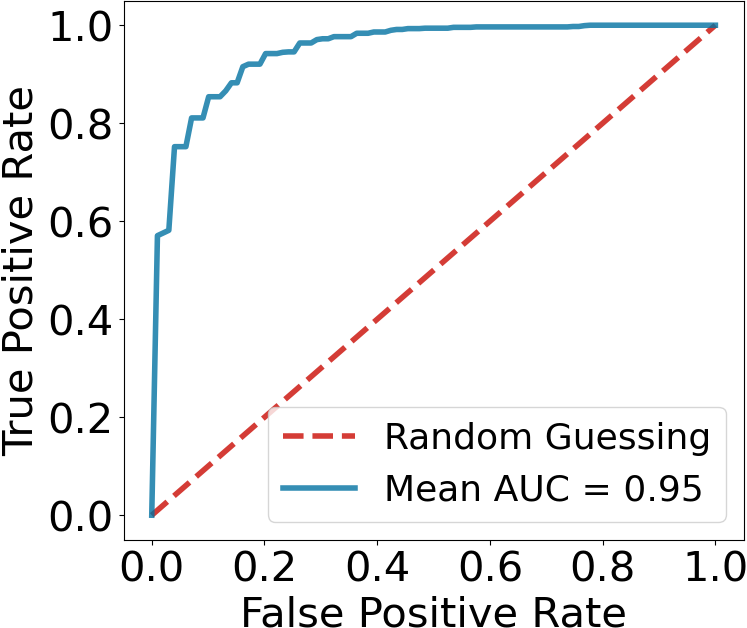}
        \vspace{-6mm}
        \caption{}\label{fig:FormStabOriginalFI_FormROC}
    \end{subfigure}
    \hfill
    \begin{subfigure}{0.24\textwidth}
        \centering
        \includegraphics[width=\linewidth]{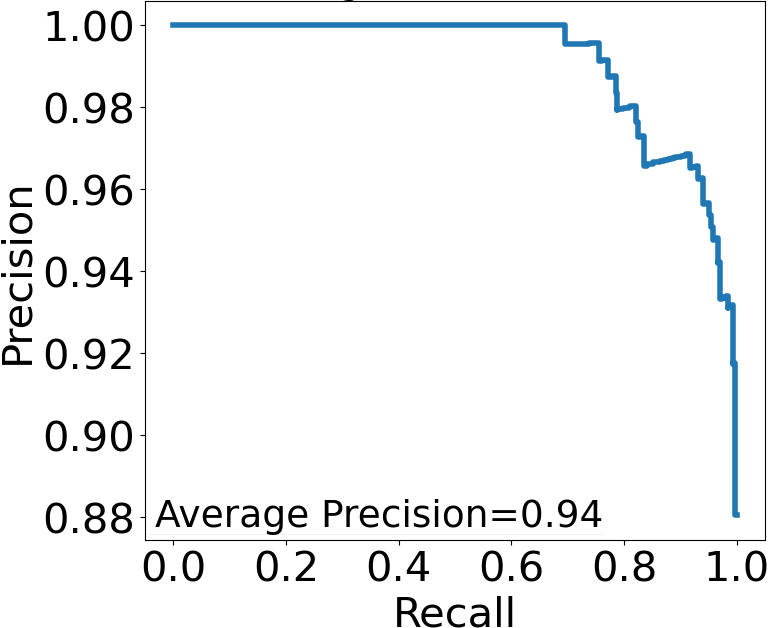}
        \vspace{-6mm}
        \caption{}\label{fig:FormStabOriginalFI_FormPRC}
    \end{subfigure}
    \hfill
    \begin{subfigure}{0.24\textwidth}
        \centering
        \includegraphics[width=\linewidth]{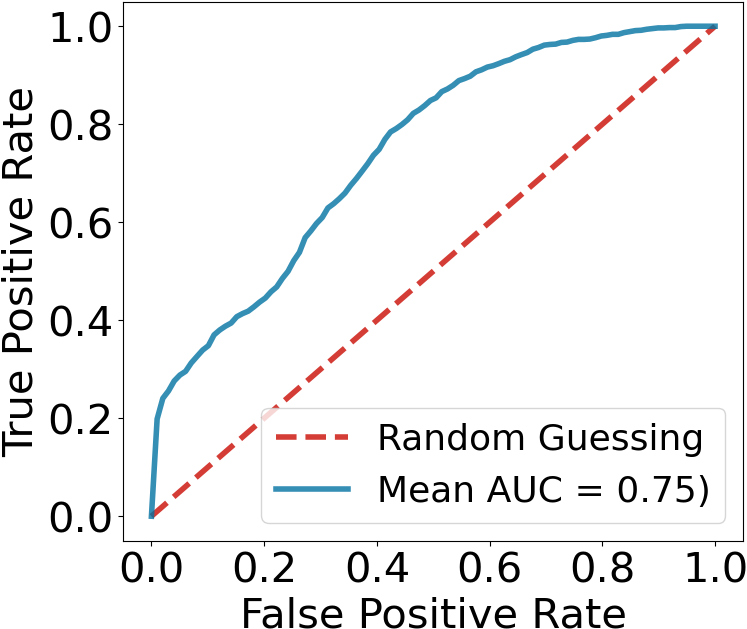}
        \vspace{-6mm}
        \caption{}\label{fig:FormStabOriginalFI_StabROC}
    \end{subfigure}
    \hfill
    \begin{subfigure}{0.24\textwidth}
        \centering
        \includegraphics[width=\linewidth]{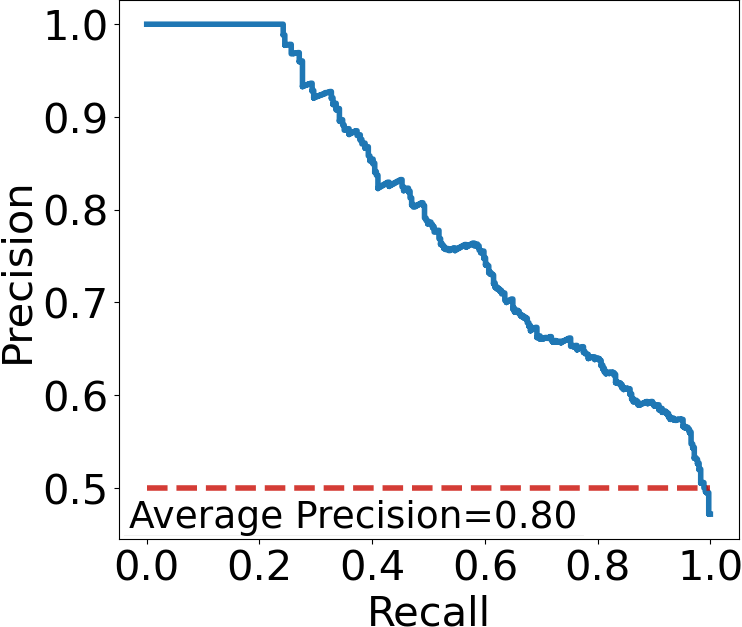}
        \vspace{-6mm}
        \caption{}\label{fig:FormStabOriginalFI_StabPRC}
    \end{subfigure}
    \vspace{-3mm}
    \caption{ (a) ROC and (b) PRC for formability classifier and respectively (c) and (d) for stability classfier using perovskite--specific features. A perfect classifier has an AUC of 1.0 for both metrics. The baseline classifier (or a random guess) is denoted by the dashed red line.}
    \label{fig:}
\end{figure}

% \vspace{-10mm}
The test and overall (training+testing) performances of the multi--output classifier for formability are shown as heatmaps of the confusion matrix in Figs. \ref{fig:FormStabOriginalFI_FormCMTest} and \ref{fig:FormStabOriginalFI_FormCMOverall}. Similarly the test and overall performance of the classifier toward stability is shown in Figs. \ref{fig:FormStabOriginalFI_StabCMTest} and \ref{fig:FormStabOriginalFI_StabCMOverall}. Using the perovskite--specific feature set, the classifier was able to predict formability with accuracies of 93.46\% (test) and 95.36\% (overall) and stability with accuracies of 78.20\% (test) and 82.67\% (overall). The Receiver Operating Characteristics (ROC)  curves for the classifier also show an Area Under the Curve (AUC) of 0.95 and 0.75 (Figs. \ref{fig:FormStabOriginalFI_FormROC}, \ref{fig:FormStabOriginalFI_StabROC}) towards formability and stability respectively against the perfect classification metric of 1.00 and and the Precision Recall Characteristics (PRC) curves have an average precision of 0.94 and 0.80 (Figs. \ref{fig:FormStabOriginalFI_FormPRC}, \ref{fig:FormStabOriginalFI_StabPRC}) respectively towards formability and stability respectively.
%%%
%%%%ROC & PRC CURVES
%%%
\subsection{Using novel generic features}
%%%
\begin{figure}[!h]%tbp]
    \centering
    \begin{subfigure}{0.24\textwidth}
        \centering
        \includegraphics[width=\linewidth]{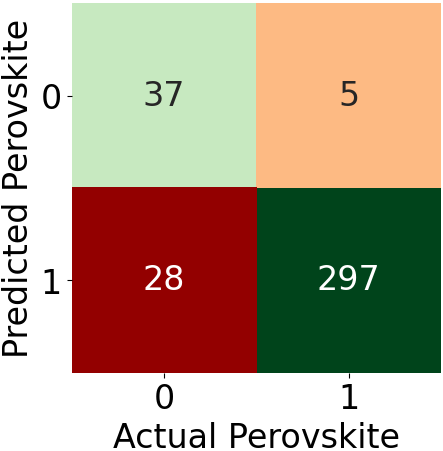}
        %\vspace{-6mm}
        \caption{}\label{fig:FormStabNewFI_FormCMTest}
    \end{subfigure}
    \hfill
    \begin{subfigure}{0.24\textwidth}
        \centering
        \includegraphics[width=\linewidth]{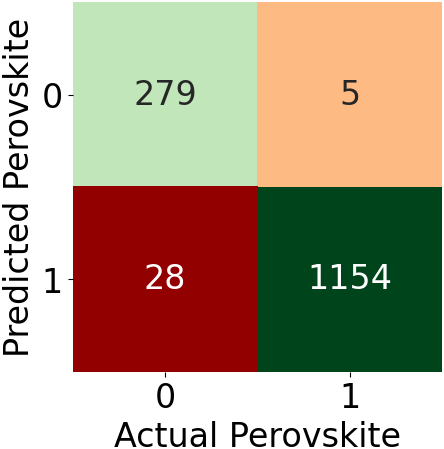}
        %\vspace{-6mm}
        \caption{}\label{fig:FormStabNewFI_FormCMOverall}
    \end{subfigure}
    \hfill
    \begin{subfigure}{0.24\textwidth}
        \centering
        \includegraphics[width=\linewidth]{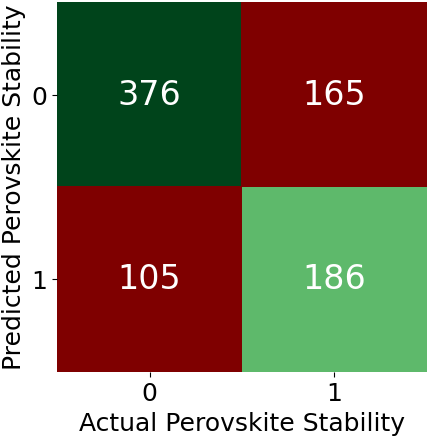}
        %\vspace{-6mm}
        \caption{}\label{fig:FormStabNewFI_StabCMTest}
    \end{subfigure}
    \begin{subfigure}{0.24\textwidth}
        \centering
        \includegraphics[width=\linewidth]{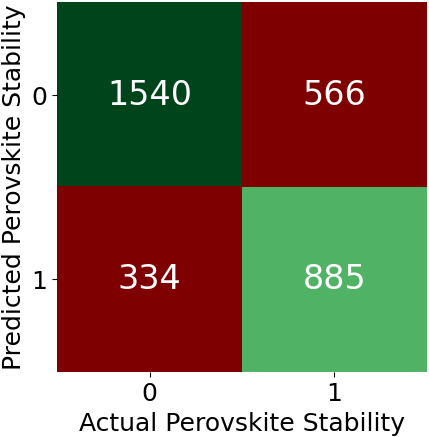}
        %\vspace{-6mm}
        \caption{}\label{fig:FormStabNewFI_StabCMOverall}
    \end{subfigure}
    %\vspace{-3mm}
    \caption{Heatmap of the confusion matrix for (a) test predictions and (b) overall predictions of formability and (c) test predictions and (d) overall predictions of stability using novel generic features. The diagonal entries (green) depict the correctly classified categories and the off--diagonal entries (red) depict the wrongly classified categories.}
    \label{fig:}
\end{figure}
\begin{figure}[!h]%tbp]
    \centering
    \begin{subfigure}{0.24\textwidth}
        \centering
        \includegraphics[width=\linewidth]{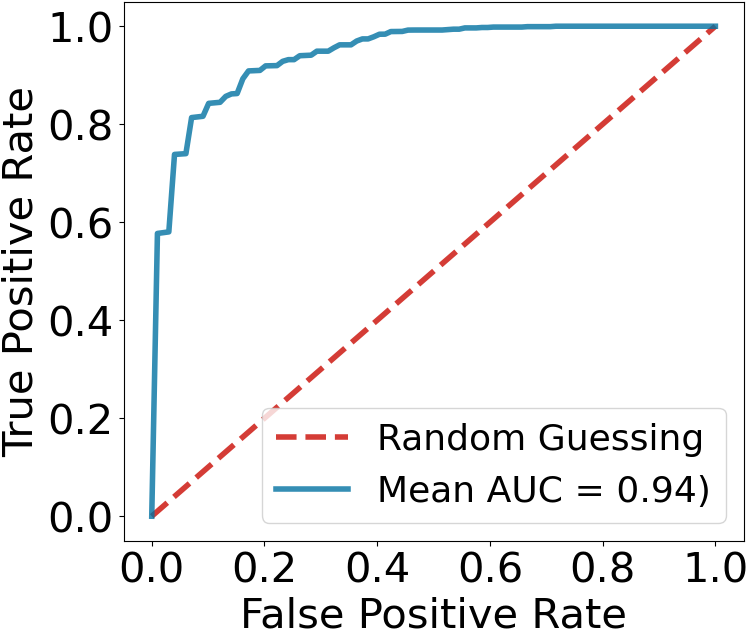}
        \vspace{-6mm}
        \caption{}\label{fig:FormStabNewFI_FormROC}
    \end{subfigure}
    \hfill
    \begin{subfigure}{0.24\textwidth}
        \centering
        \includegraphics[width=\linewidth]{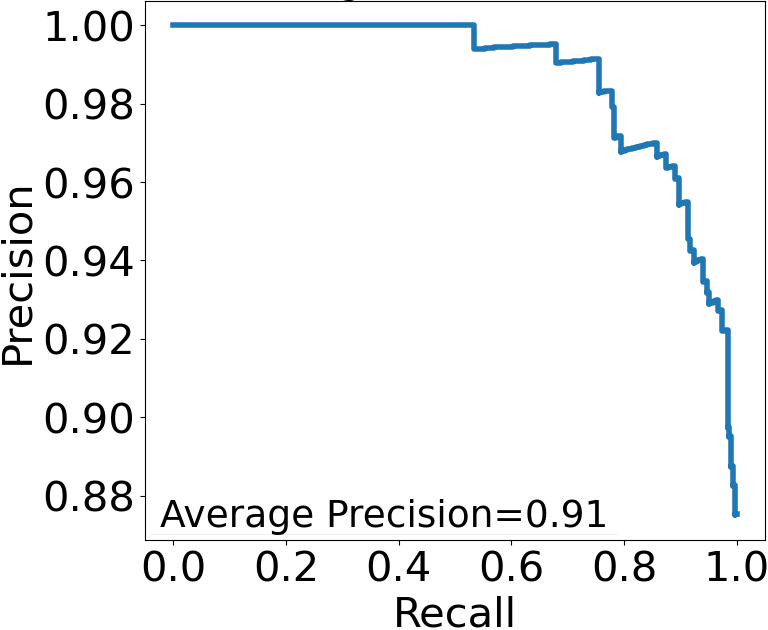}
        \vspace{-6mm}
        \caption{}\label{fig:FormStabNewFI_FormPRC}
    \end{subfigure}
    \hfill
    \begin{subfigure}{0.24\textwidth}
        \centering
        \includegraphics[width=\linewidth]{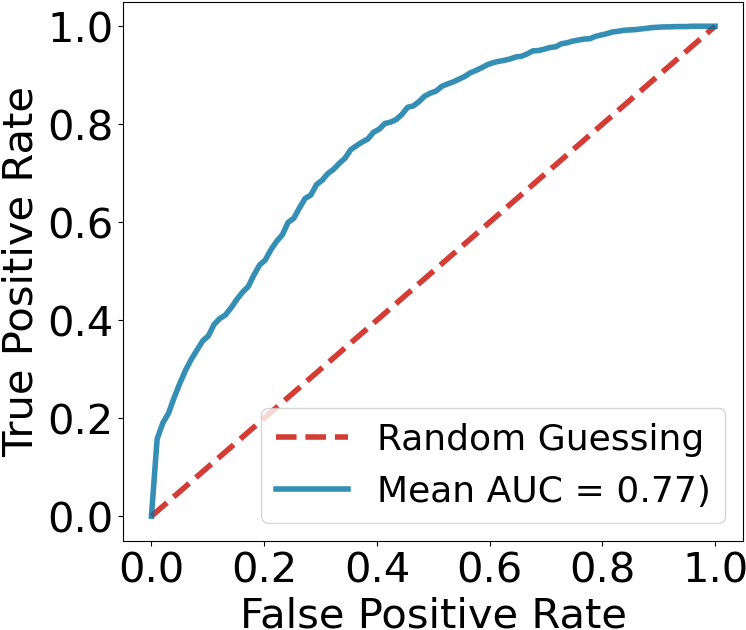}
        \vspace{-6mm}
        \caption{}\label{fig:FormStabNewFI_StabROC}
    \end{subfigure}
    \hfill
    \begin{subfigure}{0.24\textwidth}
        \centering
        \includegraphics[width=\linewidth]{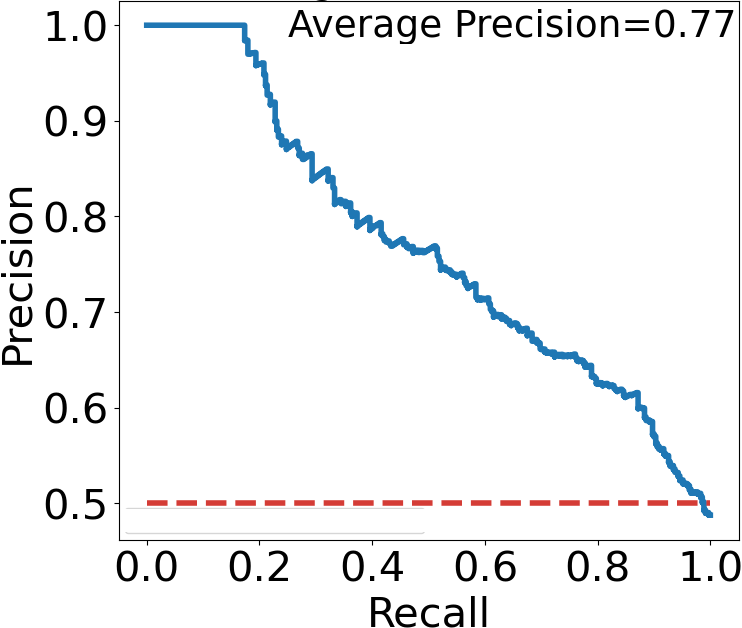}
        \vspace{-6mm}
        \caption{}\label{fig:FormStabNewFI_StabPRC}
    \end{subfigure}
    %\vspace{-3mm}
    \caption{ (a) ROC and (b) PRC for formability classifier and respectively (c) and (d) for stability classfier using novel generic features. A perfect classifier has an AUC of 1.0 for both metrics. The baseline classifier (or a random guess) is denoted by the dashed red line.}
    \label{fig:}
\end{figure}

The test and overall (training+testing) performances of the multi--output classifier as heatmaps of the confusion matrix are shown in Fig. \ref{fig:FormStabNewFI_FormCMTest} and Fig. \ref{fig:FormStabNewFI_FormCMOverall}, towards formability and in Figs. \ref{fig:FormStabNewFI_StabCMTest} and \ref{fig:FormStabNewFI_StabCMOverall}  towards stability respectively. The classifier was able to predict formability with accuracies of 89.81\% (test) and 91.13\% (overall) and stability with accuracies of 78.75\% (test) and 81.58\% (overall) using the generic feature set. The ROC curves for the classifier show AUC as 0.94 and 0.77 (Figs. \ref{fig:FormStabNewFI_FormROC}, \ref{fig:FormStabNewFI_StabROC}) towards formability and stability respectively and the PRC curves have an average precision of 0.91 and 0.77 (Figs. \ref{fig:FormStabNewFI_FormPRC}, \ref{fig:FormStabNewFI_StabPRC}) towards formability and stability respectively.
%\clearpage
%%%
%%%
\subsection{\label{sub_sec:pv_pec}Oxide perovskites for photovoltaic and photocatalytic applications}
\begin{figure}[!h]%tbp]
    \centering
    \includegraphics[width=0.63\linewidth]{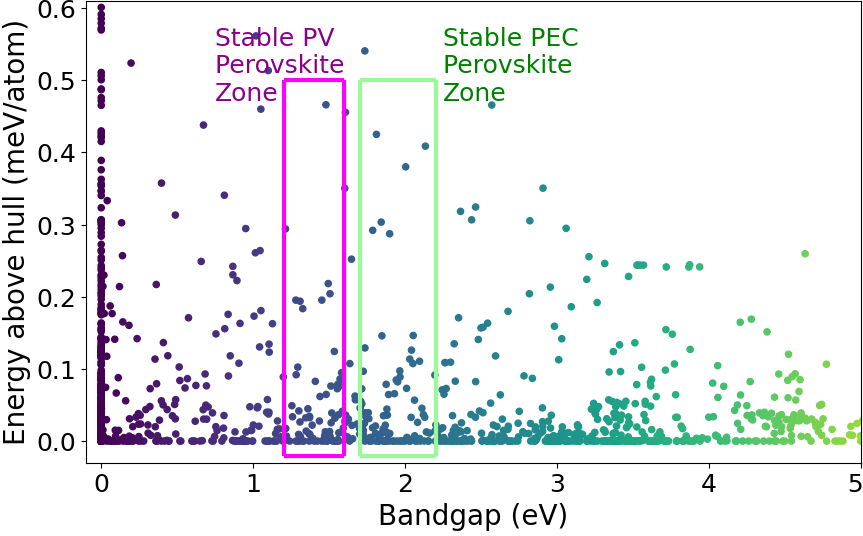}
    \caption{Stable oxide perovksites for (a) photovoltaic (magenta envelope) and (b) photocatalytic applications (green envelope).}
    \label{fig:PerovskitePhotovoltaicRange}
\end{figure}

The multioutput classifier was used to augment the formability of the oxide compounds in overall dataset. The set of all perovskite oxides that are stable were then extracted and their band gap data collated from the Materials Project database. The stable perovskite compositions were screened for the optimal range of bandgap 1.2--1.6 eV for photo--voltaic (PV) \cite{pu2021} and 1.7--2.2 eV for photo--electro--chemical (PEC) \cite{montoya2017} applications. The stable PV and PEC zones for the oxide pervoksites considered in this study are demarcated on the plot in Fig. \ref{fig:PerovskitePhotovoltaicRange}. Some of the stable perovskites identified for PV and PEC applications are listed in Table \ref{tab:perovapplntable}.
%%%
%%%
\begin{table}[!htbp]
    \centering
    \caption{Perovskites for PV and PEC applications}
    % \begin{adjustbox}{width=0.85\textwidth}
    \begin{tabular}{p{0.1\linewidth}p{0.1\linewidth}|p{0.1\linewidth}p{0.1\linewidth}} %{llr}
        \toprule
            \multicolumn{1}{c}{PV Perovskite} & \multicolumn{1}{c|}{Bandgap (eV)} & \multicolumn{1}{c}{PEC Perovskite} & \multicolumn{1}{c}{Bandgap (eV)}\\
        \midrule
            \ce{BaNiO3} & 1.5467 & \ce{Sr2MnWO6} & 1.9449 \\
            \ce{PrVO3} & 1.3016 & \ce{Sr2FeWO6} & 1.9746 \\
            \ce{PrMnO3} & 1.4020 & \ce{LiSbO3} & 2.1517 \\
            \ce{SmMnO3} & 1.4671 & \ce{Sr2FeSbO6} & 1.8090 \\    
        \bottomrule
    \end{tabular}
    % \end{adjustbox}
    \label{tab:perovapplntable} 
\end{table}
%%%
\section{Conclusions}\label{sec:conclusion}
The addition of novel elemental and thermodynamic features enabled the evaluation of feature importances of the perovskite--specific features in the database reported by Pilania and co--workers \cite{talapatra2021}. Using hierarchical clustering of features based on their Pearson's correlation coefficients and plotting the tree and permutation feature importances, we are able to visualise -- using partial--dependency plots -- the most important features that contribute to the formability and stability of single-- and double--oxide perovskites. 
\par In the case of the perovskite--specific features, the tolerance factor $t$ and the mismatch factor $\mu B$ initially appear to be the most significant factors influencing the formability and stability of oxide--perovskites. Upon hierarchical clustering of features, it becomes apparent that the energies of the Lowest--Unoccupied and the Highest--Occupied Molecular Orbitals ($LUMO^{B-}$, $HOMO^{A-}$, $LUMO^{B+}$, $LUMO^{A-}$) in addition to the Zunger's pseudopotential radius ($rad^{B-}_Z$), oxidation state (\textit{B\_OxidState}) of the B--atom and the electon affinity ($E^{A-}_{ea}$), electronegativity ($\chi^{A-}$) of the A--atom are important underlying perovskite--specific factors. In the case of the novel generic features, the $\mathit{\Omega}$ parameter and electronegativity $\chi$ initially appear to be the most important features. Subsequent to the reduction in multi--collinearity by hierarchical clustering the RSSD value of lattice parameter, $\Delta a$ and the RoM values of the lattice parameter $a$\textit{\_RoM}, radius $r$\textit{\_RoM} and density $\rho$\textit{\_RoM} are seen to exert an influence on the formability and stability of oxide--perovskites.
\par We are also able to identify the influence of different classes of features on the formability and stability of oxide perovskites and also quantify the accuracy of predictions for each class. RF classifiers were able to predict the formability of single-- and double-- oxide perovskite--type compositions with testing accuracies of 93.46\% and 89.81\% using perovskite--specific and generic features respectively. The testing accuracies towards predictions of stability were 78.20\% and 78.78\% using perovskite--specific and generic features respectively. 
\par These results indicate that a generic bank of features can perform as well as a specific set of features in predicting important perovskite parameters. This would translate into easier transferability of learnt structure--property relations to classes of materials other than oxide--perovskites. Furthermore, relying upon online repository data for ab--initio bandgap values, stable oxide perovskite compositions were identified for solar cell and photocatalytic based on optimal bandgap criteria for PV and PEC applications.
\par In order to improve the classifier accuracies and characteristics of classifier performance towards prediction of stability of perovskite compositions, the energy--above--convex--hull could be generated using ML regressors and the stability established based on experimentally reported criterion. It would also necessitate the augmentation of data to the existing database to capture more meaningful dependencies for perovskite formation and stability. Furthermore, the method for extracting structure--property linkages and prediction of formability and stability could be extended to include oxide--, halide-- and other perovskite structure compositions to identify stable application--specific compositions.
%%%
\section*{Declaration of interests}
The authors have no relevant financial or non-financial interests to disclose.
%%%
\section*{Data Statement}
The datasets generated during and/or analysed during the current study are available from the authors on reasonable request.
%%%
\bibliography{sn-bibliography.bib}% common bib file

\pagebreak
\begin{appendices}
%%%
\section{Dendograms for hierarchical clustering of features}\label{secA1}
%%%
\subsection{\label{sub_sec:clusteringFormability}Hierarchical clustering of features towards formability}
%%%FEATURE IMPORTANCE PLOTS
\begin{figure}[!h]%tbp]
    \centering
    \begin{subfigure}{0.9\textwidth}
        \centering
        \includegraphics[width=\linewidth]{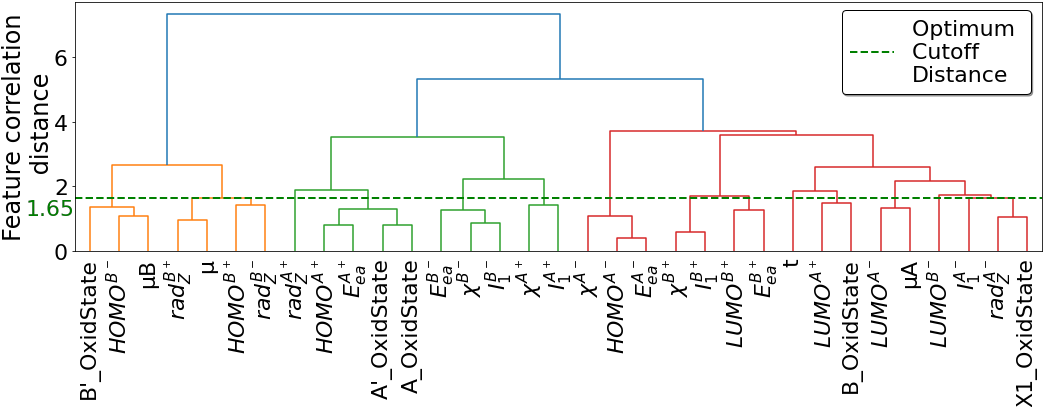}
        \vspace{-5mm}
        \caption{}\label{fig:FormOriginalFI_dendro}
    \end{subfigure}
    \centering
    \begin{subfigure}{0.9\textwidth}
        \centering
        \includegraphics[width=\linewidth]{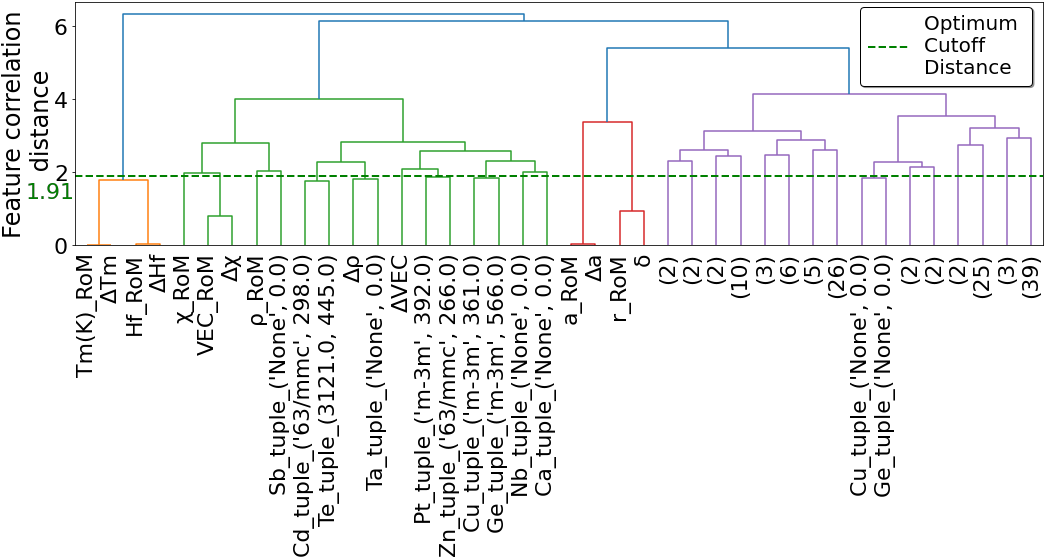}
        \vspace{-7mm}
        \caption{}\label{fig:FormNewFI_dendro}
    \end{subfigure}
    \vspace{-3mm}
    \caption{Hierarchical clustering dendrogram for Spearman rank--order correlations towards prediction of formability using (a) perovskite-specific features and (b) novel generic features. The optimal cutoff distances of 1.65 and 1.91 for maximum accuracy of predictions of formability for each feature set, is depicted by the dashed green lines in the dendrograms. Each cluster of features below these lines are replaced by the feature with the highest feature importance score among that cluster. Only the features that survive this hierarchical clustering are then considered for subsequent TFI and PFI scores with reduced collinearity among features.}
    \label{fig:}
\end{figure}
\clearpage
%%%
\vspace{-\topsep}
\subsection{\label{sub_sec:clusteringStability}Hierarchical clustering of features towards stability}
%%%
\begin{figure}[!h]%tbp]
    \centering
    \begin{subfigure}{0.9\textwidth}
        \centering
        \includegraphics[width=\linewidth]{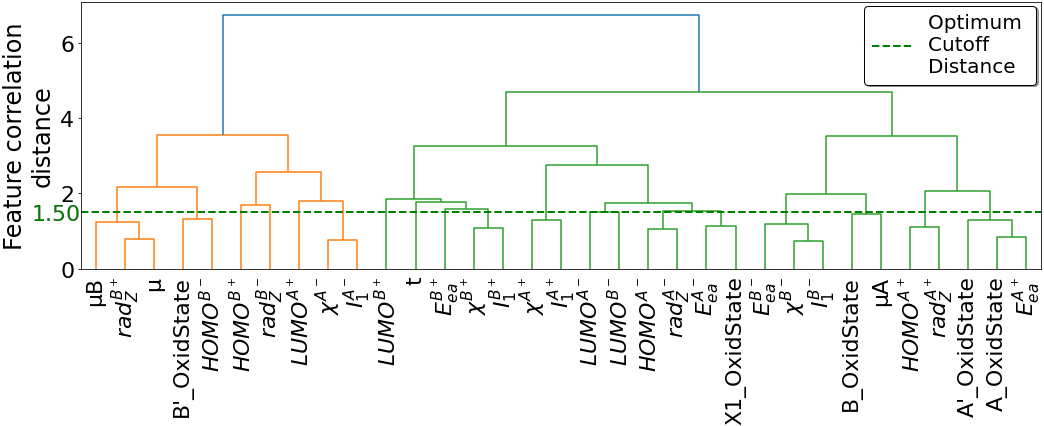}
        \vspace{-7mm}
        \caption{}\label{fig:StabOriginalFI_dendro}
    \end{subfigure}
    \centering
    \begin{subfigure}{0.9\textwidth}
        \centering
        \includegraphics[width=\linewidth]{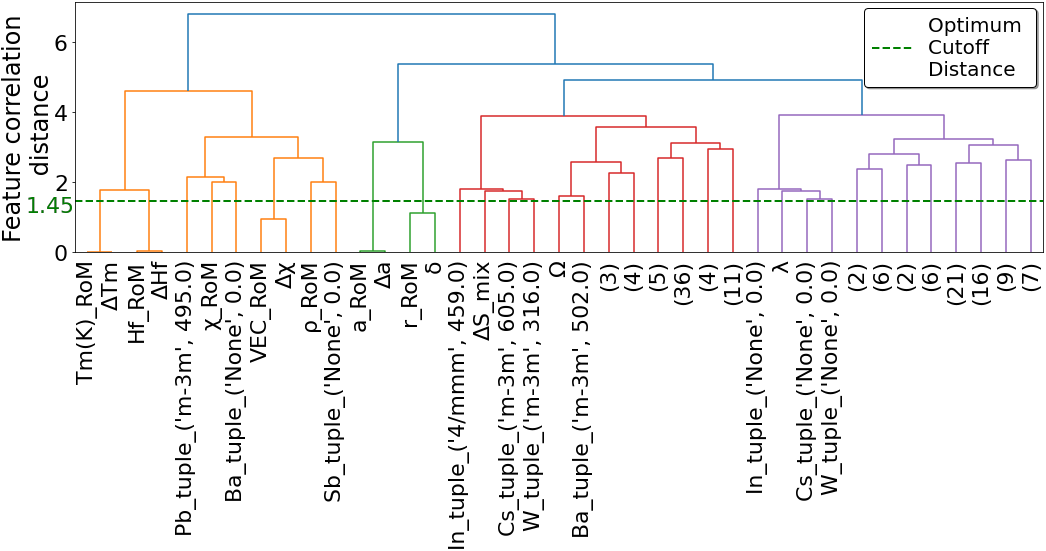}
        \vspace{-7mm}
        \caption{}\label{fig:StabNewFI_dendro}
    \end{subfigure}
    \caption{Hierarchical clustering dendrogram for Spearman rank--order correlations towards prediction of stability using (a) perovskite--specific features and (b) novel generic features. The optimal cutoff distances of 1.50 and 1.45 for maximum accuracy of predictions of stability for each feature set, is depicted by the dashed green lines in the dendrograms. Each cluster of features below these lines are replaced by the feature with the highest feature importance score among that cluster. Only the features that survive this hierarchical clustering are then considered for subsequent TFI and PFI scores with reduced collinearity among features.}
    \vspace{-5mm}
    \label{fig:}
\end{figure}
\clearpage
%%%
\subsection{\label{sub_sec:clusteringFormStab} Hierarchical clustering of features towards both formability and stability simultaneously}
%%%
\begin{figure}[!h]%tbp]
    \centering
    \begin{subfigure}{0.9\textwidth}
        \centering
        \includegraphics[width=\linewidth]{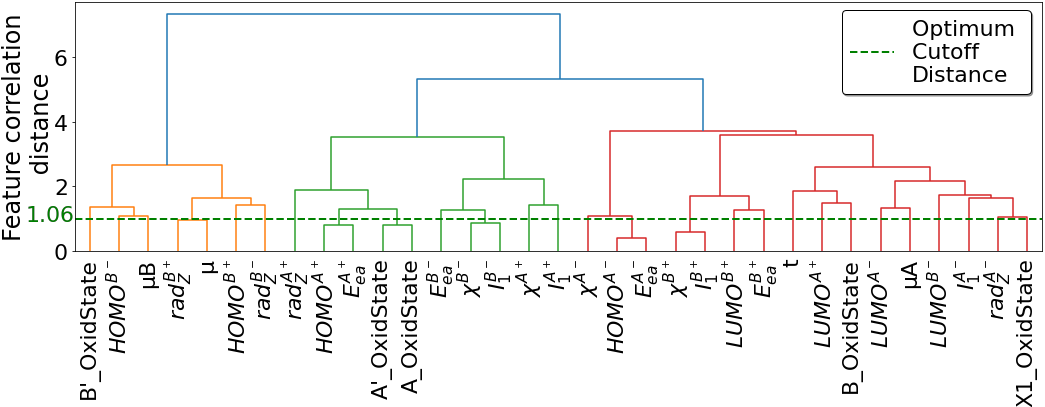}
        \vspace{-7mm}
        \caption{}\label{fig:FormStabOriginalFI_dendro}
    \end{subfigure}
    \centering
    \begin{subfigure}{0.9\textwidth}
        \centering
        \includegraphics[width=\linewidth]{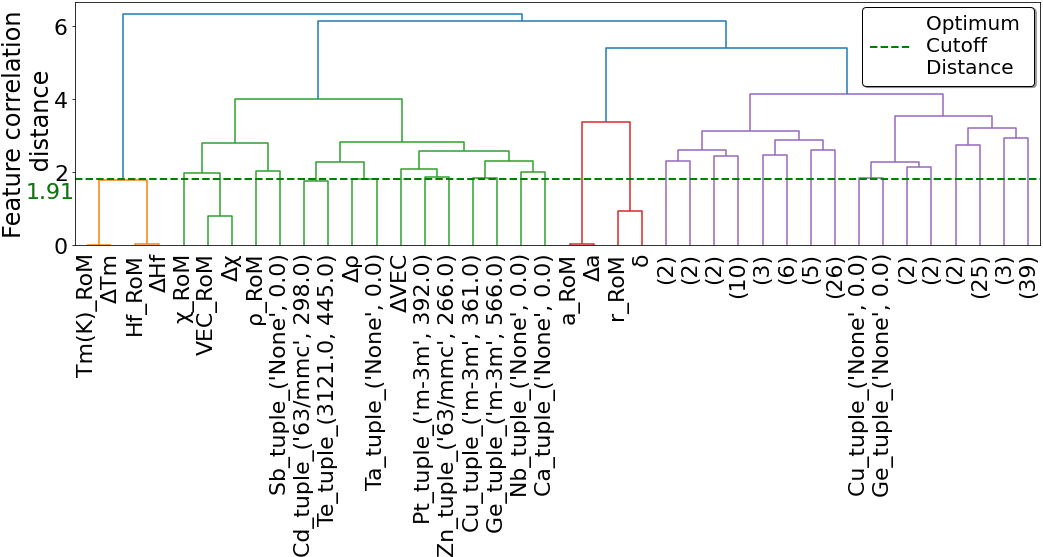}
        \vspace{-7mm}
        \caption{}\label{fig:FormStabNewFI_dendro}
    \end{subfigure}
    \caption{Hierarchical clustering dendrogram for Spearman rank--order correlations towards prediction of formability and stability simultaneously using (a) perovskite--specific features and (b) novel generic features. The optimal cutoff distances of 1.06 and 1.91 for maximum accuracy of predictions of both formability and stability for each feature set, is depicted by the dashed green lines in the dendrograms. Each cluster of features below these lines are replaced by the feature with the highest feature importance score among that cluster. Only the features that survive this hierarchical clustering are then considered for subsequent TFI and PFI scores with reduced collinearity among features.}
    %\vspace{-5mm}
    \label{fig:}
\end{figure}
%%%
\end{appendices}
%%%
\end{document}